\documentclass[aos]{imsart}

\usepackage[colorlinks,citecolor=blue,urlcolor=blue]{hyperref}
\usepackage{titletoc}
\usepackage{apacite}
\usepackage{amsfonts}
\usepackage{amsgen,amsmath,amstext,amsbsy,amsopn,amssymb,subfigure, amsthm}
\usepackage[dvips]{graphicx}
\usepackage[authoryear]{natbib}
\usepackage{blkarray}
\usepackage{algorithm}
\usepackage{algpseudocode}
\usepackage{url}
\usepackage{longtable}
\usepackage{mleftright}
\usepackage[normalem]{ulem}

\usepackage{multirow}
\usepackage[usenames]{color}

\theoremstyle{plain}

\newtheorem{Theorem}{Theorem} \newtheorem*{Theorem*}{Theorem} \newtheorem{Lemma}{Lemma} \newtheorem{Remark}{Remark} \newtheorem{Corollary}{Corollary} \newtheorem{Assumption}{Assumption} \newtheorem*{Assumption*}{Assumption} \newtheorem{Proposition}{Proposition}

\theoremstyle{definition}

\newcommand{\be}{\begin{equation}}
	\newcommand{\ee}{\end{equation}}
 \newcommand{\bs}{\begin{split}}
	\newcommand{\es}{\end{split}}
\newcommand{\bea}{\begin{eqnarray}}
	\newcommand{\eea}{\end{eqnarray}}
\newcommand{\beas}{\begin{eqnarray*}}
	\newcommand{\eeas}{\end{eqnarray*}}

\newcommand{\T}{^{\top}}

\newcommand{\EE}{\mathbb{E}}

\renewcommand{\r}{{\boldsymbol{r}}}

\DeclareMathAlphabet\mathbfcal{OMS}{cmsy}{b}{n}

\renewcommand{\i}{^{-1}}

\newcommand{\X}{{\mathbf{X}}}

\newcommand{\E}{{\mathbf{E}}}

\renewcommand{\S}{{\mathbf{S}}}

\newcommand{\Y}{{\mathbf{Y}}}
\newcommand{\Z}{{\mathbf{Z}}}

\renewcommand{\(}{\left(}
\renewcommand{\)}{\right)}

\renewcommand{\ln}{\left\|}
\newcommand{\rn}{\right\|}

\newcommand{\diag}{{\rm diag}}

\newcommand{\SVD}{{\rm SVD}}

\newcommand{\argmin}{\mathop{\rm arg\min}}

\newcommand{\RR}{\mathbb{R}}
\newcommand{\PP}{\mathbb{P}}
\newcommand{\MM}{\mathcal{M}}

\allowdisplaybreaks

\startlocaldefs
\theoremstyle{plain}    
\theoremstyle{definition}   

\endlocaldefs

\begin{document}

\begin{frontmatter}
    \title{Revisit CP Tensor Decomposition: Statistical Optimality and Fast Convergence}
    \runtitle{CP Tensor Decomposition}

    \begin{aug}
        \author[A]{\fnms{Runshi}~\snm{Tang} \ead[label=e1]{rt3075@columbia.edu}},
        \author[B]{\fnms{Julien}~\snm{Chhor}\ead[label=e2]{julien.chhor@tse-fr.eu}}, 
        \author[C]{\fnms{Olga}~\snm{Klopp}\ead[label=e3]{Olga.KLOPP@math.cnrs.fr}},
        \and
        \author[D]{\fnms{Anru R.}~\snm{Zhang}\orcid{0000-0002-8721-5252}
        	\ead[label=e4]{anru.zhang@duke.edu}}
        \address[A]{Department of Statistics, Columbia University \printead[presep={ ,\ }]{e1}}

        \address[B]{Toulouse School of Economics, University of Toulouse Capitole, Toulouse, France \printead[presep={,\ }]{e2}}

        \address[C]{ESSEC Business School, Cergy France \printead[presep={,\ }]{e3}}

        \address[D]{Department of Biostatistics \& Bioinformatics and Department of Computer Science, Duke University \printead[presep={,\ }]{e4}}
    \end{aug}

    \begin{abstract}
        Canonical Polyadic (CP) tensor decomposition is a fundamental technique for analyzing high-dimensional tensor data.
        While the Alternating Least Squares (ALS) algorithm is widely used for computing CP decomposition due to its simplicity and empirical success, its theoretical foundation, particularly regarding statistical optimality and convergence behavior, remains underdeveloped, especially in noisy, non-orthogonal, and higher-rank settings.

        In this work, we revisit CP tensor decomposition from a statistical perspective and provide a comprehensive theoretical analysis of ALS under a signal-plus-noise model.
        We establish non-asymptotic, minimax-optimal error bounds for tensors of general order, dimensions, and rank, assuming suitable initialization.
        To enable such initialization, we propose Tucker-based Approximation with Simultaneous Diagonalization (TASD), a robust method that improves stability and accuracy in noisy regimes.
        Combined with ALS, TASD yields a statistically consistent estimator.
        We analyze the convergence dynamics of ALS, showing that for general rank it converges quadratically above a coherence-dependent transition scale and, when further refinement is required below that scale, subsequently converges linearly.
        In the rank-one setting, ALS with an appropriately chosen initialization attains optimal error within just one or two iterations.
    \end{abstract}



\end{frontmatter}

\section{Introduction}\label{sec_intro}

Tensor data analysis has emerged as a rapidly advancing field with wide-ranging applications in physics, signal processing, and deep learning.
Due to the inherently high dimensionality and complex structure of tensors, dimensionality reduction is both a practical necessity and a theoretical challenge.
One widely used strategy, inspired by matrix analysis, is low-rank approximation and decomposition, which aims to capture the core structure of tensor data while significantly reducing computational and storage demands.
Tensor rank is a central concept in tensor decomposition and can be defined in several ways.
In this work, we focus on the Canonical Polyadic (CP) rank.
Formally, the CP rank of a tensor $\X \in \RR^{p_1\times \dots \times p_d}$ is the smallest non-negative integer $R$ such that $\X$ can be written as
\be
\label{eq_cpd} \X = \sum_{r = 1}^R \lambda_r a_{1,r} \circ \cdots \circ a_{d,r},
\ee
where $a_{k,r}\in\RR^{p_k}$ are {\it loading vectors} with unit Euclidean norm, $\lambda_r\in\RR$ are scalar coefficients, and $\circ$ denotes the tensor outer product. 
The expression \eqref{eq_cpd} is referred to as a CP decomposition (CPD) of $\X$.
CPD is a classical tensor decomposition method dating back to \cite{Hitchcock1927}, \cite{Hitchcock1927_2}.
A key advantage of CP decomposition is its uniqueness under mild conditions \citep{kruskal1977three}, setting it apart from matrix factorizations and other tensor decomposition approaches, which generally lack this property.
By decomposing a tensor into rank-one components, CP decomposition enables the discovery of latent factors and hidden patterns or structures in multiple-axes data analysis across many fields, including chemometrics~\citep{bro2009modeling}, psychometrics~\citep{carroll1970analysis}, signal processing~\citep{qian2018tensor}, neuroscience~\citep{zhang2023cocaine}, computer vision~\citep{acar2011all}, sequencing data analysis~\citep{shi2024tempted}, moment tensor analysis \citep{anandkumar2014tensor}, and density estimation \citep{zheng2020nonparametric}.

In practice, due to measurement errors, the observed tensor is rarely exactly of low CP rank. Motivated by this, we consider the following signal-plus-noise model for tensor CPD:
\be
\label{eq_signal_plus_noise}
\Y=\X+\Z\in\RR^{p_1\times\cdots\times p_d},
\ee
where $\Z$ is a centered random noise tensor, and the signal tensor $\X$ has CP rank $R$ and admits a decomposition of the form \eqref{eq_cpd}. Given $\Y$, our goal is to estimate the magnitudes $\{|\lambda_r|\}_{r=1}^R$, the loading vectors $\{a_{k,r}\}_{k=1,\ldots,d;\,r=1,\ldots,R}$ up to the usual componentwise sign and common-permutation indeterminacies, and the signal tensor $\X$.

A widely adopted approach for computing CPD is the Alternating Least Squares (ALS) algorithm, which has demonstrated strong empirical performance \citep{harshman1970foundations, carroll1970analysis, koldaTensorDecompositionsApplications2009}.
ALS proceeds by iteratively updating the loading vectors: at the $(t+1)$st iteration, for a given mode $k$, it fixes $\{\hat a_{i,r}^{(t)}: i\in[d]\setminus\{k\},\,r\in[R]\}$ from the previous iteration and solves the following least-squares problem to update the mode-$k$ loading vectors:
$$
    (\hat b_{k,1}^{(t+1)}, \cdots, \hat b_{k,R}^{(t+1)}) = \argmin_{b_{k,r}, r\in [R]}\ln\Y-\sum_{r = 1}^R \hat a_{1,r}^{(t)} \circ
    \cdots \circ b_{k,r} \circ \cdots \circ \hat a_{d,r}^{(t)}\rn,\ \text{and}\ \hat a_{k,r}^{(t+1)} = \frac{\hat
        b_{k,r}^{(t+1)}} {\|\hat b_{k,r}^{(t+1)}\|}.
$$
The loss surface is known to be highly non-convex with exponentially many local minima \citep{arous2019landscape} when the signal to noise ratio is not sufficiently strong, making convergence to the true loading vectors theoretically uncertain.

Despite its widespread adoption, the ALS algorithm is still poorly understood from a theoretical perspective.
In particular, three fundamental questions remain largely unresolved:

\vskip.05cm
\noindent \fbox{ \parbox{0.98\textwidth}{
        {\it 1.
                Does ALS have statistical guarantees?
            }
    }}
\vskip.05cm

Prior works have provided theoretical insights under simplified assumptions.
For example, when $R=1$, the problem reduces to Tensor PCA, which has been analyzed from a statistical perspective \citep{richard2014statistical, anandkumar2017homotopy, huang2022power, wu2024sharp}.
Other studies focus on orthogonal tensors with $d=3$~(\citealp{anandkumar2014tensor}; \citealp{anandkumar2014guaranteed}), where loading vectors across each mode are assumed orthogonal or nearly orthogonal.
These approaches often estimate loading vectors sequentially by leveraging guarantees in the rank-1 setting.
However, this sequential estimation strategy can fail in the general (non-orthogonal) case, where the sum of the first $k \leq R$ rank-one CP components in \eqref{eq_cpd} does not necessarily yield the best rank-$k$ approximation \citep{kolda2001orthogonal, stegemanSubtractingBestRank12010}.
Thus, the theoretical guarantees derived for special cases do not extend to the general CPD setting or to ALS in its full generality.

\vskip.05cm
\noindent \fbox{ \parbox{0.98\textwidth}{
        {\it 2.
                How can we effectively initialize ALS to ensure it converges to the desired estimator?
            }
    }}
\vskip.05cm

Due to the existence of local minima, initialization is crucial for the success of ALS.
When $R = 1$, methods like Higher-Order Singular Value Decomposition (HOSVD) can serve as consistent estimators and are effective initializations \citep{zhang2018tensor, richard2014statistical}.
However, to the best of our knowledge, there are currently no initialization methods with statistical guarantees for general rank $R$.
The existence of local minima highlights the need for a robust warm-start strategy to ensure convergence to meaningful solutions.

Classical algorithms, such as the well-known Simultaneous Diagonalization (SimDiag) method or Jennrich’s algorithm, are capable of exactly recovering the loading vectors in the absence of noise \citep{leurgans1993decomposition, domanov2014canonical}.
Yet, they are known to be highly sensitive to noise \citep{beltranPencilbasedAlgorithmsTensor2019}, which severely limits their utility in practice as initialization procedures.
Therefore, developing a robust, noise-tolerant initialization method for ALS remains an open and important problem.

\vskip.05cm
\noindent \fbox{ \parbox{0.98\textwidth}{
        {\it 3.
                Why does ALS converge fast, especially when the tensor is near orthogonal?
            }
    }}
\vskip.05cm

Empirically, ALS often converges in a small number of iterations — especially when the underlying tensor has near-orthogonal structure \citep{sharan2017orthogonalized}.
This is also evident in our simulation study.
In the rank-one case (Supplement Figure \ref{fig_3}), an ``all or nothing'' phenomenon is observed: ALS typically converges within one or two iterations -- if it converges at all.
In contrast, in the general-rank case (Supplement Figure \ref{fig_3_1}), the high coherence of the loading matrix significantly impedes ALS convergence.
Recent studies have attempted to characterize the fast convergence for the rank-one setting.
For example, \cite{huang2022power} showed that ALS reaches optimal error bounds in $\log p$ iterations, while \cite{wu2024sharp} demonstrated $\log\log p$ convergence to any given constant.

However, these analyses do not extend to general $R$.
It remains unclear whether ALS enjoys linear or even quadratic convergence in higher-rank settings and under what structural assumptions such fast convergence arises.

\subsection{Our Contributions}\label{sec_contributions}

This paper makes several key contributions to the theoretical understanding of CPD and the development of robust initialization strategies for CP tensor decomposition in noisy settings:

\medskip
\noindent{\bf Local Convergence Guarantee for ALS with Optimal Error Bound.}
We develop a general theoretical framework establishing the local statistical convergence of ALS.
Specifically, after a common relabeling of the components, we derive non-asymptotic error bounds for the sign-invariant loading-vector estimation error $\min_{s\in\{-1,1\}}\|\hat a_{k,r}^{(t)}-s a_{k,r}\|$ that apply to arbitrary tensor order $d$, mode dimensions $p_k$, and CP rank $R$, assuming a suitable initialization.
We further show that this bound is minimax optimal by constructing a matching lower bound.

\medskip
\noindent{\bf TASD: A Robust Initialization Method.}
To address the instability of classical pencil-based methods, we propose a hybrid initialization strategy—Tucker-based Approximation with Simultaneous Diagonalization (TASD).
Our method first applies Tucker decomposition to compress the tensor, followed by SimDiag applied to the low-dimensional core, significantly enhancing robustness to noise.
We analyze the statistical performance of TASD and show that, when used to initialize ALS, the resulting estimator (TASD-ALS) is statistically consistent.
The code of our method is publicly available at \url{https://github.com/RunshiTang/ALS-TASD-Fast-Reliable-CP-Decomposition}.

\medskip
\noindent{\bf Convergence Rate of ALS.}
We characterize the convergence behavior of ALS under proper initialization.
In the rank-one setting ($R = 1$), we prove that ALS achieves the optimal error bound within one or two iterations.
For general rank $R$, ALS converges quadratically above a coherence-dependent transition scale and, when further refinement is required below that scale, subsequently converges linearly.
These theoretical findings are corroborated by our simulation results.
This provides the first rigorous justification for the empirically observed fast convergence of ALS in near-orthogonal regimes.

\medskip
\noindent{\bf Technical Contributions.}
Our general-rank ALS analysis jointly controls all CP components through the Khatri--Rao representation of the updates, separating quadratic same-component errors from coherence-weighted cross-component interactions.
This avoids sequential rank-one extraction and allows non-orthogonal loading matrices.
We also establish noise bounds uniformly over the data-dependent ALS iterates.
For TASD, we extend HOOI perturbation theory to arbitrary tensor order and analyze randomized simultaneous diagonalization using small-ball and eigenvector perturbation arguments, thereby avoiding a separation condition on the coefficient magnitudes $|\lambda_r|$.

\medskip
To the best of our knowledge, this is the first work to provide comprehensive statistical guarantees for ALS under general CPD with sub-Gaussian noise.
Table~\ref{table_comparison} benchmarks our rank-one results against the most relevant prior work.
See Section~\ref{sec_rank_one} for further discussion.
In the general rank setting, we compare our full procedure with existing ALS variants both theoretically (Sections~\ref{sec_local} and~\ref{sec_overall}) and empirically (Section~\ref{sec_simulation}).
Across all comparisons, our analysis operates under weaker assumptions and yields sharper error bounds and faster convergence rate.

\begin{table}[!ht]
    \centering
    \begin{tabular}{l|p{2cm}|c|c|c}
        \hline
         & Initialization                             & Required $\lambda$           & Iterations       & Error                                                 \\ \hline\hline
        \multirow{2}{*}{\cite{richard2014statistical}}
         & Unfolding & $p^{\lceil d/2 \rceil /2}$ & NA               & \multirow{3}{1.5cm}{\centering $p^{1/2} / |\lambda|$} \\ \cline{2-4}
         &                                            & $p^{d/2}$                    & NA               &                                                       \\ \cline{1-1}\cline{3-4}
        \cite{huang2022power}
         & Random                                     & $p^{(d-1)/2 + \varepsilon}$      & $\log p$       &                                                       \\ \cline{1-1}\cline{3-5}
        \cite{wu2024sharp}
         &                                            & $p^{(d-1)/2}(\log p)^{-C}$       & $\log\log p$ & $\forall \delta>0$                                    \\ \cline{1-1}\cline{2-5}
        Our Theorem \ref{thm_overall_R1}
         & HOSVD \eqref{eq_initialization_formula_R1} & $p^{d/4}$                    & 1 or 2           & $p^{1/2} / |\lambda|$                                 \\
        \hline
    \end{tabular}
    \caption{Comparison of results for rank-one tensor decomposition.}
    \label{table_comparison}
\end{table}

\subsection{Related Literature}
Tensor decomposition has been extensively studied across statistics, machine learning, and signal processing.
While a comprehensive or exhaustive literature review is beyond the scope of this paper, we highlight several lines of research most relevant to our study, particularly those concerning Canonical Polyadic Decomposition (CPD), Alternating Least Squares (ALS), initialization strategies, and statistical modeling.

Various extensions of the ALS algorithm have been proposed to improve performance or accommodate structural constraints.
One such variant is CANDELINC~\citep{douglas1980candelinc}, which constrains the column space of one or more factor matrices.
\cite{sorber2015structured} introduced a generalized framework incorporating structured least-squares loss functions.
In~\cite{sharan2017orthogonalized}, a modified ALS procedure was proposed that orthogonalizes the factor matrices at each iteration; the authors proved global convergence in the order-3, near-orthogonal, noiseless setting.
Other practical adaptations include randomized methods for large-scale tensor decomposition~\citep{vervliet2015randomized, battaglino2018practical} and generalized CP frameworks (GCP) that support flexible loss functions, optimized via gradient-based methods~\citep{hong2020generalized, kolda2020stochastic}.

Effective initialization is crucial for avoiding poor local minima in ALS and many other gradient-based methods.
The semi-algebraic approach for approximate CPD, known as SECSI~\citep{roemerSemialgebraicFrameworkApproximate2013}, performs Tucker compression followed by simultaneous diagonalization, but lacks statistical guarantees.
SimDiag has also been analyzed under fixed-dimensional, low-noise regimes~\citep{kuleshov2015tensor}.
In the orthogonal setting, \cite{wang2017tensor} proposed collapsing a tensor into a matrix and applying the matrix power method to obtain an initial estimate.
Whitening-based techniques~\citep{souloumiac2009joint, le2011ica} aim to transform the problem into an approximately orthogonal form but often suffer from ill-conditioning in high dimensions~\citep{huang2013fast}.
Perturbation analyses have been developed to assess the robustness of various decompositions; see, e.g.,~\cite{auddy2023perturbation} for orthogonally decomposable tensors.

Although this paper focuses on the undercomplete regime ($R \leq p_k$), the overcomplete setting ($R > p_k$) has been widely studied.
Notable approaches include decomposition via fourth-order cumulants~\citep{de2007fourth}, sum-of-squares (SoS) methods~\citep{barak2015dictionary, ge2015decomposing, ma2016polynomial}, and hybrid techniques that combine spectral initialization with SoS optimization~(\citealp{hopkins2016fast}; \citealp{hopkins2019robust}).
Issues of identifiability and uniqueness in overcomplete settings have also been explored~\citep{koiran2025efficient}.

Several works extend CPD to incorporate structural assumptions.
\cite{cohen2017dictionary} considered cases where one factor belongs to a known dictionary.
\cite{sun2017provable} studied CPD under sparsity constraints, while~\cite{liu2020tensor, cai2022nonconvex, xia2021statistically, zhao2015bayesian} analyzed tensor completion from partially observed or noisy low-rank data.
These efforts aim to enhance interpretability or address practical challenges in high-dimensional and incomplete data settings.

Recent works have extended classical CPD to accommodate infinite-dimensional or smooth domain structures.
Examples include functional tensor decomposition~\citep{timmerman2002three, han2021guaranteed, larsen2024tensor, tang2024tensor}.
A growing body of work has focused on statistical models that formalize the decomposition problem under noise.
The tensor spiked covariance model~\citep{han2022tensor, tang2025mode}, an extension of the classical spiked covariance model for vectors and matrices, offers a unified framework for analyzing the statistical behavior of CPD under structured noise.
Beyond traditional applications, tensor methods have also been used in areas such as network analysis~\citep{jing2021community, lei2024computational}, neuroimaging and sequencing data analysis~\citep{ma2023tensor, xu2024tensor, shi2024tempted}, time-series modeling~\citep{chen2022analysis, wang2022high}, semi-parametric factor models~\citep{chen2024semi}, recommender systems~\citep{bi2018multilayer}, etc.

While our work focuses on CPD, other tensor formats such as Tucker~\citep{zhang2018tensor, luo2021sharp, agterberg2024statistical}, tensor-train~\citep{zhou2022optimal}, and tensor ring decompositions~\citep{chen2020tensor, chen2024one} have also received attention for specific applications.

For a broader introduction to tensor methods, we refer readers to several surveys that cover theoretical, algorithmic, and applied perspectives~\citep{koldaTensorDecompositionsApplications2009, sidiropoulos2017tensor, sun2014tensors, bi2021tensors, auddy2024tensors}.

\subsection{Organization}

The rest of this paper is organized as follows.
Section~\ref{sec_notation} introduces the notation, reviews basic tensor operations, and outlines the main objective of this work.
Section~\ref{sec_rank_one} focuses on the rank-one setting as a simplified case, and Section~\ref{sec_general_rank} extends the analysis to general-rank scenarios.
In Section~\ref{sec_algorithm_ALS_TASD}, we present the proposed algorithms and explain the underlying intuition.
Section~\ref{sec_local} establishes the local convergence properties of ALS.
The initialization method, TASD, is analyzed in Section~\ref{sec_tasd}, and its integration with ALS is examined in Section~\ref{sec_overall}.
Section~\ref{sec_simulation} presents numerical experiments that support the theoretical results.
Finally, Section~\ref{sec_discussion} concludes with a discussion of our major results and remaining challenges.

\section{Preliminaries, Notation, and Objectives}\label{sec_notation}

We use $[d]$ to denote the set $\{1,\ldots,d\}$ for a positive integer $d$. For mode dimensions $p_1,\ldots,p_d$, define $p_{\min}:=\min_{k\in[d]}p_k$ and $p_{\max}:=\max_{k\in[d]}p_k$. Whenever Tucker ranks $r_1,\ldots,r_d$ are considered, we similarly define $r_{\max}:=\max_{k\in[d]}r_k$.
For sequences $\{x_n\}$ and $\{y_n\}$, we write $x_n \lesssim y_n$ to indicate that there exists an absolute constant $c$ such that $x_n \leq c y_n$.
For a vector $a$, $\|a\|$ denotes its $\ell_2$ norm.

Let $I$ denote the identity matrix of appropriate dimension, determined by context.
For a matrix $A \in \mathbb{R}^{p \times q}$, we use $\sigma_i(A)$ to denote the $i$th singular value of $A$, ordered such that $\sigma_1(A) \geq \sigma_2(A) \geq \cdots \geq \sigma_{\min(p, q)}(A) \geq 0$.
We write $\mathrm{SVD}_r(A) = [u_1, \ldots, u_r]$ for the matrix formed by the top $r$ left singular vectors, where $u_i$ corresponds to $\sigma_i(A)$.
The notation $A^\dagger$ denotes the Moore-Penrose pseudoinverse of $A$.
For a full-column-rank matrix
$A\in\mathbb{R}^{p\times q}$, define its condition number by
$\kappa(A)
    :=
    {\sigma_1(A)} / {\sigma_q(A)}.$
The spectral norm of $A$, denoted $\|A\|$, is equal to $\sigma_1(A)$, and the Frobenius norm is given by $\|A\|_F = \sqrt{\mathrm{tr}(A A^\top)}$.
For matrices $A \in \mathbb{R}^{p_1 \times p_2}$ and $B \in \mathbb{R}^{p_3 \times p_4}$, their Kronecker product $A \otimes B \in \mathbb{R}^{p_1 p_3 \times p_2 p_4}$ is defined element-wise as
\[
    (A \otimes B)_{p_3(r-1)+v,\, p_4(s-1)+w} = A_{rs} B_{vw}, \quad r \in [p_1],\ s \in [p_2],\ v \in [p_3],\ w \in [p_4].
\]
The Khatri-Rao product $A \odot B$ of matrices $A = [a_1, \ldots, a_p]$ and $B = [b_1, \ldots, b_p]$ is defined as $A \odot B = [a_1 \otimes b_1, \ldots, a_p \otimes b_p]$.
An order-$d$ tensor $\mathcal{A} \in \mathbb{R}^{p_1 \times \cdots \times p_d}$ is a multi-dimensional array, where the entry $(i_1, \ldots, i_d)$ maps to $\mathcal{A}_{i_1, \ldots, i_d} \in \mathbb{R}$.
Given a matrix $B\in\mathbb{R}^{q_k\times p_k}$, the mode-$k$
product of $\mathcal A$ with $B$, denoted by
$\mathcal A\times_k B$, is a tensor in $\mathbb{R}^{
        p_1\times\cdots\times p_{k-1}
        \times q_k
        \times p_{k+1}\times\cdots\times p_d
    }$
defined entrywise by
\[
    (\mathcal A\times_k B)_{
        i_1,\ldots,i_{k-1},\ell,i_{k+1},\ldots,i_d
    }
    =
    \sum_{j=1}^{p_k}
    \mathcal A_{
        i_1,\ldots,i_{k-1},j,i_{k+1},\ldots,i_d
    }
    B_{\ell j},
    \qquad
    \ell\in[q_k].
\]
The mode-$k$ unfolding of $\mathcal{A}$, denoted $\mathcal{M}_k(\mathcal{A})$, is a matrix in $\mathbb{R}^{p_k \times \prod_{h \neq k} p_h}$, obtained by flattening the tensor along its $k$th mode.
The column index is determined by lexicographic ordering of the remaining indices.

For two tensors $\mathcal{A} \in \mathbb{R}^{p_1 \times \cdots \times p_d}$ and $\mathcal{B} \in \mathbb{R}^{q_1 \times \cdots \times q_k}$, their outer product $\mathcal{A} \circ \mathcal{B}$ is a tensor in $\mathbb{R}^{p_1 \times \cdots \times p_d \times q_1 \times \cdots \times q_k}$, defined as $(\mathcal{A} \circ \mathcal{B})_{i_1, \ldots, i_d, j_1, \ldots, j_k} = \mathcal{A}_{i_1, \ldots, i_d} \mathcal{B}_{j_1, \ldots, j_k}$.

The Tucker rank of $\mathcal{A}$ is defined as $\mathrm{rank}_{\mathrm{Tucker}}(\mathcal{A}) = (r_1, \ldots, r_d)$, where $r_k = \mathrm{rank}(\mathcal{M}_k(\mathcal{A}))$.
The corresponding Tucker decomposition is given by $\mathcal{A} = \mathcal{S} \times_1 U_1 \times_2 \cdots \times_d U_d$, where each $U_k \in \mathbb{R}^{p_k \times r_k}$ consists of the top left singular vectors of $\mathcal{M}_k(\mathcal{A})$, and $\mathcal{S} \in \mathbb{R}^{r_1 \times \cdots \times r_d}$ is the core tensor.
The Frobenius norm of a tensor $\mathcal{X}$ is denoted by $\|\mathcal{X}\|_F$.
Throughout the paper, we assume tensor order $d \geq 3$. 
All other notation will be introduced when it first appears.

This paper focuses on the statistical model for tensor CP decomposition:
\be
\label{eq_main_model} \Y = \X + \Z \in \RR^{p_1 \times \cdots \times p_d}, \quad \X = \sum_{r = 1}^R \lambda_r a_{1,r} \circ
\cdots \circ a_{d,r},
\ee
where each $\lambda_r\in\RR$ and $a_{k,r}\in\RR^{p_k}$ satisfies $\|a_{k,r}\|=1$.
The entries $Z_{i_1,\ldots,i_d}$ of $\Z$ are i.i.d., mean-zero, and sub-Gaussian with variance proxy $\sigma^2$, in the sense that $\EE\exp\left(tZ_{i_1,\ldots,i_d}\right)\leq\exp\left(\frac{\sigma^2t^2}{2}\right)$ for all $t\in\RR$.
Generalizing the concepts from matrix SVD and PCA, the vectors $a_{k,r}$'s are referred to as {\it loading vectors} for mode $k$ and rank $r$, while the scalars $\lambda_r$'s are referred to as singular values.
For each mode $k$, define the loading matrix $A_k:=[a_{k,1},\ldots,a_{k,R}]\in\RR^{p_k\times R}$.
Denote $\lambda_{\max}:=\max_{r\in[R]}|\lambda_r|$ and $\lambda_{\min}:=\min_{r\in[R]}|\lambda_r|$.
For $R\geq2$, define $\xi:=\max_{k\in[d]}\max_{\substack{r,s\in[R]\\r\neq s}}|a_{k,r}\T a_{k,s}|$, and set $\xi:=0$ when $R=1$.
The primary objective of this paper is to estimate $\{|\lambda_r|\}_{r\in[R]}$, the loading matrices $\{A_k\}_{k\in[d]}$ up to the usual componentwise sign and common-permutation indeterminacies, and the signal tensor $\X$, given $\Y$.

\section{Tensor Decomposition -- Rank-One Case}\label{sec_rank_one}

In this section, we focus on a simplified setting for a clearer presentation and comparison with existing results in the literature.
When $R = 1$, \eqref{eq_main_model} reduces to
\be
\label{eq_main_model_R1} \Y = \X + \Z \in \RR^{p_1 \times \cdots \times p_d}, \quad \X = \lambda a_{1} \circ \cdots \circ a_{d}.
\ee
In this case, the ALS algorithm can be described as follows: in the $(t+1)$st iteration, we update $\hat{a}_k$ via
\be
\label{eq_update_formula_R1}
\hat b_k^{(t+1)}
=
\argmin_{b\in\RR^{p_k}}
\left\|
    \Y
    -
    \hat a_1^{(t)}
    \circ\cdots\circ
    \hat a_{k-1}^{(t)}
    \circ b\circ
    \hat a_{k+1}^{(t)}
    \circ\cdots\circ
    \hat a_d^{(t)}
\right\|_F,
\qquad
\hat a_k^{(t+1)}
=
\frac{\hat b_k^{(t+1)}}
     {\|\hat b_k^{(t+1)}\|}.
\ee
In fact, $\hat b_k^{(t+1)}$ has the following closed-form solution \citep[Section 3.4]{koldaTensorDecompositionsApplications2009}:
\[
    \hat b_k^{(t+1)} = \MM_k(\Y) \left[ \left( \otimes_{i\neq k}\hat a_i^{(t)} \right)^\top \right]^\dagger.
\]
This procedure is also known as the tensor power iteration \citep{richard2014statistical}.
ALS can be effectively initialized by evaluating the first singular vector of the unfolded tensor $\Y$:
\be
\label{eq_initialization_formula_R1} \hat a_k^{(0)} = \SVD_1\(\MM_k(\Y)\).
\ee
We summarize the overall algorithm for rank-one tensor decomposition in Algorithm \ref{algorithm_als_r1}.
\begin{algorithm}[h]
    \caption{R1-ALS: {A}lternating {L}east {S}quares for Rank-one Tensor Decomposition}
    \label{algorithm_als_r1}
    \begin{algorithmic}[1]
        \algrenewcommand\algorithmicensure{\textbf{Run:}}
        \algrenewcommand\algorithmicrequire{\textbf{Input:}}
        \Require {Order-$d$ Tensor $\Y$}
        \State{Calculate $\hat a_k^{(0)} = \SVD_1\(\MM_k(\Y)\)$ for $k\in[d]$.}
        \For{$t = 0, \cdots$}
        \For{$k = 1, \cdots d$}
        \State{ Update $\hat a_k^{(t+1)}$ by $\hat b_k^{(t+1)} = \MM_k(\Y) \left[ \left( \otimes_{i\neq k}\hat a_i^{(t)} \right)^\top \right]^\dagger$ and $\hat a_k^{(t+1)} = \hat b_k^{(t+1)}/\|\hat b_k^{(t+1)}\|$. }
        \EndFor
        \EndFor\\
        \Return $\hat a_{k}^{(t)}$ for $k\in[d]$
    \end{algorithmic}
\end{algorithm}

It is important to note that model \eqref{eq_main_model_R1} is invariant under simultaneous sign flips of any even number of the $a_k$'s, and the sign of $\lambda$ can be absorbed into any one loading vector.
Thus, without imposing a sign convention, only $|\lambda|$ and the loading directions are identifiable.
After obtaining the loading estimates $\{\hat a_k^{(t)}\}_{k\in[d]}$, we estimate $|\lambda|$ by one additional mode-$1$ contraction:
\be
\label{eq_estimate_lambda_R1}
\widehat{|\lambda|}^{(t)}
=
\left\|
\MM_1(\Y)
\left(
\hat a_2^{(t)}
\otimes\cdots\otimes
\hat a_d^{(t)}
\right)
\right\|
=
\|\hat b_1^{(t+1)}\|.
\ee
The same construction can be based on any fixed mode $k\in[d]$.
To account for both possible directions of the loading vectors, we define the error at the $t$th iteration as
\be
\label{eq_error_formula_R1} \varepsilon_t = \max_{i\in [d]}\left\{\sqrt{2-2|\hat a_i^{(t)\top} a_i|}\right\} = \max_{i\in
    [d]}\left\{\min\left\{\|\hat a_i^{(t)} - a_i\|, \|\hat a_i^{(t)} + a_i\|\right\}\right\}.
\ee
We have the following result for ALS in the rank-one case:
\begin{Theorem}[Rank-One Case: Iteration Error]\label{thm_overall_R1}
    In model \eqref{eq_main_model_R1}, let $\hat a_k^{(0)}$ be calculated by \eqref{eq_initialization_formula_R1}, $\hat a_k^{(t)}$ be calculated iteratively by \eqref{eq_update_formula_R1}, and define $\varepsilon_t$ by \eqref{eq_error_formula_R1}.
    There exist absolute constants $C_1$, $C_2$, and $C_3$, and an absolute integer $C_4\geq3$, such that if $p_{\min} \geq C_1 \log d$ and $|\lambda| \geq C_1\sigma p_{\max}^{d/4}$, then with probability at least $1-\exp\(-C_2 p_{\min}\)$, we have
    \[
        \varepsilon_0 \leq C_3 \left( \frac{\sigma^2 p_{\max}}{\lambda^2} + \frac{\sigma^4 p_{\max}^d}{\lambda^4} \right)^{1/2}.
    \]
    Moreover, the following holds for $t\geq2$ with probability at least $1-\exp\(-C_2 p_{\min}\)$:
    \be
    \label{eq_bound_et_r1} \varepsilon_t \leq C_3 \frac{\sigma\sqrt{p_{\max}}}{|\lambda| }.
    \ee
    If, in addition, $d\leq C_4$, then \eqref{eq_bound_et_r1} holds for every $t\geq1$ with the same probability.
\end{Theorem}

\begin{Corollary}[Rank-One Case: Estimation Error of $|\lambda|$]
\label{corollary_lambda_R1}
In the setting of Theorem~\ref{thm_overall_R1}, let $\widehat{|\lambda|}^{(t)}$ be defined by \eqref{eq_estimate_lambda_R1}.
Then, with probability at least $1-C\exp(-cp_{\min})$ for some absolute constants $C,c>0$, we have
\[
    \left|
        \widehat{|\lambda|}^{(t)}
        -
        |\lambda|
    \right|
    \lesssim
    \sigma\sqrt{p_{\max}}
\]
for every $t\geq2$.
If, in addition, $d\leq C_4$, then the same conclusion holds for every $t\geq1$ with the same probability.
\end{Corollary}

Theorem~\ref{thm_overall_R1} demonstrates that, for $R=1$, ALS can achieve the desired loading-vector accuracy within two iterations when initialized via \eqref{eq_initialization_formula_R1}, while Corollary~\ref{corollary_lambda_R1} shows that $|\lambda|$ can then be estimated with error of order $\sigma\sqrt{p_{\max}}$ using one additional tensor contraction.
To compare our results with existing literature, we assume $p_k = p$ and $\sigma = 1$, and present the comparison in Table \ref{table_comparison}.

\begin{Remark}[Theory of Rank-One Tensor Decomposition]
    We highlight several insights regarding Theorem \ref{thm_overall_R1}.

    \cite{zhang2018tensor} proved that the required signal strength $|\lambda| \gtrsim \sigma p_{\max}^{d/4}$ is optimal for the existence of a computationally efficient estimator under the hypergraphic planted clique assumption.
    They further showed that $\frac{\sigma \sqrt{p_{\max}}}{|\lambda|}$ is the information-theoretic limit, which implies that the resulting error bound of Theorem \ref{thm_overall_R1} is rate-optimal.

    A widely used alternative to spectral initialization \eqref{eq_initialization_formula_R1} is random initialization, where the initial vectors $\hat{a}_k^{(0)}$ are sampled uniformly from the unit sphere.
    However, this approach typically requires a stronger signal strength $|\lambda| \gtrsim p^{d/2}$ to ensure convergence of ALS to the true loading vectors \citep{richard2014statistical}.
    In contrast, Theorem \ref{thm_overall_R1} shows that with warm-start initialization via \eqref{eq_initialization_formula_R1}, ALS achieves a two-step convergence under a significantly weaker signal condition $|\lambda| \gtrsim p^{d/4}$.
\end{Remark}

\section{Tensor Decomposition -- General Rank Case}\label{sec_general_rank}

In this section, we consider the CPD model~\eqref{eq_main_model} for a general rank $R$.
We begin by presenting the general algorithms in Section~\ref{sec_algorithm_ALS_TASD}, followed by theoretical analyses of ALS, TASD, and the overall algorithm in Sections~\ref{sec_local}, \ref{sec_tasd}, and \ref{sec_overall}, respectively.

\subsection{Algorithm: ALS and TASD}\label{sec_algorithm_ALS_TASD}

Recall the CPD model~\eqref{eq_main_model}: $\Y = \X + \Z \in \RR^{p_1 \times \cdots \times p_d}, \X = \sum_{r = 1}^R \lambda_r a_{1,r} \circ \cdots \circ a_{d,r}.$ The ALS algorithm updates the $k$th mode for a general rank-$R$ tensor according to the following rule:
\be
\label{eq_update_formula} \hat B_k^{(t+1)} = \argmin_{B \in \RR^{p_k \times R}} \left\| \MM_k(\Y) - B \left( \odot_{i \neq k}
\hat A_i^{(t)} \right)^\top \right\|_F, \quad \hat A_k^{(t+1)} = \operatorname{Normalize}(\hat B_k^{(t+1)}).
\ee
Here, $\hat A_k^{(t+1)}$ is obtained by normalizing each column of $\hat B_k^{(t+1)}$ to have unit $\ell_2$ norm.
The update $\hat B_k^{(t+1)}$ admits a closed-form solution~\citep[Section 3.4]{koldaTensorDecompositionsApplications2009}:
\be
\label{eq_update_formula_closed_form} \hat B_k^{(t+1)} = \MM_k(\Y) \left[ \left( \odot_{i\neq k}\hat A_i^{(t)} \right)^\top
\right]^\dagger.
\ee
Although the signs of individual $\lambda_r$'s and $a_{k,r}$'s are not identifiable, we can estimate $|\lambda_r|$ and the low-rank signal tensor $\X$ as follows:
\be
\label{eq_estimate_lambda_X}
\widehat{|\lambda_r|}
=
\|\hat b_{1,r}^{(t+1)}\|,
\qquad
\hat\X
=
\sum_{r=1}^R
\hat b_{1,r}^{(t+1)}
\circ
\hat a_{2,r}^{(t)}
\circ\cdots\circ
\hat a_{d,r}^{(t)}.
\ee
where $\hat b_{1,r}^{(t+1)}$ and $\hat a_{k,r}^{(t)}$ denote the $r$th columns of $\hat B_1^{(t+1)}$ and $\hat A_k^{(t)}$, respectively.

This procedure is summarized in Algorithm~\ref{algorithm_als}.
\begin{algorithm}[h]
    \caption{ALS: {A}lternating {L}east {S}quares}
    \label{algorithm_als}
    \begin{algorithmic}[1]
        \algrenewcommand\algorithmicensure{\textbf{Run:}}
        \algrenewcommand\algorithmicrequire{\textbf{Input:}}
        \Require {Order-$d$ Tensor $\Y$, Target CP Rank $R$, and Initialization $\hat A_{k}^{(0)}$ for $k\in[d]$}
        \For{$t = 0, \cdots$}
        \For{$k = 1, \cdots d$}
        \State{Update $\hat A_{k}^{(t+1)}$ by \eqref{eq_update_formula}: $\hat B_k^{(t+1)} = \MM_k(\Y)[(\odot_{i\neq k}\hat A_i^{(t)})\T]^\dagger$, $\hat A_k^{(t+1)} = \operatorname{Normalize}(\hat B_k^{(t+1)})$.}
        \EndFor
        \EndFor
        \State{
            Calculate
            $\widehat{|\lambda_r|}=\|\hat b_{1,r}^{(t+1)}\|$
            and
            $\hat\X
            =
            \sum_{r=1}^R
            \hat b_{1,r}^{(t+1)}
            \circ
            \hat a_{2,r}^{(t)}
            \circ\cdots\circ
            \hat a_{d,r}^{(t)}$.
            }
            \State{
                \Return
                the normalized loading matrices
                $\{\hat A_k^{(t)}\}_{k\in[d]}$,
                the magnitude estimates
                $\{\widehat{|\lambda_r|}\}_{r\in[R]}$,
                and
                $\hat\X$.
            }
    \end{algorithmic}
\end{algorithm}

\begin{Remark}[Selection of the CP Rank]\label{remark_rank_selection}
    Algorithm~\ref{algorithm_als} and the TASD procedure introduced below take the target CP rank $R$ as an input, and our theoretical analysis assumes that $R$ is known.
    In practice, $R$ may be selected in a preliminary step.
    Under the full-column-rank conditions imposed in our general-rank theory, the signal tensor $\X$ has Tucker rank $(R,\ldots,R)$.
    Thus, one natural approach is to estimate $R$ from the singular-value profiles of the mode unfoldings of $\Y$, for example using a noise-calibrated threshold or a scree-plot criterion.
    Alternatively, TASD--ALS may be fitted over a range of candidate ranks, with the rank selected using a penalized reconstruction criterion.
    Establishing theoretical guarantees for rank selection and for TASD--ALS with an estimated rank is beyond the scope of the present paper.
\end{Remark}

It is important to note that Algorithm~\ref{algorithm_als} requires an initialization $\hat A_k^{(0)}$ for each $k \in [d]$.
However, unlike the rank-one case, the tensor unfolding initialization in~\eqref{eq_initialization_formula_R1} generally fails for higher-rank tensors, particularly when some of the $\lambda_r$'s are identical or $\{a_{k, r}\}_{r=1}^R$'s are non-orthogonal.
To address this challenge, we propose a Tucker-based Approximation with Simultaneous Diagonalization (TASD) to provide a warm start for ALS.

To illustrate the core idea of TASD, we first consider the noiseless setting, where $\Z = 0$, so the model~\eqref{eq_main_model} becomes:
\be
\label{eq_cpd_d3} \X = \sum_{r = 1}^R \lambda_r a_{1,r} \circ \cdots \circ a_{d,r}.
\ee
Assume each loading matrix $A_k$ has linearly independent columns.
We randomly generate vectors $\left\{w_{ik} \in \RR^{p_k}: i = 1,2;~k \in \{3,\dots,d\}\right\}$ and contract the last $(d-2)$ modes of $\X$ with them, and unfold the resulting tensors.
This yields the matrices:
$$
    M_i:= \MM_1(\X \times_{k=3}^d w_{ik}\T) = \sum_{r=1}^R \lambda_r \prod_{k=3}^d \langle a_{k,r}, w_{ik} \rangle a_{1,r}
    a_{2,r}^\top = A_1 D_i A_2^\top \in \RR^{p_1 \times p_2}, \quad i = 1,2,
$$
where $D_i \in \RR^{R \times R}$ is diagonal, with entries $(D_i)_{rr} = \lambda_r \prod_{k=3}^d \langle a_{k,r}, w_{ik} \rangle$.
From this, we compute
\be
\label{eq_M1M2} M_1 M_2^\dagger = A_1 D_1 A_2^\top (A_1 D_2 A_2^\top)^\dagger = A_1 D_1 D_2^\dagger A_1^\dagger,
\ee
which implies $(M_1 M_2^\dagger) A_1 = A_1 D_1 D_2^\dagger$, i.e., the columns of $A_1$ are eigenvectors of $M_1 M_2^\dagger$.

If the vectors $w_{ik}$ are drawn independently with i.i.d.\ standard Gaussian entries, then, with probability one with respect to these Gaussian contraction vectors, all diagonal entries of $D_1$ and $D_2$ are nonzero and the diagonal entries of $D_1D_2^\dagger$ are pairwise distinct.
Consequently, $M_1M_2^\dagger$ has distinct nonzero eigenvalues, and its eigen-decomposition recovers $A_1$ up to column scaling and permutation.
Here the probability statement concerns only the internal randomization of SimDiag; the noiseless tensor $\X$ itself is treated as fixed. 
This procedure is known as Simultaneous Diagonalization (SimDiag) \citep{harshman1970foundations,leurgans1993decomposition}.

While SimDiag works well in the noiseless setting, it lacks robustness in the presence of noise, especially when the ambient dimension $p_k$ is large—see discussion in Section~\ref{sec_intro}.
To overcome this limitation, we propose the TASD method, which combines Tucker decomposition with SimDiag for greater stability.

Specifically, we begin by performing a Tucker decomposition of $\X$ in~\eqref{eq_cpd_d3} with target Tucker rank $(R, \ldots, R)$.
This can be achieved using the Higher-Order Orthogonal Iteration (HOOI) algorithm~(see Supplement Algorithm~\ref{algorithm_hooi}; cf.~\cite{de2000best}):
$$
    \X = \S \times_{k=1}^d U_k,
$$
where $\S$ is the Tucker core.
The CPD of $\X$ can also be written as
$$
    \X = \Lambda \times_{k=1}^d A_k,
$$
where $\Lambda$ is the order-$d$ superdiagonal tensor with $\Lambda_{r,\ldots,r}=\lambda_r$ for $r\in[R]$. 
Let the singular value decomposition of $A_k$ be $A_k = U_{A_k} \Sigma_{A_k} V_{A_k}^\top$.
Then there exists an orthogonal matrix $O_k \in \RR^{R \times R}$ such that $U_k = U_{A_k} O_k$.
This implies
$$
    \S = \Lambda \times_{k=1}^d (O_k^\top \Sigma_{A_k} V_{A_k}^\top).
$$
We then apply SimDiag to the Tucker core $\S$, which recovers $O_k^\top \Sigma_{A_k} V_{A_k}^\top$ (up to column scaling and permutation).
Finally, we recover the loading matrices by computing
$$
    U_k O_k^\top \Sigma_{A_k} V_{A_k}^\top = U_{A_k} O_k O_k^\top \Sigma_{A_k} V_{A_k}^\top = A_k.
$$
A key advantage of TASD is that it applies SimDiag to an order-$d$ Tucker core containing $R^d$ entries instead of the full ambient tensor containing $\prod_{k=1}^d p_k$ entries.
When $R\ll p_k$ for every mode, this compression substantially reduces the dimension on which simultaneous diagonalization is performed and contributes to the noise robustness of TASD.
At the same time, explicitly storing and contracting the dense core scales as $R^d$, so the present implementation is primarily intended for low-rank settings and may become computationally expensive when either $R$ or $d$ is large.

For a target mode $h\in[d]$, let $h^+$ denote the next mode cyclically, i.e., $h^+=h+1$ if $h<d$ and $h^+=1$ if $h=d$.
This procedure is summarized in Algorithm \ref{algorithm_tasd}.
\begin{algorithm}[h]
    \caption{TASD: {T}ucker-based {A}pproximation with {S}imultaneous {D}iagonalization, Part I}
    \label{algorithm_tasd}
    \begin{algorithmic}[1]
        \algrenewcommand\algorithmicensure{\textbf{Output:}}
        \algrenewcommand\algorithmicrequire{\textbf{Input:}}
        \Require {Order-$d$ Tensor $\Y$, Target mode $h\in[d]$, and Target CP Rank $R$}
        \State{Obtain $\hat U_k$ and $\hat \S$ from Tucker decomposition, e.g., HOOI (Supplement Algorithm \ref{algorithm_hooi}), of $\Y$ with Tucker rank $(R,\ldots,R)$;}
        \State{Generate $w_{ik}\in\RR^R$ with i.i.d. standard normal entries for $i\in\{1,2\}$ and $k\notin\{h,h+1\}$;}
        \State{Calculate $\hat M_i = \MM_h(\hat\S \times_{k \notin \{h,h+1\}} w_{ik}\T)$ for $i \in \{1,2\}$;}
        \State{Obtain eigenvectors $\hat V_h$ from the eigen-decomposition of $\hat M_1(\hat M_2)^\dagger$;}\\
        \Return The normalized estimate of the $h$th loading matrix
        $\hat A_h
        =
        \operatorname{Normalize}\!\left(
            \hat U_h\operatorname{Re}(\hat V_h)
        \right)$.
    \end{algorithmic}
\end{algorithm}

\begin{Remark}[TASD vs. Existing Work]
    The semi-algebraic framework for approximate CP decompositions via simultaneous matrix diagonalizations (SECSI; \cite{roemerSemialgebraicFrameworkApproximate2013}) adopts a procedure similar to TASD, namely, applying eigen-decomposition to matrices derived from the Tucker decomposition to estimate the CP components.
    SECSI has gained increasing attention in the signal processing community \citep{becker2015brain, naskovskaExtensionSemialgebraicFramework2016, gherekhloo2021tensor}.
    However, to the best of our knowledge, no statistical guarantees have been established for SECSI.

    The key difference between SECSI and TASD lies in how the matrices of the form $M_1 M_2^\dagger = A_1 D_1 D_2^\dagger A_1^\dagger$ (as in~\eqref{eq_M1M2}) are constructed.
    SECSI computes these matrices through a fully deterministic procedure (see Equation (17) in \cite{roemerSemialgebraicFrameworkApproximate2013}).
    As a result, there exist input tensors for which the diagonal entries of $D_1 D_2^\dagger$ may coincide, making the associated eigenvectors not uniquely identifiable and problematic for eigen-decomposition.
    In contrast, TASD uses randomly generated vectors $w_{ik}$, which, with high probability, ensures that $D_1 D_2^\dagger$ has distinct diagonal entries, thereby avoiding such issues.
\end{Remark}

As shown later in Theorem~\ref{thm_tasd}, the loading matrices can be consistently estimated by TASD, up to a permutation of their columns.
To combine the estimated loadings from different modes, it is necessary to identify a permutation that aligns the column order across all estimated loading matrices.
\cite{roemerSemialgebraicFrameworkApproximate2013} suggests searching all possible combinations of the permutations of $\hat A_h, h\in [d]$ such that the least square estimation of $\argmin_{\lambda_r}\|\sum_{r = 1}^R \lambda_r a_{1,\pi_1(r)} \circ \cdots \circ a_{d,\pi_d(r)} - \Y\|$ has descending order.
In practice, for small values of $R$ and $d$, this alignment can be achieved by the exhaustive search mentioned above.
However, this approach has time complexity $(R!)^{d-1}$, which becomes infeasible for large $R$ or $d$.

To overcome this challenge, we propose another procedure.
Given an estimated loading matrix $\hat A_h$ from TASD for some mode $h$, we solve the least squares problem:
\[
    \argmin_{A \in {\RR^{R \times \prod_{k\neq h}p_k}}}\|\MM_h(\Y) - \hat A_h A\|_F,
\]
which admits the closed form solution
\be
\label{eq_closed_solution_ls} \hat A = (\hat A_h)^\dagger\MM_h(\Y).
\ee
This step is motivated by the structure of exact low-rank tensors.
When $\Y$ is exactly rank-$R$, we have $\MM_h(\Y) = A_h \diag(\lambda) (\odot_{k \ne h} A_k)\T$, where $\diag(\lambda)$ is diagonal with entries $\lambda_r$.
In this case, each row of $\hat A$ approximates a row of $(\odot_{k \ne h} A_k)\T$, up to scaling.
Thus, each row of $\hat A$ corresponds to a rank-one tensor, which can be efficiently decomposed using R1-ALS (Algorithm~\ref{algorithm_als_r1}).
By retaining the row index $r$ when the corresponding rank-one tensor is decomposed, all loading vectors recovered from the $r$th row inherit the component ordering of the $r$th column of $\hat A_h$.
Consequently, TASD Part~II produces factor matrices with a common component ordering.
This procedure is summarized in Algorithm~\ref{algorithm_tasd_ls}.

\begin{algorithm}[h]
    \caption{TASD: Part II}
    \label{algorithm_tasd_ls}
    \begin{algorithmic}[1]
        \algrenewcommand\algorithmicensure{\textbf{Output:}}
        \algrenewcommand\algorithmicrequire{\textbf{Input:}}
        \Require {Order-$d$ Tensor $\Y$ , Target CP Rank $R$ and $\hat A_h$ for some $h$}
        \State{Calculate $\hat A$ as in \eqref{eq_closed_solution_ls};}
        \State{For each $r\in[R]$, fold the $r$th row of $\hat A$ into an order-$(d-1)$ tensor $\Y_r\in\RR^{p_1\times\cdots\times p_{h-1}\times p_{h+1}\times\cdots\times p_d}$, using the ordering induced by the mode-$h$ unfolding;}
        \State{For each $r\in[R]$, apply Algorithm~\ref{algorithm_als_r1} once to $\Y_r$ to obtain $\hat a_{k,r}$ for every $k\in[d]\setminus\{h\}$;}
        \State{For each $k\in[d]\setminus\{h\}$, let $\hat A_k=[\hat a_{k,1},\ldots,\hat a_{k,R}]$;}\\
        \Return $\hat A_k$ for $k\neq h$, with the column index $r$ inherited from the $r$th row of $\hat A$ and hence aligned with the $r$th column of $\hat A_h$.
    \end{algorithmic}
\end{algorithm}

\subsection{Theory of ALS}\label{sec_local}

We develop the theory for ALS (Algorithm \ref{algorithm_als}) in this section.
By Kruskal’s Theorem \citep{kruskal1977three}, when $d \geq 3$ and each loading matrix $A_k$ has linearly independent columns, the CPD $\X = \sum_{r = 1}^R \lambda_r\, a_{1,r} \circ \cdots \circ a_{d,r}$ in model \eqref{eq_cpd} is unique up to a permutation of the loadings (indexed by $r$) and a sign flip of $\lambda_r$ and $a_{k,r}$.
Since the result below is local, we assume throughout this subsection that the columns of all initialization matrices $\hat A_k^{(0)}$, $k\in[d]$, have first been jointly relabeled so that the $r$th column of every $\hat A_k^{(0)}$ corresponds to the $r$th true CP component.
This common component ordering is kept fixed throughout the ALS iterations.
To account for the remaining sign ambiguity, we define the following error metric:
\be
\label{eq_error_formula}
\varepsilon_t
=
\max_{k\in[d]}
\max_{r\in[R]}
\min_{s\in\{-1,1\}}
\left\|
    \hat a_{k,r}^{(t)}
    -
    s a_{k,r}
\right\|.
\ee
This error reduces to \eqref{eq_error_formula_R1} when $R=1$.
The following theorem characterizes the local convergence of the ALS algorithm.
\begin{Theorem}[Convergence of ALS]\label{thm_local}
    In model~\eqref{eq_main_model}, assume that $R\geq2$ and that each $A_k$ has full column rank.
    Let $\hat A_k^{(t)}$ be calculated iteratively by Algorithm~\ref{algorithm_als} and let $\varepsilon_t$ be defined as in \eqref{eq_error_formula}.
    Define the transition scale
    \be
    \label{eq_transition_scale} \tau_{\mathrm{ALS}} :=
    \begin{cases}
        \xi,            & d=3, \\[2mm]
        (R-1)\xi^{d-2}, & d>3.
    \end{cases}
    \ee
    Assume the following hold for some absolute constant $C_1$:
    \bea
    1. & & \varepsilon_0 \frac{\lambda_{\max}}{\lambda_{\min}} (d+R) \lesssim C_1\i; \label{eq_assumption_1_general_R} \\
    2. & &\text{if $d >3$, } \xi \leq \(C_1 dR^{\frac{1}{d-3}}\)\i; \text{ if $d =3$, $\xi \leq (C_1 R^{1/2})\i$; }
    \label{eq_assumption_2_general_R} \\
    3. & &\lambda_{\min} \geq C_1 \sigma \(p_{\max}^{1/2} d^2 \log d \lor \frac{\lambda_{\max}}{\lambda_{\min}}
    \sqrt{dp_{\max}\log d} (d+R)\)\label{eq_assumption_3_general_R}.
    \eea
    Then, if $\frac{\sigma\sqrt{p_{\max}}}{\lambda_{\min}} \geq \tau_{\mathrm{ALS}}$, we have $\varepsilon_t \leq C_2\frac{\sigma\sqrt{p_{\max}}}{\lambda_{\min}}$ for all $t\geq T$ with some $T \leq C_2 \log\log\left(\frac{\lambda_{\min}}{\sigma\sqrt{p_{\max}}}\right)$, with probability at least $1-C_3dR\exp(-c_3p_{\max})$, for some absolute constants $C_2,C_3,c_3>0$.

    If $\frac{\sigma\sqrt{p_{\max}}}{\lambda_{\min}} < \tau_{\mathrm{ALS}}$, and we additionally have $\frac{\lambda_{\max}}{\lambda_{\min}}Rd\xi^{d-2} \leq c$ for some absolute constant $c$, then $\varepsilon_t \leq C_2\frac{\sigma\sqrt{p_{\max}}}{\lambda_{\min}}$ for all $t\geq T$ with some $T \leq C_2 \log\log\left(\frac{\lambda_{\min}}{\sigma\sqrt{p_{\max}}}\right) + C_2 \log\left(\frac{\tau_{\mathrm{ALS}}\lambda_{\min}}{\sigma\sqrt{p_{\max}}}\right)$, with probability at least $1-C_3dR\exp(-c_3p_{\max})$, for some absolute constants $C_2,C_3,c_3>0$.
\end{Theorem}

\begin{Remark}[Interpretation of the ALS Assumptions]\label{remark_als_assumptions}
Condition~\eqref{eq_assumption_1_general_R} places the initialization inside the local basin of attraction.
Together with the coherence condition~\eqref{eq_assumption_2_general_R}, it keeps the Gram matrices of the estimated Khatri--Rao factors uniformly invertible and controls deterministic interference between different CP components.
The need to control this interference is consistent with the empirical deterioration of ALS for highly correlated loading vectors \citep{naskovskaExtensionSemialgebraicFramework2016}.
Condition~\eqref{eq_assumption_3_general_R} ensures that the stochastic contraction errors are small enough for the iterates to remain in the local region.
The additional condition in the second conclusion of Theorem~\ref{thm_local} makes the coefficient of the subsequent linear recurrence strictly smaller than one.

If $d$ and $\lambda_{\max}/\lambda_{\min}$ are treated as constants, the assumptions simplify, up to constants depending on $d$, to
\[
    \varepsilon_0\lesssim\frac{1}{d+R},
    \qquad
    \lambda_{\min}\gtrsim\sigma\sqrt{p_{\max}}(d+R),
\]
together with $\xi\lesssim R^{-1/2}$ when $d=3$ and $\xi\lesssim R^{-1/(d-3)}$ when $d>3$.
If $R$ is also treated as a constant, they reduce further to a sufficiently accurate constant-quality initialization, sufficiently small constant coherence, and $\lambda_{\min}\gtrsim\sigma\sqrt{p_{\max}}$.

The separate treatment of $d=3$ is algebraic rather than a distinction between statistical models.
When $d=3$, a direct expansion of the deterministic numerator in the local recurrence gives leading terms of orders $R\varepsilon_t^2$ and $R\xi\varepsilon_t$, so their transition occurs at $\varepsilon_t\asymp\xi$.
When $d>3$, the additional modes introduce products of coherence factors, and the corresponding cross-component contribution is of order $Rd\xi^{d-2}\varepsilon_t$.
Together with the sufficient coherence condition in~\eqref{eq_assumption_2_general_R}, this yields the transition scale $(R-1)\xi^{d-2}$.
\end{Remark}

When $\varepsilon_t\gtrsim\tau_{\mathrm{ALS}}$, the quadratic bound applies.
Once $\varepsilon_t\lesssim\tau_{\mathrm{ALS}}$, the linear bound applies until the algorithm reaches the statistical accuracy $\sigma\sqrt{p_{\max}}/\lambda_{\min}$.
In the exactly orthogonal case $\xi=0$, we have $\tau_{\mathrm{ALS}}=0$, so only the quadratic regime is needed.
If the final statistical error is below $\tau_{\mathrm{ALS}}$, the algorithm may undergo a subsequent linear-convergence phase.
This leads to the following two convergence patterns:
\begin{enumerate}
    \item a quadratic-only regime when
          \[
              \tau_{\mathrm{ALS}} \leq \frac{\sigma\sqrt{p_{\max}}}{\lambda_{\min}};
          \]
        \item a quadratic-then-linear regime when
          \[
              \tau_{\mathrm{ALS}} > \frac{\sigma\sqrt{p_{\max}}}{\lambda_{\min}}, \qquad \xi^{d-2} \lesssim \frac{\lambda_{\min}}
              {Rd\lambda_{\max}}.
          \]
\end{enumerate}

The assumption \eqref{eq_assumption_3_general_R} imposes a requirement on the minimum signal strength $\lambda_{\min}$.
By \eqref{eq_update_formula_closed_form}, we have the following expression for $\hat{B}_k^{(t+1)}$:
\begin{equation}\label{eq_formula_bhat}
    \hat{B}_k^{(t+1)}  = A_k \, \text{diag}(\lambda) \left( \odot_{i \neq k} A_i \right)\T \left[ \left( \odot_{i \neq k}
    \hat{A}_i^{(t)} \right)\T \right]^\dagger + \MM_k(\Z) \left[ \left( \odot_{i \neq k} \hat{A}_i^{(t)} \right)\T
    \right]^\dagger,
\end{equation}
where $\lambda = (\lambda_1, \ldots, \lambda_R)$ and $\text{diag}(\lambda) \in \RR^{R \times R}$ denotes the diagonal matrix with $[\text{diag}(\lambda)]_{rr} = \lambda_r$.
Even when $\Z = 0$, immediate convergence is not guaranteed.
This is because, when estimating the loading vector of rank $r$, the loading vectors from other ranks contribute as noise.
This highlights a crucial difference between the case $R = 1$ and the more general case $R > 1$.

Moreover, we establish the following information-theoretic lower bound for loading estimation. The lower bound is proved even over an equal-strength orthogonal subclass and identifies the optimal terminal statistical error rate in Theorem~\ref{thm_local}. The initialization condition in Theorem~\ref{thm_local} serves the separate purpose of ensuring that ALS enters its local convergence region.
\begin{Theorem}[Lower Bound]\label{thm_lower_bound}
    Assume that $d\geq3$, $1\leq R\leq p_{\min}$, and $p_{\min}\geq3$.
    In model~\eqref{eq_main_model}, suppose that the entries of $\Z$ are independently distributed as $\mathcal{N}(0,\sigma^2)$.
    For $\lambda>0$, let $\mathcal{F}_{R,\lambda}^{\mathrm{orth}}$ denote the class of tensors admitting a CP decomposition
    $
        \X = \lambda \sum_{r=1}^R a_{1,r}\circ\cdots\circ a_{d,r},
    $
    such that  
    $A_h^\top A_h=I_R$ 
    and $\operatorname{rank}(A_h)=R$ for every $h\in[d]$, where
    $
        A_h=[a_{h,1},\ldots,a_{h,R}].
    $
    Define
    \[
        \mathcal{L}_k(\hat A_k,A_k) := \min_{\pi\in\mathfrak{S}_R} \max_{r\in[R]} \min_{s\in\{-1,1\}} \left\| \hat a_{k,r} - s
        a_{k,\pi(r)} \right\|,
    \]
    where $\mathfrak{S}_R$ denotes the set of permutations of $[R]$.
    Then, for every $k\in[d]$,
    \[
        \inf_{\hat A_k} \sup_{\X\in\mathcal{F}_{R,\lambda}^{\mathrm{orth}}} \EE_{\X} \mathcal{L}_k(\hat A_k,A_k) \geq c \left( 1 \land
        \frac{\sigma\sqrt{p_k}}{\lambda} \right),
    \]
    where the infimum is taken over all measurable estimators $\hat A_k=\hat A_k(\Y)$ taking values in $\RR^{p_k\times R}$, and $c>0$ is an absolute constant.
\end{Theorem}

\begin{Remark}[Uniform Lower Bound over Modes]\label{remark_uniform_lower_bound}
Under the conditions of Theorem~\ref{thm_lower_bound}, let $k_*\in[d]$ satisfy $p_{k_*}=p_{\max}$.
Since $\max_{k\in[d]}\mathcal{L}_k(\hat A_k,A_k)\geq\mathcal{L}_{k_*}(\hat A_{k_*},A_{k_*})$, Theorem~\ref{thm_lower_bound} yields
\[
    \inf_{\hat A_1,\ldots,\hat A_d}
    \sup_{\X\in\mathcal{F}_{R,\lambda}^{\mathrm{orth}}}
    \EE_{\X}
    \max_{k\in[d]}
    \mathcal{L}_k(\hat A_k,A_k)
    \geq
    c
    \left(
        1
        \land
        \frac{\sigma\sqrt{p_{\max}}}{\lambda}
    \right).
\]
Thus, after taking the maximum over modes in the loss, the lower bound also involves $p_{\max}$ and matches the terminal upper-bound rate in Theorem~\ref{thm_local} under the same uniform loss, even when the mode dimensions are unbalanced.

A mode-wise upper bound of order $\sigma\sqrt{p_k}/\lambda_{\min}$ is not established by our analysis. As \eqref{eq_update_formula_closed_form} and \eqref{eq_formula_bhat} show, the ALS update for mode $k$ depends on estimates from all other modes through their Khatri--Rao product, so their errors propagate into the least-squares design, pseudoinverse, and noise term. 
This error coupling is absent in one-shot matrix SVD or least-squares analyses.
\end{Remark}

We also have the following upper bounds on the estimation error of $\widehat{|\lambda_r|}$ and $\hat \X$.
\begin{Corollary}[Estimation Error of $\widehat{|\lambda_r|}$ and $\hat \X$]
    \label{corollary_lambda_X}
    In the setting of Theorem~\ref{thm_local}, assume that ALS has been performed for $t \geq T+1$ iterations.
    Let $\widehat{|\lambda_r|}$ and $\hat \X$ be calculated according to \eqref{eq_estimate_lambda_X}.
    Then, on the same high-probability event as in Theorem~\ref{thm_local}, and hence with probability at least $1-CdR\exp(-cp_{\max})$ for some absolute constants $C,c>0$,
    \[
        \max_{r\in[R]} \left| \widehat{|\lambda_r|}-|\lambda_r| \right| \lesssim \sigma\sqrt{p_{\max}},
    \]
    and
    \[
        \|\hat \X-\X\|_F \lesssim d\sigma\sqrt{R p_{\max}}.
    \]
\end{Corollary}

\begin{Remark}[Comparison with Existing Work]
    We finally compare our results to rank-one-ALS-based methods proposed by \cite{anandkumar2014tensor} and \cite{anandkumar2014guaranteed}.
    In \cite{anandkumar2014tensor}, the authors proposed an algorithm for estimating the loading vectors in \eqref{eq_main_model} by iteratively applying rank-one ALS and sequentially subtracting the extracted signal.
    This method is statistically guaranteed under the assumption that the loading vectors are exactly orthogonal.
    Subsequently, \cite{anandkumar2014guaranteed} extended this algorithm to handle the non-orthogonal case and established convergence guarantees.
    However, their statistical model assumes a prior that the loading vectors in \eqref{eq_cpd} are independently and uniformly distributed on the unit sphere, which can be potentially violated in practice, for example, when the signal is non-negative.
    Moreover, the algorithm in \cite{anandkumar2014guaranteed} is only proved to converge linearly, and the final error bound includes an extra $\sqrt{d \log d}$ factor when generalizing to the general order $d$ case, which is avoided in our analysis.
    We lastly remark that all these rank-one-ALS-based approaches can fail when the sum of the first $k \leq R$ rank-one CP components in \eqref{eq_cpd} does not yield the best rank-$k$ approximation \citep{kolda2001orthogonal}, which is observed in our simulation study (Section \ref{sec_simulation}).
\end{Remark}

\subsection{Theory of TASD}\label{sec_tasd}

In this section, we present the theoretical analysis of TASD, as introduced in Section~\ref{sec_algorithm_ALS_TASD}.
We begin with the following proposition, which formalizes exact recovery in the noiseless setting.
\begin{Proposition}[TASD: Noiseless Setting]\label{prop_tasd_noiseless}
    Consider the model defined in \eqref{eq_main_model} with $\Z=0$, and suppose that every loading matrix $A_k$ has full column rank.
    Fix a target mode $h\in[d]$.
    In exact arithmetic, with probability one with respect to the Gaussian contraction vectors generated in Algorithm~\ref{algorithm_tasd}, TASD exactly recovers $A_h$ up to a permutation of its columns and columnwise sign changes.
\end{Proposition}

We are now ready to establish the statistical guarantee for TASD in the presence of noise.
\begin{Theorem}[Theory of TASD]\label{thm_tasd}
    Consider the model defined in~\eqref{eq_main_model}, where $R\geq2$ and each loading matrix $A_k$ has full column rank.
    Define
    $
        \lambda_{\operatorname{Tucker}} = \min_{k\in[d]} \sigma_R\left(\MM_k(\X)\right) $,
    $
        \bar\lambda_{\operatorname{Tucker}} = \max_{k\in[d]} \sigma_1\left(\MM_k(\X)\right) $, and
    $
        \kappa_T = \bar\lambda_{\operatorname{Tucker}} / \lambda_{\operatorname{Tucker}}.
    $
    Fix $h\in[d]$ and a confidence level $\delta\in(0,1/2)$.
    Let $\hat A_h=[\hat a_{h,1},\ldots,\hat a_{h,R}]$ be the output of TASD (Algorithm~\ref{algorithm_tasd}), where $w_{ik}\in\RR^R$, $i\in\{1,2\}$ and $k\notin\{h,h^+\}$, are generated independently with i.i.d.\ standard normal entries and the Tucker decomposition is performed using HOOI initialized with HOSVD.

    There exists a constant $C_\delta\geq1$, depending only on $\delta$, such that the following conclusion holds if
    \begin{enumerate}
        \item
              $
                  p_{\min} \geq C\log\left(\frac{d}{\delta}\right);
              $
        \item
              $
                \lambda_{\operatorname{Tucker}}
                  \geq
                  Cd\sigma p_{\max}^{d/4}
                  \lor
                  \sigma dC^dR^{\frac{d-1}{2}}
                  \lor
                  Cd\sigma\sqrt{p_{\max}R};
              $
        \item
              $
                  \lambda_{\min} \gtrsim \sigma\kappa_T \frac{\lambda_{\max}}{\lambda_{\min}} C_\delta^d d^{3d-5} R^{\frac{5d}{2}-4} (\log
                  R)^{\frac{d-2}{2}} (R\lor\log d)^{d-2} \sqrt{p_{\max}\lor R^{d-1}};
              $

        \item
              $
                  (R-1)\xi \leq 1-c_0
              $
              for some fixed absolute constant $c_0\in(0,1)$.
    \end{enumerate}
    Then, with probability at least $1-\delta$, there exists a permutation $\pi_h\in\mathfrak S_R$ such that
    \begin{align*}
         & \max_{r\in[R]} \min_{s\in\{-1,1\}} \left\| \hat a_{h,r} - s a_{h,\pi_h(r)} \right\|                                   \\
         & \qquad\lesssim \sigma \frac{ \sqrt{p_{\max}\lor R^{d-1}} }{ \lambda_{\operatorname{Tucker}} }                         \\
         & \qquad\quad+ \sigma\kappa_T \frac{\lambda_{\max}}{\lambda_{\min}} \frac{ C_\delta^d d^{3d-5} R^{\frac{5d}{2}-4} (\log
            R)^{\frac{d-2}{2}} (R\lor\log d)^{d-2} \sqrt{p_{\max}\lor R^{d-1}} }{ \lambda_{\min} }.
    \end{align*}
\end{Theorem}

\begin{Remark}[Dependence on the confidence level]\label{remark_tasd_confidence}
In Theorem~\ref{thm_tasd}, one may take $C_\delta=C_*\delta^{-3}\log(e/\delta)$, where $C_*\geq1$ is a sufficiently large constant depending only on the fixed coherence margin $c_0$ in condition~4 of Theorem~\ref{thm_tasd}.
We justify this choice in Supplement Section~\ref{sec_proof_remark_tasd_confidence}.
Unlike the usual exponential tail bounds in high-dimensional probability, our bound depends polynomially on $1/\delta$.
This is because the small-ball argument used to analyze the random contractions yields lower bounds on eigenvalue gaps that decay polynomially in $\delta$, and the reciprocals of these gaps enter the perturbation bounds.

A related issue arises in tensor power iteration with random initialization.
The convergence guarantee of \citet[Theorem~6]{richard2014statistical} requires a sufficiently large initial alignment with the signal, and their discussion of a uniformly random initialization gives a positive success probability at the stated signal scale.
In both settings, concentration of the observation noise cannot ensure that the additional random quantities are bounded away from zero with exponentially small failure probability, giving rise to a polynomial dependence on $1/\delta$.
\end{Remark}

\begin{Remark}[Sources of the Rank Dependence in TASD]\label{remark_tasd_rank_dependence}
The dependence on $R$ in Theorem~\ref{thm_tasd} has several distinct sources.
First, computationally, an explicit order-$d$ Tucker core of multilinear rank $(R,\ldots,R)$ contains $R^d$ entries, and the contractions used by SimDiag act on this core.

Second, the term $\sqrt{R^{d-1}}$ in Theorems~\ref{thm_tasd} and~\ref{thm_general_tucker} comes from controlling a mode unfolding of the projected $R\times\cdots\times R$ noise core, whose matrix dimensions are $R\times R^{d-1}$.
It is therefore the natural dimension term in our spectral-norm perturbation analysis, although we do not establish a lower bound proving that it is unavoidable.
The condition $(R-1)\xi\leq1-c_0$ has a different, structural role: it keeps the loading matrices and the associated Khatri--Rao products uniformly well-conditioned.

Third, the larger polynomial factors in Theorem~\ref{thm_tasd}, including $R^{\frac{5d}{2}-4}(R\lor\log d)^{d-2}$ up to logarithmic factors, arise from the randomized SimDiag perturbation argument.
The proof simultaneously lower-bounds the minimum of $R$ random contraction magnitudes and the minimum of $\binom{R}{2}$ random eigenvalue gaps, upper-bounds the Gaussian contraction-vector norms, and combines these controls with perturbation bounds for pseudoinverses and eigenvectors.
Some deterioration with $R$ is therefore expected for this randomized construction, but the specific polynomial powers obtained from our worst-case small-ball and union-bound arguments are only sufficient bounds.
Improving both the computational and statistical dependence on $R$, especially in high-rank regimes, remains an open problem.
\end{Remark}

Theorem~\ref{thm_tasd} provides a statistical guarantee that the loading matrices can be consistently estimated by TASD, up to a permutation of their columns.
As outlined in Section \ref{sec_algorithm_ALS_TASD}, we can exhaustively search for the best permutation for small $R$ and $d$, or apply Algorithm \ref{algorithm_tasd_ls} for large $R$ or $d$.
We numerically show in Supplement Section \ref{sec_add_sim_set2} that the additional error introduced by Algorithm \ref{algorithm_tasd_ls} is negligible. 

To ensure accurate estimation of $\hat A_h$ in TASD, a high-quality approximation of the Tucker decomposition is essential.
A widely used method for Tucker decomposition is the Higher-Order Orthogonal Iteration (HOOI).
The statistical properties of this approach have been analyzed in the $d = 3$ case in~\cite{zhang2018tensor}.
Here, we extend this theory to tensors of arbitrary order $d \geq 3$ and present the formal guarantee below.



\begin{Theorem}[Estimation Error of HOOI for General Order-$d$ Tensor Decomposition]\label{thm_general_tucker}
    Assume $\Y=\X+\Z=\S\times_{k=1}^dU_k+\Z\in\RR^{p_1\times\cdots\times p_d}$, where $U_k\in\RR^{p_k\times r_k}$ has orthonormal columns for every $k\in[d]$ and $\S\in\RR^{r_1\times\cdots\times r_d}$.
    Denote $\lambda_{\operatorname{Tucker}}:=\min_{k\in[d]}\sigma_{r_k}(\MM_k(\X))$, and let $\hat U_k^{(t)}$ denote the HOOI iterates initialized by HOSVD as in Supplement Algorithm~\ref{algorithm_hooi}.
    There exist absolute constants $C_1,C_2,C_3,c_2>0$ such that, if $\lambda_{\operatorname{Tucker}}\geq C_1d\sigma p_{\max}^{d/4}\lor d\sigma C_1^dr_{\max}^{\frac{d-1}{2}}\lor C_1d\sigma\sqrt{p_{\max}r_{\max}}$ and $p_{\min}>C_1\log d$, then there exists a nonnegative integer $$T\leq C_3\left[1+\log\left(1+\frac{\sigma p_{\max}^{d/2}}{\lambda_{\operatorname{Tucker}}\left(\sqrt{p_{\max}}+\sqrt{r_{\max}^{d-1}}\right)}\right)\right]$$ for which the following holds.
    Set $\hat U_k:=\hat U_k^{(T)}$ for every $k\in[d]$.
    Then, with probability at least $1-C_2d\exp(-c_2p_{\min})$, we have
    \be
    \label{eq_bound_U_hat_general_d}
        \max_{k\in[d]}
        \left\|
            U_kU_k\T-\hat U_k\hat U_k\T
        \right\|
        \leq
        C_3\sigma
        \frac{
            \sqrt{p_{\max}}+\sqrt{r_{\max}^{d-1}}
        }{
            \lambda_{\operatorname{Tucker}}
        },
    \ee
    and
    \be
    \label{eq_bound_UZ_hat_general_d}
        \max_{h\in[d]}
        \left\|
            \MM_h\left(
                \Z\times_{k=1}^d\hat U_k\T
            \right)
        \right\|
        \leq
        C_3\sigma
        \left(
            \sqrt{p_{\max}}+\sqrt{r_{\max}^{d-1}}
        \right).
    \ee
\end{Theorem}

\begin{Remark}
    Compared to the results in~\cite{zhang2018tensor}, our Theorem~\ref{thm_general_tucker} requires weaker assumptions.
    In particular, it does not impose the conditions $p_{\max} \lesssim p_{\min}$, or $r_{\max}^2 \lesssim p_{\max}$.
    Under the dimension-balancing and rank conditions of \cite{zhang2018tensor}, and for $d=3$, Theorem 5 recovers the same signal-strength and subspace-error orders as their HOOI result.
\end{Remark}

\subsection{Overall Theory}\label{sec_overall}
By combining Theorems~\ref{thm_local} and~\ref{thm_tasd}, we establish a global convergence guarantee for ALS when initialized with TASD.
\begin{Corollary}[Overall Theory for General Rank CP Tensor Decomposition]\label{corollary_global}
    Fix a confidence level $\delta\in(0,1/2)$ and set $\delta_0:=\delta/(2d)$.
    Let $C_{\delta_0}$ be the confidence-dependent constant appearing in Theorem~\ref{thm_tasd}.
    In the same setting as Theorem~\ref{thm_tasd}, assume that TASD is applied to each mode, the resulting loading matrices have been correctly permuted, and
    \begin{enumerate}
        \item $p_{\min}\geq C\log\left(2d^2/\delta\right)$,
        \item
        $
            \lambda_{\operatorname{Tucker}}
            \geq
            d\sigma
            \left[
                Cp_{\max}^{d/4}
                \lor
                C^dR^{\frac{d-1}{2}}
                \lor
                C\frac{\lambda_{\max}}{\lambda_{\min}}
                (d+R)
                \sqrt{p_{\max}\lor R^{d-1}}
            \right];
        $
        \item $\lambda_{\min}\gtrsim \sigma\kappa_T\left(\frac{\lambda_{\max}}{\lambda_{\min}}\right)^2 C_{\delta_0}^d d^{3d-5}(d+R)R^{\frac{5d}{2}-4}(\log R)^{\frac{d-2}{2}}(R\lor\log d)^{d-2}\sqrt{p_{\max}\lor R^{d-1}}$, and
        \item $(R-1)\xi\leq 1-c_0,\qquad \xi\leq\begin{cases}
                      (CdR^{1/(d-3)})^{-1}, & d>3, \\
                      (CR^{1/2})^{-1},      & d=3.
                  \end{cases}$
    \end{enumerate}
    Then the following holds with probability at least $1-\delta$ for ALS initialized by TASD.
    Let $\tau_{\mathrm{ALS}}$ be defined as in \eqref{eq_transition_scale}.
    Then:
    \begin{itemize}
        \item If $\sigma\sqrt{p_{\max}}/\lambda_{\min}\geq\tau_{\mathrm{ALS}}$, then $\varepsilon_t\lesssim\sigma\sqrt{p_{\max}}/\lambda_{\min}$ for all $t > T$ with some $T\lesssim\log\log\left(\lambda_{\min}/(\sigma\sqrt{p_{\max}})\right)$.

        \item If $\sigma\sqrt{p_{\max}}/\lambda_{\min}<\tau_{\mathrm{ALS}}$, and we additionally have $(\lambda_{\max}/\lambda_{\min})Rd\xi^{d-2}\leq c$ for some absolute constant $c$, then $\varepsilon_t\lesssim\sigma\sqrt{p_{\max}}/\lambda_{\min}$ for all $t > T$ with some $T\lesssim\log\log\left(\lambda_{\min}/(\sigma\sqrt{p_{\max}})\right)+\log\left(\tau_{\mathrm{ALS}}\lambda_{\min}/(\sigma\sqrt{p_{\max}})\right)$.
    \end{itemize}
\end{Corollary}

\begin{Remark}[A Simplified TASD--ALS Regime]\label{remark_simplified_tasd_als}
Consider the balanced setting $p_k\asymp p$ and suppose that $d$, $R$, $\delta$, $\lambda_{\max}/\lambda_{\min}$, and $\kappa_T$ are treated as constants.
Also suppose that the loading matrices are uniformly well-conditioned, as ensured by $(R-1)\xi\leq1-c_0$.
In this regime, the local ALS conditions in Theorem~\ref{thm_local} reduce to a sufficiently accurate constant-quality initialization and $\lambda_{\min}\gtrsim\sigma\sqrt p$, and the resulting terminal loading-estimation error is of order $\sigma\sqrt p/\lambda_{\min}$.

The TASD conditions in Theorem~\ref{thm_tasd} reduce to $\lambda_{\operatorname{Tucker}}\gtrsim\sigma p^{d/4}$ and $\lambda_{\min}\gtrsim\sigma\sqrt p$.
Under the assumed conditioning and bounded coefficient ratio, $\lambda_{\operatorname{Tucker}}$ and $\lambda_{\min}$ are comparable up to constants depending only on the fixed conditioning parameters.
Consequently, the combined TASD--ALS guarantee applies at the sufficient signal scale $\lambda_{\min}\gtrsim\sigma p^{d/4}$, while its terminal statistical error remains of order $\sigma\sqrt p/\lambda_{\min}$.
The stronger $p^{d/4}$ scale is therefore used to construct a polynomial-time warm start and should not be interpreted as the information-theoretic threshold for the terminal estimation error.
\end{Remark}

\begin{Remark}[Statistical Optimality and Computational Signal Strength]
Theorem~\ref{thm_lower_bound} is proved over the equal-strength orthogonal class $\mathcal{F}_{R,\lambda}^{\mathrm{orth}}$. 
Combining Corollary~\ref{corollary_global} with Theorem~\ref{thm_lower_bound} therefore shows that
$
    {\sigma\sqrt{p_{\max}}} / {\lambda}
$
is the minimax-optimal terminal loading-estimation rate in the high-signal regime covered by TASD--ALS.

The signal-strength requirements in Corollary~\ref{corollary_global} play a different role from the information-theoretic lower bound, as they guarantee that the polynomial-time TASD initializer enters the local convergence region of ALS.
They should not be interpreted as an information-theoretic threshold for the terminal estimation error.

For weaker signals, whether the information-theoretic rate is achievable in polynomial time remains open under our analysis. In the related tensor-SVD setting, \cite{zhang2018tensor} identified an intermediate signal regime where statistically optimal estimation is possible but polynomial-time consistent estimation is computationally hard under the hypergraphic planted clique hypothesis. This suggests that the stronger signal requirement for our polynomial-time procedure may reflect a genuine statistical--computational gap, although we do not pursue computational lower bounds in this work.
\end{Remark}

\begin{Remark}
    Recently, \cite{han2022tensor} proposed a method for tensor CP decomposition, called Iterative Concurrent Orthogonalization (ICO), initialized by Composite PCA (CPCA).
    Their approach relies on performing SVD on a matrix derived from tensor unfolding.
    As a result, the quantity $G := \min_{i\neq j \in [R]} ||\lambda_i| - |\lambda_j||$ plays a central role in their theoretical analysis.
    Specifically, under certain conditions, their final estimation error contains the multiplicative factor $\alpha^{1-d} \cdot \frac{\sigma \sqrt{p_{\max}}}{\lambda_{\min}},$ where $\alpha = \sqrt{1 - \sqrt{\xi}} - \frac{(\sqrt{R} + 1)(\sqrt{\xi} \lambda_{\max} + \sigma \sqrt{p_{\max}})}{G}.$

    In contrast, our estimation bound in Corollary~\ref{corollary_global} does not depend on this gap-dependent factor $\alpha$, and its terminal rate matches the minimax lower-bound rate established in Theorem~\ref{thm_lower_bound}.
    This distinction is further illustrated by the loss curves in our simulation study.
    Moreover, while ICO with CPCA does not recover the loading vectors when the gap $G$ is small, our method ALS+TASD remains effective even when $G = 0$.

\end{Remark}

\section{Simulation Studies}\label{sec_simulation}

We evaluate TASD, ALS initialized by TASD (TASD--ALS), and several competing procedures under Gaussian instances of model~\eqref{eq_main_model}. All loading vectors are normalized to have unit Euclidean norm before the signal tensor is formed. The reported loading loss minimizes over a common permutation of the CP components and then takes the maximum sign-invariant loading-vector error over all modes and components. Detailed implementation settings are collected in Supplement Section~\ref{sec_simulation_details}.

\subsection{Comparison with Existing Methods}\label{sec_sim_method_comparison}

\begin{figure}[htbp]
    \centering
    \includegraphics[width=\linewidth]{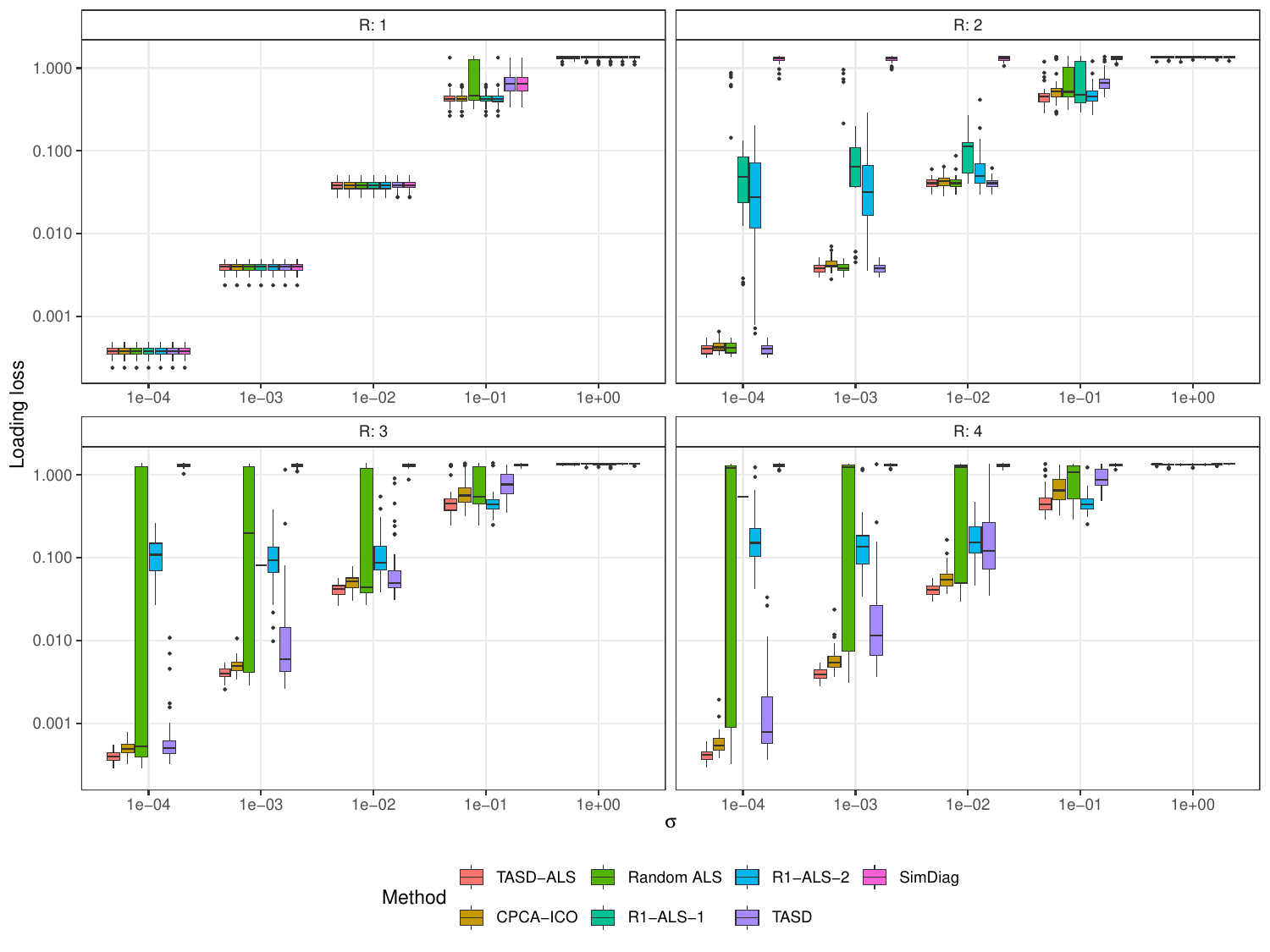}
    \caption{Loading-loss comparison over ranks $R\in\{1,2,3,4\}$ and positive noise levels $\sigma\in\{10^{-4},10^{-3},10^{-2},10^{-1},1\}$. Each boxplot is based on the finite outputs among 50 independent repetitions; numerical nonreturns are not assigned artificial loss values.}
    \label{fig_1}
\end{figure}

We first consider order-$3$ tensors with dimensions $(p_1,p_2,p_3)=(15,12,10)$, ranks $R\in\{1,2,3,4\}$, coefficients $\lambda_r=r$, and i.i.d. Gaussian noise. We compare TASD--ALS and TASD with CPCA--ICO, Random ALS, two rank-one-ALS procedures, and SimDiag over 50 independent repetitions. Figure~\ref{fig_1} shows that TASD--ALS is accurate and stable in the low- and moderate-noise regimes, particularly for $R\in\{3,4\}$, while direct SimDiag becomes unstable as soon as noise is introduced. ALS refinement improves TASD substantially for the larger ranks, and Random ALS becomes unreliable in the more difficult rank-$4$ settings. At $\sigma=1$, all methods have large loading error.

\subsection{Direct Verification of the Statistical Rate}\label{sec_sim_statistical_rate}

\begin{figure}[htbp]
    \centering
    \includegraphics[width=\linewidth]{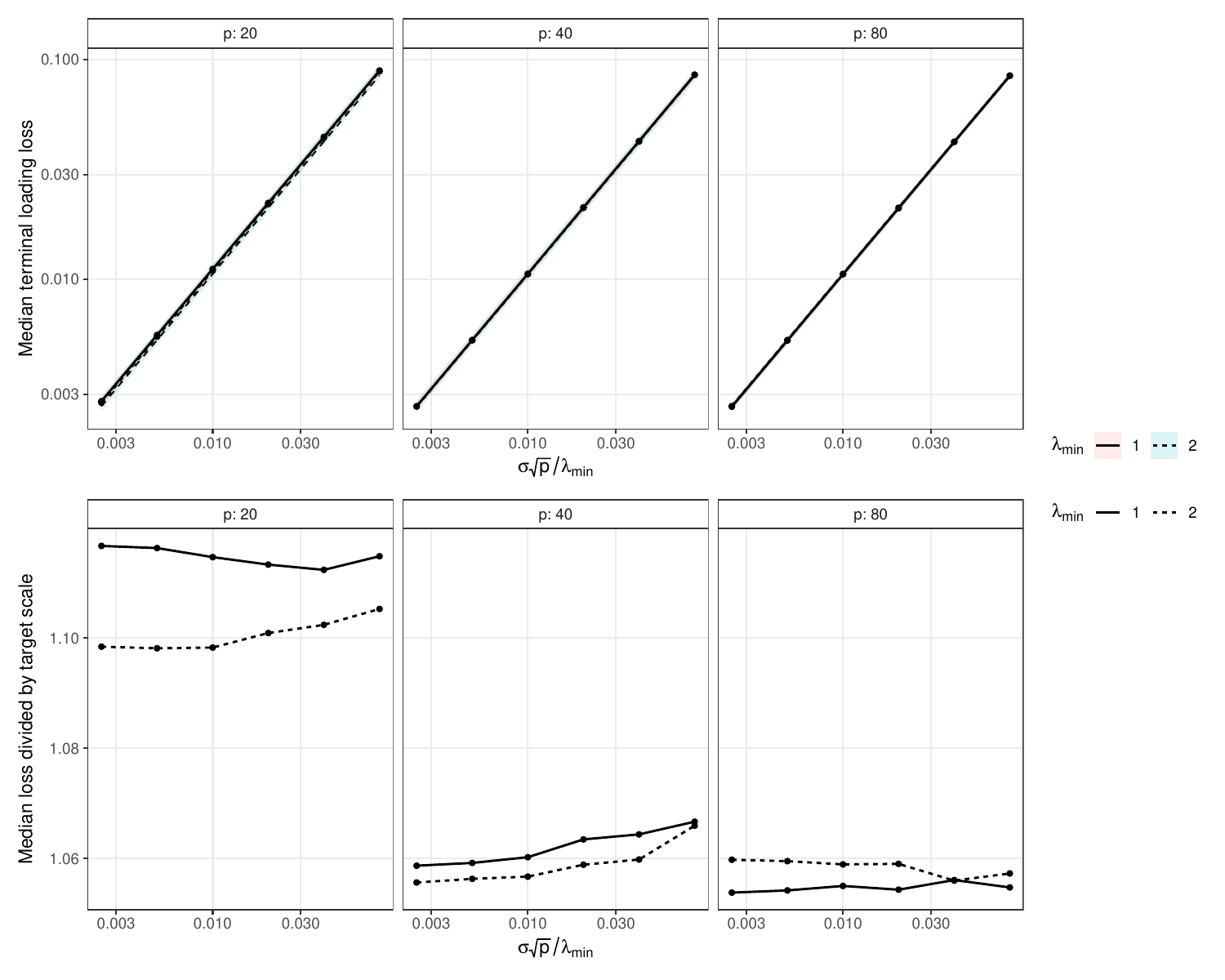}
    \caption{Direct verification of the terminal loading-estimation rate. The upper panels plot the median terminal loss against $\sigma\sqrt p/\lambda_{\min}$, with a dashed slope-one reference line. The lower panels plot the median loss divided by this target scale. Each setting uses 100 independent repetitions.}
    \label{fig_sim_statistical_rate}
\end{figure}

To isolate the local statistical behavior of ALS, we use a controlled warm start with loading loss $0.12$. We set $d=3$, $R=3$, coherence $\xi=0.05$, $p\in\{20,40,80\}$, and vary the target scale $\sigma\sqrt p/\lambda_{\min}$ over $\{0.0025,0.005,0.01,0.02,0.04,0.08\}$. Figure~\ref{fig_sim_statistical_rate} shows an essentially slope-one relationship on logarithmic axes. The fitted slopes range from $0.999$ to $1.002$, and the loading loss divided by $\sigma\sqrt p/\lambda_{\min}$ remains nearly constant. These results directly support the terminal rate $\sigma\sqrt p/\lambda_{\min}$ predicted by Theorem~\ref{thm_local} within the local convergence region.

\subsection{Quadratic and Linear Convergence Regimes}\label{sec_sim_convergence_regimes}

\begin{figure}[htbp]
    \centering
    \includegraphics[width=\linewidth]{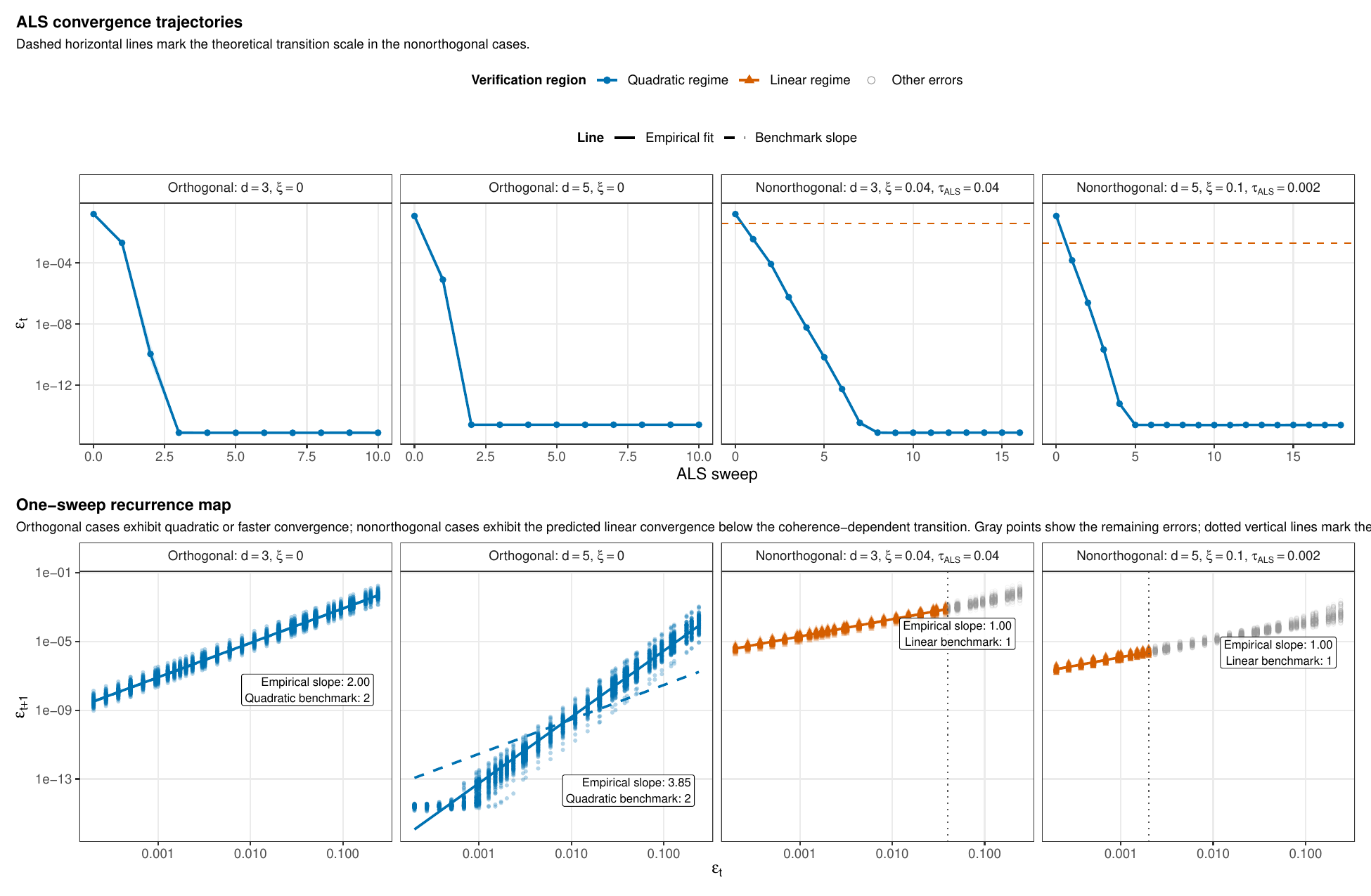}
    \caption{ALS convergence trajectories and one-sweep recurrence maps in noiseless rank-$3$ problems. The orthogonal cases verify quadratic or faster contraction. In the nonorthogonal cases, the colored points below the dotted transition line verify the predicted linear regime; gray points are outside the region used for the linear-slope fit.}
    \label{fig_sim_convergence_regimes}
\end{figure}

We next remove stochastic noise and study the deterministic ALS recurrence from controlled initializations. We consider orthogonal and equicorrelated rank-$3$ tensors for $d\in\{3,5\}$ and evaluate both complete trajectories and one-sweep maps over a dense grid of input errors. Figure~\ref{fig_sim_convergence_regimes} shows a fitted slope of $2.00$ for the orthogonal order-$3$ case and a faster slope of $3.85$ for the orthogonal order-$5$ case. In the nonorthogonal cases, the fitted slopes below the coherence-dependent transition scale are $1.00$ for both tensor orders. Thus, the experiment supports quadratic or faster contraction in the orthogonal regime and the predicted linear contraction below the transition scale in the nonorthogonal regime.

\section{Discussion}\label{sec_discussion}

This work advances the theoretical understanding of CPD and offers rigorous insight into the behavior of the alternating least squares (ALS) and Tucker-based Approximation with Simultaneous Diagonalization (TASD) algorithms.
Motivated by long-standing questions on CP tensor decomposition, particularly in general, non-orthogonal, higher-rank scenarios, we make three major contributions.
First, we establish non-asymptotic, minimax-optimal error bounds for ALS under general CP models with arbitrary rank and tensor order, assuming a suitable initialization; these upper bounds are matched by minimax lower bounds, demonstrating that ALS attains the statistically optimal rate rather than merely an order-wise upper limit.
Second, we introduce Tucker-based approximation with simultaneous diagonalization (TASD) and prove that TASD yields statistically consistent estimates in low-rank regimes and, when used to warm-start ALS, leads to provably accurate recovery.
Empirically, TASD is remarkably more robust to noise than traditional methods.
Third, we characterize a quadratic convergence regime and, when further refinement is required below a coherence-dependent transition scale, a subsequent linear regime, thereby formalizing empirical observations of rapid convergence in near-orthogonal settings.
More specifically, in the rank-one case, we prove that ALS with spectral initialization achieves the statistically optimal error rate, even within just one or two iterations.

Despite these advances, several limitations and open questions remain, requiring further investigation.
Our convergence proof requires the initialization error to lie below a specific threshold, and whether this threshold is minimax-sharp is not yet clear.
The current TASD implementation explicitly forms a Tucker core containing $R^d$ entries, and its perturbation guarantee has conservative high-order polynomial dependence on $R$.
Developing an initializer with improved computational and statistical rank dependence remains an important open problem.
Finally, extending the present framework to structured tensor models, such as binary, binomial, or non-negative tensors, would further broaden its applicability to modern data-analytic tasks.

\begin{acks}[Acknowledgments]
    The authors would like to express their deepest gratitude to Professor Alexandre B.
    Tsybakov for inspiring discussions throughout the course of this work. Anru Zhang is the corresponding author.
\end{acks}

\begin{funding}
    Olga Klopp gratefully acknowledges financial support for her research by Franco-American commission for educational exchange (Fulbright) grant and the  U.
    CY Initiative (grant ``Investissements d'Avenir'' ANR-16-IDEX-0008).
    This paper has also been funded by the Agence Nationale de la Recherche under grant ANR-17-EURE-0010 (Investissements d'Avenir program).
    Anru Zhang was supported in part by NSF Grant CAREER-2203741.
\end{funding}

\begin{supplement}
    \stitle{Supplementary material for ``Revisit CP Tensor Decomposition: Statistical Optimality and Fast Convergence''}
    \sdescription{We collect the additional simulation studies and all technical proofs in this supplementary material.}
\end{supplement}


\bibliographystyle{imsart-nameyear} 
\bibliography{bib}       

\clearpage
\setcounter{page}{1}
\appendix

\begin{center}
    \textbf{\Large{}{}{}{}{}{} Supplementary material for ``Revisit CP Tensor Decomposition: Statistical Optimality and Fast Convergence''}{\Large{}{}}\\
    {\huge{}{} }{\huge\par}
    \par\end{center}

\begin{center}
    \large{}{}{}Runshi Tang\footnote{Department of Statistics, Columbia University}, ~ Julien Chhor\footnote{Toulouse School of Economics, University of Toulouse Capitole, Toulouse, France}, ~ Olga Klopp\footnote{ESSEC Business School, Cergy France},~ and ~ Anru R.
    Zhang\footnote{Department of Biostatistics \& Bioinformatics and Department of Computer Science, Duke University}{\large\par}
    \par\end{center}

We collect the additional simulation studies and all technical proofs in these supplementary materials.

\startcontents[supplement]

\begin{center}
    \textbf{\large SUPPLEMENT: CONTENTS}
\end{center}
\markboth{Supplement}{Supplement}

\printcontents[supplement]{}{1}{\setcounter{tocdepth}{2}}
\clearpage

\section{Additional Simulations}\label{sec_additional_simulations}

Detailed implementation settings are collected in Supplement Section~\ref{sec_simulation_details}.

\subsection{Signal Strength and the Two Reference Exponents}\label{sec_sim_signal_strength}

We set $d=3$, $R=3$, $p_1=p_2=p_3=p\in\{20,30,50\}$, $\lambda_r=r$, and $\sigma=0.25p^{-\alpha}$ for $\alpha\in\{0.25,0.30,\ldots,1\}$. Within each repetition, the signal, standardized noise tensor, and TASD randomization are held fixed across $\alpha$. The vertical lines in Supplement Figure~\ref{fig_2} mark the order-level references $\alpha=1/2$ for the terminal ALS error and $\alpha=3/4$ for the sufficient TASD warm-start scale. The loading losses decrease smoothly after $\alpha=1/2$, while the recovery probabilities rise rapidly around $\alpha\approx0.6$--$0.7$. There is no sharp finite-sample break at $\alpha=3/4$, which is consistent with that exponent being a sufficient order-level condition rather than an exact phase-transition point. Across the displayed range, ALS refinement consistently improves the TASD initialization.

\begin{figure}[htbp]
    \centering
    \includegraphics[width=\linewidth]{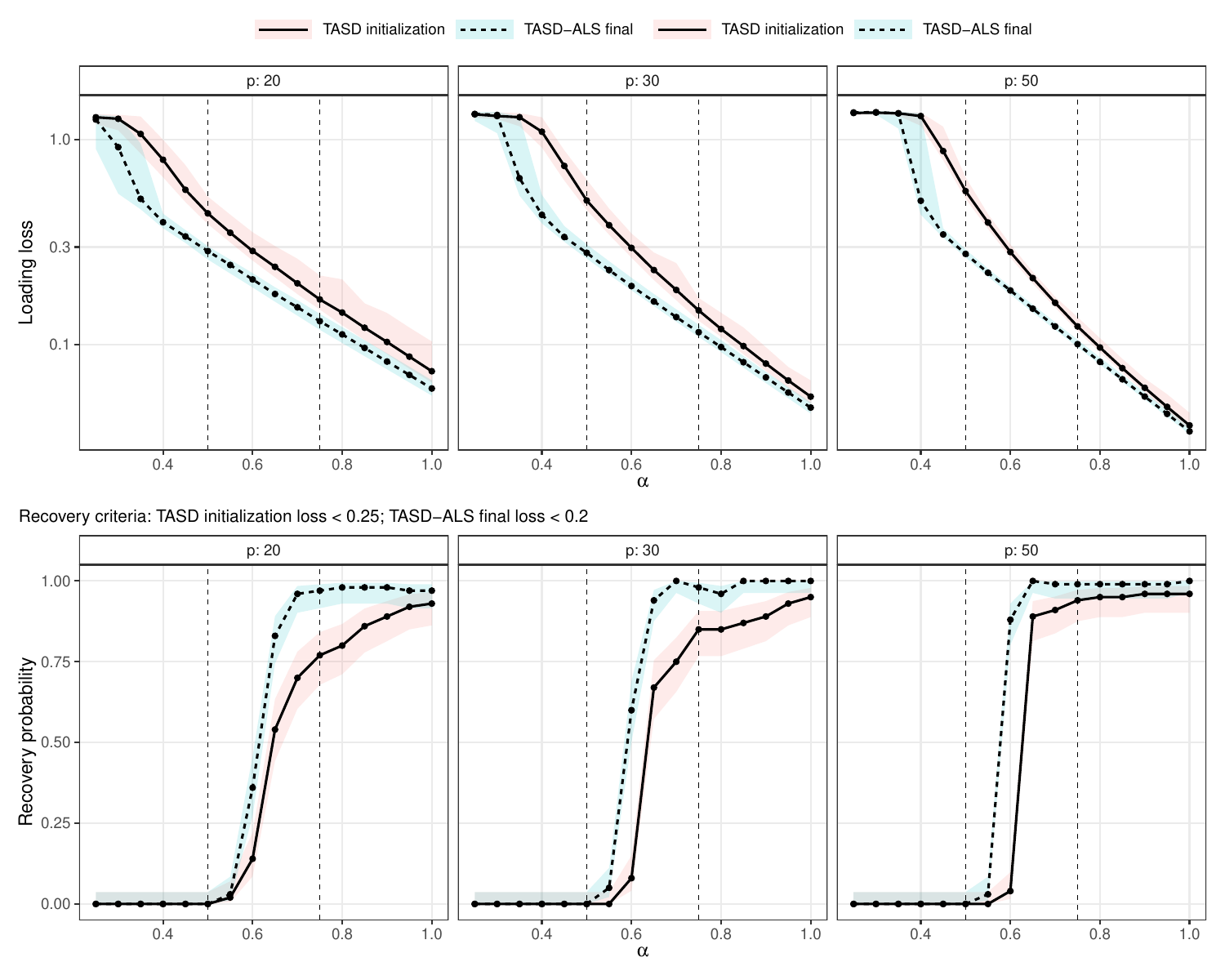}
    \caption{TASD initialization and TASD--ALS performance as a function of $\alpha$. The upper panels show median loading losses with interquartile bands, and the lower panels show recovery probabilities using initialization loss below $0.25$ and final loss below $0.20$. Dashed vertical lines mark the reference exponents $1/2$ and $3/4$.}
    \label{fig_2}
\end{figure}

\subsection{Rank-One Convergence Speed}\label{sec_sim_rank_one_speed}

\begin{figure}[htbp]
    \centering
    \includegraphics[width=\linewidth]{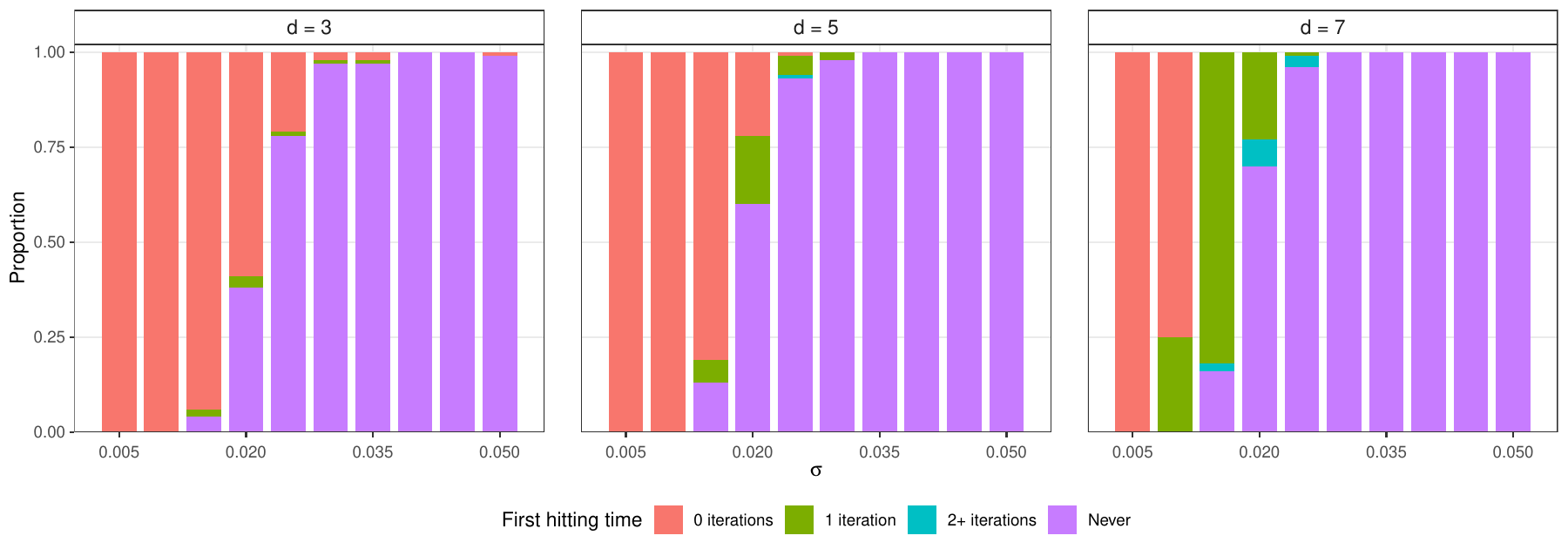}
    \caption{First hitting time for the rank-one loading loss to fall below $0.05$, based on 100 repetitions per setting. ``$0$ iterations'' means that the spectral initialization already satisfies the criterion, and ``Never'' means that the criterion is not reached within 20 ALS sweeps.}
    \label{fig_3}
\end{figure}

For the rank-one setting, we take $p_k=5$, $d\in\{3,5,7\}$, $\lambda=1$, and $\sigma\in\{0.005,0.010,\ldots,0.050\}$. Supplement Figure~\ref{fig_3} reports the first iteration at which the loading loss falls below $0.05$. In the recoverable low-noise regime, the spectral initialization is often already accurate, and the remaining successful runs require at most one or two ALS sweeps. As $d$ or $\sigma$ increases, successful recovery requires a stronger signal-to-noise ratio, producing the observed all-or-nothing pattern.

\subsection{Effect of Loading Coherence}\label{sec_sim_coherence}

\begin{figure}[htbp]
    \centering
    \includegraphics[width=\linewidth]{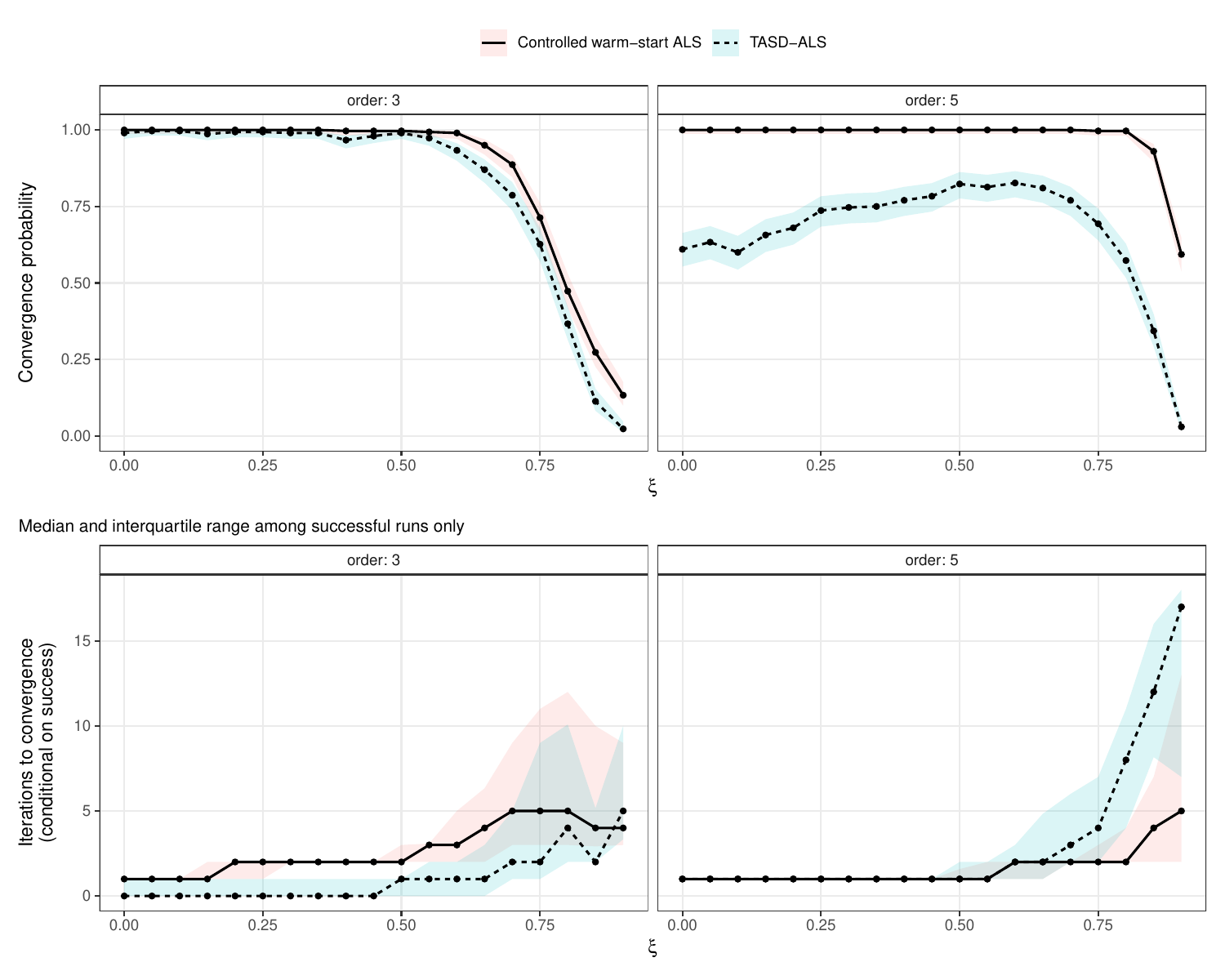}
    \caption{Effect of coherence on controlled warm-start ALS and the complete TASD--ALS pipeline. The upper panels show convergence probabilities with 95\% Wilson intervals. The lower panels show the median and interquartile range of the first hitting time conditional on successful recovery.}
    \label{fig_3_1}
\end{figure}

We set $p_k=5$, $R=3$, $\lambda_r=r$, $\sigma=0.01$, $d\in\{3,5\}$, and vary the exact pairwise coherence over $\xi\in\{0,0.05,\ldots,0.9\}$. We compare ALS from a controlled loading error of $0.10$ with the complete TASD--ALS pipeline, using common bases, noise, perturbations, and TASD randomization across values of $\xi$. Supplement Figure~\ref{fig_3_1} shows that controlled warm-start ALS remains highly reliable until the loading matrices become strongly coherent, whereas TASD--ALS can be less stable because coherence also affects the randomized initialization. In particular, the nonmonotone behavior for $d=5$ at small and moderate $\xi$ is primarily an initialization effect rather than a monotonicity property of local ALS. Both procedures deteriorate once $\xi$ becomes large, well beyond the sufficient coherence regime of the theory.

\subsection{Larger Dimensions, Larger Ranks, and Unbalanced Modes}\label{sec_sim_scaling_robustness}

\begin{figure}[htbp]
    \centering
    \includegraphics[width=\linewidth]{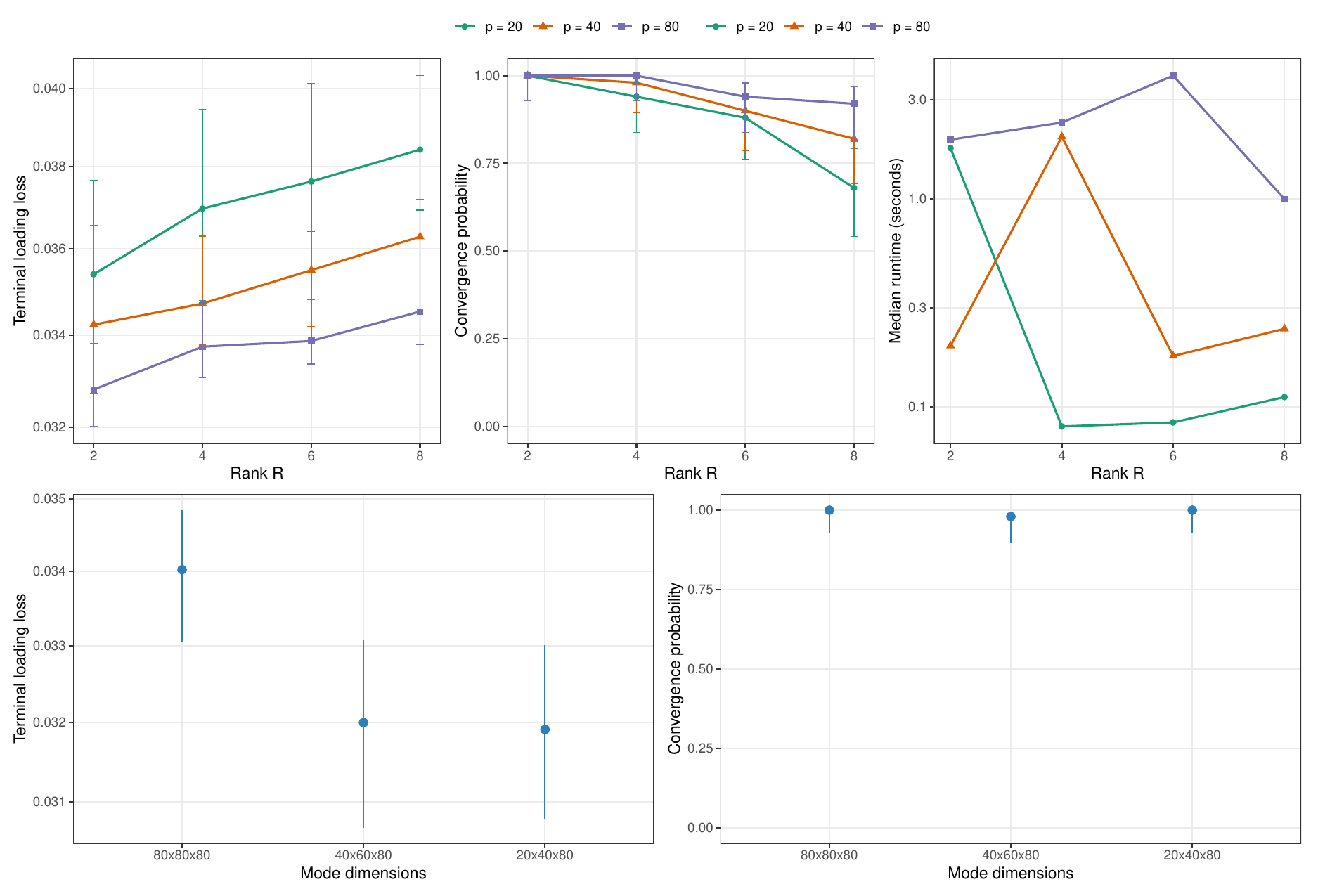}
    \caption{Robustness to larger dimensions, larger ranks, and unbalanced mode dimensions. The upper panels report terminal loading loss, convergence probability, and runtime in balanced designs. The lower panels compare terminal loss and convergence probability across the three unbalanced designs. Loss error bars show interquartile ranges, and probability error bars are 95\% Wilson intervals.}
    \label{fig_sim_scaling_robustness}
\end{figure}

To examine robustness beyond the small baseline designs, we consider balanced order-$3$ tensors with $p\in\{20,40,80\}$ and $R\in\{2,4,6,8\}$. We hold $(R-1)\xi=0.30$ and $\sigma\sqrt p/\lambda_{\min}=0.03$ fixed, so that the loading-matrix conditioning and nominal terminal statistical scale remain comparable across ranks and dimensions. We also compare dimensions $(80,80,80)$, $(40,60,80)$, and $(20,40,80)$ at $R=4$. Supplement Figure~\ref{fig_sim_scaling_robustness} shows that the terminal loss remains in a narrow range as $p$ and $R$ increase, although the probability of successful convergence decreases with rank, especially when $p=20$. The unbalanced designs have nearly identical terminal accuracy and convergence probability, indicating no evident instability from dimension imbalance in this range.

\subsection{Equal-Strength Components}\label{sec_add_sim_set1}

\begin{figure}[htbp]
    \centering
    \includegraphics[width=\linewidth]{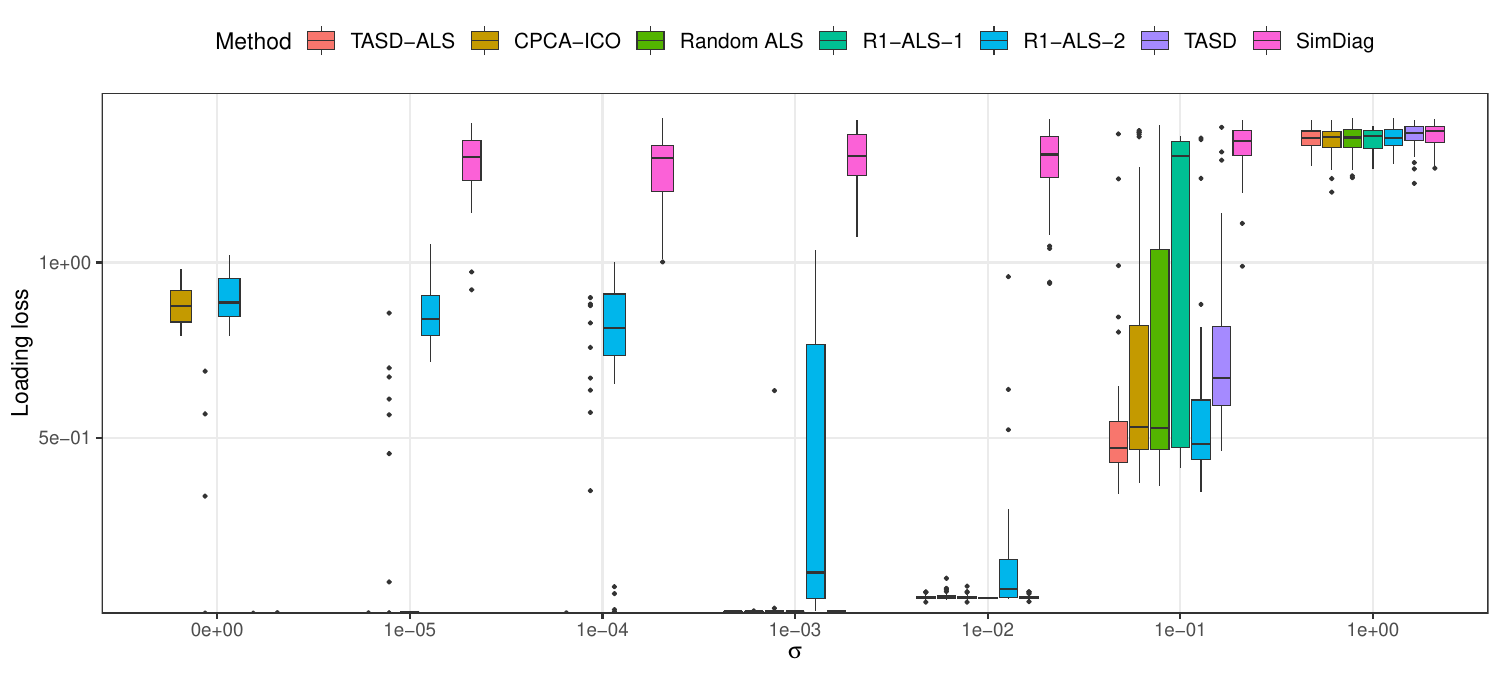}
    \caption{Method comparison for equal-strength rank-$2$ components, $\lambda_1=\lambda_2=1$. Each boxplot is based on finite outputs among 50 independent repetitions.}
    \label{fig_4}
\end{figure}

We repeat the method comparison with $d=3$, dimensions $(15,12,10)$, $R=2$, and $\lambda_1=\lambda_2=1$, so the coefficient eigengap is zero. Supplement Figure~\ref{fig_4} shows that TASD and TASD--ALS recover the loadings exactly in the noiseless setting and remain accurate for small noise, without requiring separation between the coefficient magnitudes. CPCA--ICO fails in the exactly noiseless equal-strength case, while direct SimDiag becomes unstable under even very small noise. The rank-one procedures also remain unreliable because a best rank-one approximation need not coincide with either true CP component.

\subsection{TASD Part II}\label{sec_add_sim_set2}

Finally, we compare direct three-mode TASD with TASD Part~II for order-$3$ tensors with dimensions $(15,15,15)$, $R\in\{2,3,4,5,6\}$, and $\sigma\in\{0.1,0.5,1\}$. Conditional on returning finite loading matrices, the two constructions have similar loading loss, while TASD Part~II is typically faster. 

\begin{figure}[htbp]
    \centering
    \includegraphics[width=\linewidth]{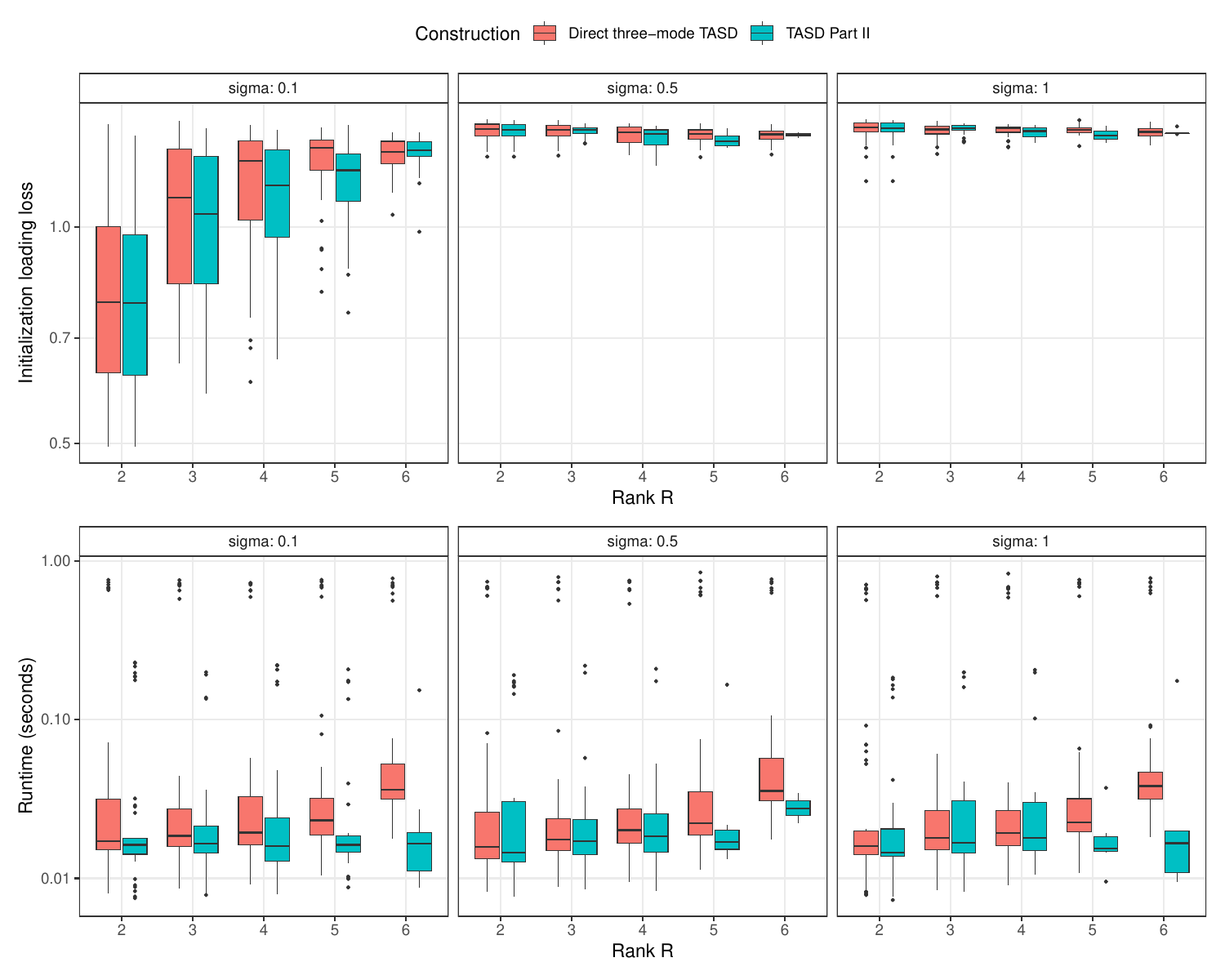}
    \caption{Initialization loss and runtime for direct three-mode TASD and TASD Part~II. Boxplots include only finite outputs from 50 repetitions per setting.}
    \label{fig_7}
\end{figure}

\section{Simulation Details}\label{sec_simulation_details}

\subsection{Common Data Generation, Loss, and Reporting Rules}\label{sec_sim_details_common}

This subsection provides the common details for the simulation studies in Section~\ref{sec_simulation}.

All experiments use the Gaussian special case of model~\eqref{eq_main_model}. We generate
\[
    \Y=\X+\sigma\Z,
    \qquad
    \X=\sum_{r=1}^R\lambda_r a_{1,r}\circ\cdots\circ a_{d,r},
\]
where the entries of $\Z$ are independent standard normal random variables. For random loading matrices, the entries of an unnormalized vector are drawn independently from $\mathrm{Unif}[-1,1]$ and the vector is then normalized to unit Euclidean norm before $\X$ is formed. Thus, the factor norms and the coefficients $\lambda_r$ are not confounded.

For coherence-controlled experiments, let $Q_k\in\RR^{p_k\times R}$ have orthonormal columns and define $G_\xi=(1-\xi)I_R+\xi\mathbf{1}\mathbf{1}\T$. If $C_\xi$ is the upper-triangular Cholesky factor satisfying $C_\xi\T C_\xi=G_\xi$, we set $A_k=Q_kC_\xi$. Consequently, $A_k\T A_k=G_\xi$, so every loading vector has unit norm and every off-diagonal inner product is exactly $\xi$ up to floating-point error.

For estimated loading matrices $\widehat A_k=[\hat a_{k,1},\ldots,\hat a_{k,R}]$, we report
\[
    \mathcal L(\widehat A,A)
    :=
    \min_{\pi\in\mathfrak S_R}
    \max_{k\in[d]}
    \max_{r\in[R]}
    \min_{s\in\{-1,1\}}
    \|\hat a_{k,r}-s a_{k,\pi(r)}\|.
\]
The same permutation is used in every mode. Numerical errors that prevent a method from returning finite loading matrices are not assigned artificial loss values. Boxplots and loss summaries use finite outputs only, whereas the recovery-probability experiments count numerical errors as failures. Unless otherwise stated, ALS is run for at most 20 full sweeps.

\subsection{Details for the Method Comparison}\label{sec_sim_details_method_comparison}

This subsection provides details for the simulation study in Section~\ref{sec_sim_method_comparison}.

Experiment~1 uses $d=3$, $(p_1,p_2,p_3)=(15,12,10)$, $R\in\{1,2,3,4\}$, $\lambda_r=r$, and $\sigma\in\{0,10^{-4},10^{-3},10^{-2},10^{-1},1\}$. Each configuration has 50 independent repetitions. The compared methods are TASD--ALS, TASD, CPCA--ICO, Random ALS, R1--ALS--1, R1--ALS--2, and SimDiag. Random ALS, TASD--ALS, CPCA--ICO, and the deflation baseline use at most 20 iterations or sweeps. R1--ALS--1 generates 50 rank-one candidates and retains candidates whose maximum cross-mode absolute inner product with a previously selected component is at most $0.5$.

The displayed figure omits $\sigma=0$ because it focuses on the positive noise grid. TASD--ALS returned finite outputs in all 800 runs with $\sigma\leq10^{-2}$ and in 1098 of the 1200 runs overall. R1--ALS--1 returned $R$ separated candidates in only 540 of 1200 runs, so its displayed boxplots are strongly conditional on numerical completion. These completion differences should be kept in mind when comparing the boxplots.

\subsection{Details for the Signal-Strength Experiment}\label{sec_sim_details_signal_strength}

This subsection provides details for the simulation study in Supplement Section~\ref{sec_sim_signal_strength}.

Experiment~2 uses $d=3$, $R=3$, $p\in\{20,30,50\}$, $\lambda=(1,2,3)$, $\sigma=0.25p^{-\alpha}$, and $\alpha\in\{0.25,0.30,\ldots,1\}$. For each $p$, we generate 100 independent signal and standardized-noise pairs. Within a repetition, the same signal tensor, standardized noise tensor, and TASD random seed are reused for every $\alpha$, so variation across $\alpha$ is paired. TASD--ALS is run for at most 20 sweeps.

The recovery indicators use loading-loss thresholds $0.25$ for TASD initialization and $0.20$ for the final TASD--ALS estimate. These thresholds are descriptive finite-sample criteria rather than constants appearing in the theory. Of the 4800 TASD--ALS fits, 310 produced a rank-deficient SimDiag eigenvector matrix; these fits are counted as recovery failures and omitted from the median-loss curves. At $p=30$, for example, the median final loss decreases from $0.281$ at $\alpha=0.50$ to $0.163$ at $\alpha=0.65$ and $0.115$ at $\alpha=0.75$, while the corresponding final recovery probabilities are $0$, $0.94$, and $0.98$. The gradual change supports an order-level interpretation of $1/2$ and $3/4$ rather than exact finite-sample thresholds.

\subsection{Details for the Statistical-Rate Experiment}\label{sec_sim_details_statistical_rate}

This subsection provides details for the simulation study in Section~\ref{sec_sim_statistical_rate}.

Experiment~3 uses $d=3$, $R=3$, $p\in\{20,40,80\}$, exact coherence $\xi=0.05$, and coefficients $\lambda_r=\lambda_{\min}\{1,1.25,1.5\}_r$ for $\lambda_{\min}\in\{1,2\}$. The initialization is constructed to have loading loss exactly $0.12$. For each combination of $p$ and $\lambda_{\min}$, we vary $s:=\sigma\sqrt p/\lambda_{\min}$ over $\{0.0025,0.005,0.01,0.02,0.04,0.08\}$ by setting $\sigma=s\lambda_{\min}/\sqrt p$. The same signal, perturbation directions, and standardized noise are reused across the six values of $s$ within a repetition. Each setting has 100 repetitions, and ALS is run for exactly 20 sweeps.

All 3600 fits completed successfully. Regressing the logarithm of the median terminal loss on $\log s$ gives slopes $0.999$--$1.002$ across the six $(p,\lambda_{\min})$ combinations. The median ratio $\mathcal L(\widehat A,A)/s$ ranges from approximately $1.05$ to $1.12$. Because the signal and noise are jointly rescaled when $\lambda_{\min}$ changes, the overlap of the two $\lambda_{\min}$ curves is expected; the primary evidence for the rate is the slope-one behavior over the six target scales and three dimensions.

\subsection{Details for the Convergence-Regime Experiment}\label{sec_sim_details_convergence_regimes}

This subsection provides details for the simulation study in Section~\ref{sec_sim_convergence_regimes}.

Experiment~4 uses noiseless rank-$3$ tensors with equal coefficients. The four cases are $(d,p,\xi)=(3,30,0)$, $(5,10,0)$, $(3,30,0.04)$, and $(5,10,0.10)$. The corresponding nonzero transition scales are $\tau_{\mathrm{ALS}}=0.04$ for $d=3$ and $\tau_{\mathrm{ALS}}=2(0.10)^3=0.002$ for $d=5$. Complete trajectories start from loading losses $0.16$ for $d=3$ and $0.12$ for $d=5$. The one-sweep map uses 36 input errors between $2\times10^{-4}$ and $0.24$, with additional points near both nonzero transition scales. Each case has 100 independent repetitions.

For the orthogonal order-$3$ case, the fitted log--log slope of the one-sweep map is $2.001$. The orthogonal order-$5$ case has fitted slope $3.847$, indicating faster-than-quadratic contraction in this special configuration; the theorem only requires an upper bound of quadratic order. Below the transition scale, the fitted slopes are $1.001$ for the nonorthogonal order-$3$ case and $0.999$ for the nonorthogonal order-$5$ case. The experiment therefore directly verifies the quadratic benchmark in an orthogonal general-rank case and the linear benchmark below the transition in nonorthogonal cases. It does not establish an exact slope-two law throughout the finite-sample nonorthogonal region above the transition.

\subsection{Details for the Rank-One Convergence-Speed Experiment}\label{sec_sim_details_rank_one_speed}

This subsection provides details for the simulation study in Supplement Section~\ref{sec_sim_rank_one_speed}.

Experiment~5 uses $R=1$, $p_k=5$, $d\in\{3,5,7\}$, $\lambda=1$, and $\sigma\in\{0.005,0.010,\ldots,0.050\}$. The rank-one TASD initialization reduces to the modewise leading-singular-vector initialization. We perform 100 repetitions per configuration and run at most 20 ALS sweeps. The first hitting time is the smallest $t\geq0$ for which the loading loss is below $0.05$.

For $d=3$, every run succeeds at $\sigma\leq0.01$, and successful runs not already below the threshold at initialization require at most one sweep. For $d=5$, successful runs require at most two sweeps, while for $d=7$ the successful runs at $\sigma\in\{0.015,0.020,0.025\}$ require one or two sweeps. The recovery probability then falls rapidly as $\sigma$ increases. This experiment is a qualitative illustration of the one- or two-step phenomenon; because $p_k=5$, it should not be used as evidence for large-dimensional scaling.

\subsection{Details for the Coherence Experiment}\label{sec_sim_details_coherence}

This subsection provides details for the simulation study in Supplement Section~\ref{sec_sim_coherence}.

Experiment~6 uses $p_k=5$, $R=3$, $\lambda=(1,2,3)$, $\sigma=0.01$, $d\in\{3,5\}$, and $\xi\in\{0,0.05,\ldots,0.9\}$. For each tensor order, 300 independent repetitions are generated. Within a repetition, the Stiefel bases, standardized noise, controlled-initialization perturbations, and TASD random seed are reused across the full coherence grid. The realized coherence differs from the requested value by at most $1.1\times10^{-15}$. Controlled warm-start ALS begins at loading loss $0.10$, and both procedures are run for at most 20 sweeps. Recovery means reaching loading loss below $0.05$.

Numerical TASD failures are counted as failures; there are 826 such method-level failures among the 22,800 pipeline records, all caused by a rank-deficient SimDiag eigenvector matrix. For $d=3$, controlled warm-start ALS has recovery probability at least $0.99$ through $\xi=0.60$, while TASD--ALS has probability $0.93$ at $\xi=0.60$ and then declines. For $d=5$, controlled warm-start ALS remains essentially certain through $\xi=0.80$, whereas TASD--ALS rises from $0.61$ at $\xi=0$ to approximately $0.83$ near $\xi=0.5$--$0.6$ before declining to $0.03$ at $\xi=0.90$. The paired controlled-initialization comparison shows that this low-to-moderate-$\xi$ nonmonotonicity is associated with TASD and finite-sample numerical conditioning rather than with a monotone deterioration of local ALS. Values of $\xi$ near $0.9$ are stress tests outside the sufficient coherence assumptions of the theory.

\subsection{Details for the Scaling and Imbalance Experiment}\label{sec_sim_details_scaling}

This subsection provides details for the simulation study in Supplement Section~\ref{sec_sim_scaling_robustness}.

Experiment~7 first considers balanced order-$3$ tensors with $p\in\{20,40,80\}$, $R\in\{2,4,6,8\}$, equal coefficients, and 50 repetitions. We set $\xi=0.30/(R-1)$ and $\sigma=0.03/\sqrt p$, so $(R-1)\xi=0.30$ and $\sigma\sqrt p/\lambda_{\min}=0.03$ are held fixed. Thus, the experiment studies increasing rank under approximately fixed loading-matrix conditioning rather than at fixed pairwise coherence. Recovery means terminal loss below $0.05$ after at most 20 sweeps.

The median terminal losses range from $0.0328$ to $0.0384$. Recovery probabilities at $R=8$ are $0.68$, $0.82$, and $0.92$ for $p=20$, $40$, and $80$, respectively. There are 34 numerical TASD failures among the 600 balanced runs, and these are counted as failures. The runtime panel is descriptive and is not used to assert monotone complexity in $R$, because implementation and system overhead visibly affect the timings.

The unbalanced comparison fixes $R=4$, $\xi=0.10$, $p_{\max}=80$, and $\sigma=0.03/\sqrt{80}$, and uses dimensions $(80,80,80)$, $(40,60,80)$, and $(20,40,80)$ with 50 repetitions each. The median terminal losses are $0.0340$, $0.0320$, and $0.0319$, and the recovery probabilities are $1.00$, $0.98$, and $1.00$, respectively. This directly supports stability under the displayed degree of mode imbalance.

\subsection{Details for the Equal-Strength Experiment}\label{sec_sim_details_equal_strength}

This subsection provides details for the simulation study in Supplement Section~\ref{sec_add_sim_set1}.

Experiment~8 uses $d=3$, $(p_1,p_2,p_3)=(15,12,10)$, $R=2$, $\lambda_1=\lambda_2=1$, and $\sigma\in\{0,10^{-5},10^{-4},10^{-3},10^{-2},10^{-1},1\}$. Each setting has 50 repetitions, and the same seven methods as in Experiment~1 are compared. R1--ALS--1 uses 50 candidates and separation threshold $0.1$.

TASD and TASD--ALS complete all repetitions through $\sigma=0.01$ and have nearly identical small-noise losses. At $\sigma=0$, their losses are at numerical precision, whereas CPCA--ICO has median loss $0.877$ because the coefficient eigengap is zero. Direct SimDiag is exact at $\sigma=0$ but has median loss about $1.30$ as soon as nonzero noise is introduced. R1--ALS--1 returns two separated candidates in only 24 of 350 total runs, so its displayed boxes are conditional on a small subset of repetitions. The experiment therefore supports the absence of a coefficient-gap requirement for TASD, not a claim that every competing procedure fails under all nonzero noise levels.

\subsection{Details for TASD Part II}\label{sec_sim_details_tasd_partII}

This subsection provides details for the simulation study in Supplement Section~\ref{sec_add_sim_set2}.

Experiment~9 uses order-$3$ tensors with dimensions $(15,15,15)$, $R\in\{2,3,4,5,6\}$, $\lambda_r=r$, $\sigma\in\{0.1,0.5,1\}$, and 50 repetitions per configuration. Direct three-mode TASD applies the two-sided three-mode SimDiag construction to the Tucker core, whereas TASD Part~II estimates one mode by SimDiag and obtains the remaining modes from the rowwise rank-one decompositions described in Algorithm~\ref{algorithm_tasd_ls}. Both constructions use the same random seed within a repetition.

The direct construction returns finite loading matrices in all 750 runs. TASD Part~II returns finite loadings in 355 of 750 runs. Its completion probability is $0.96$ for $(R,\sigma)=(2,0.1)$, but it decreases to $0.04$ and $0.10$ for $(R,\sigma)=(6,0.5)$ and $(6,1)$, respectively. Conditional on completion, its median loading loss is close to that of direct three-mode TASD and its median runtime is usually lower.

\section{Additional Notations}

We denote $\mathbb{O}_{p, r}:=\left\{{U} \in \mathbb{R}^{p \times r}: {U}^{\top} {U}={I}\right\}$ as the set of all $p$-by-$r$ semi-orthonormal matrices, i.e., matrices with orthonormal columns. 
For $U\in\mathbb{O}_{p,r}$, let
$U_\perp\in\mathbb{O}_{p,p-r}$ denote any matrix whose columns form
an orthonormal basis for the orthogonal complement of
$\operatorname{col}(U)$.
For a unit vector $u\in\mathbb{R}^p$, define
$u_\perp\in\mathbb{O}_{p,p-1}$ analogously.
For any matrix $A$, let
\[
    P_A:=AA^\dagger
\]
denote the orthogonal projector onto $\operatorname{col}(A)$.
In this work, we employ the $\sin \Theta$ distance to characterize the distance between subspaces.
For any $U, V \in \mathbb{O}_{p, r}$, we define $\|\sin \Theta(U, V)\| = \|U_\perp^\top V\| = \|UU^\top - VV^\top\|$.
The $\alpha$-to-$\beta$ operator norm of matrix $A \in \RR^{p\times q}$ is defined as $\|A\|_{\alpha \rightarrow \beta} = \max\{\|Ax\|_\beta: \|x\|_\alpha = 1\}$.

\section{Algorithm for Tensor Tucker Decomposition}

\begin{algorithm}[h]
    \caption{HOOI initialized by HOSVD \citep{de2000best}}
    \label{algorithm_hooi}
    \begin{algorithmic}[1]
        \algrenewcommand\algorithmicensure{\textbf{Run:}}
        \algrenewcommand\algorithmicrequire{\textbf{Input:}}
        \Require {Order-$d$ Tensor $\Y$, Target Tucker Rank $(r_1, \cdots, r_d)$}

        \For{$k = 1, \cdots d$}
        \State{Calculate $\hat U^{(0)}_k = \SVD_{r_k}(\MM_k(\Y))$;}
            \EndFor

            \For{$t = 1, \cdots$}
            \For{$k = 1, \cdots d$}
            \State{Update $\hat U^{(t)}_k = \SVD_{r_k}(\MM_k(\Y \times_{h\neq k} \hat U^{(t-1)\top}_h)) $.}
        \EndFor
        \EndFor\\
        \Return $\hat U^{(t)}_k$
    \end{algorithmic}
\end{algorithm}

\section{Proof of Theorem \ref{thm_overall_R1}}\label{sec_proof_thm_overall_R1}

\subsection{Proof of Initialization}

Let's assume $\sigma = 1$  without loss of generality.
Consider the initialization of the first mode for now.
By \eqref{eq_initialization_formula_R1}, we have
\[
    \hat a_1^{(0)} = \SVD_1(\MM_1(\Y)) = \SVD_1(\lambda a_1 (a_2\otimes \cdots \otimes a_d)\T + \MM_1(\Z)).
\]
If $\hat a_1^{(0)\top}a_1>0$, then
\[
    \left\|
        \hat a_1^{(0)}-a_1
    \right\|
    =
    \sqrt{
        2\left(
            1-
            \sqrt{
                1-
                \left\|
                    \sin\Theta\left(
                        \hat a_1^{(0)},a_1
                    \right)
                \right\|^2
            }
        \right)
    }
    \leq
    2
    \left\|
        \sin\Theta\left(
            \hat a_1^{(0)},a_1
        \right)
    \right\|.
\]
Otherwise,
\[
    \left\|
        \hat a_1^{(0)}+a_1
    \right\|
    =
    \sqrt{
        2\left(
            1-
            \sqrt{
                1-
                \left\|
                    \sin\Theta\left(
                        \hat a_1^{(0)},a_1
                    \right)
                \right\|^2
            }
        \right)
    }
    \leq
    2
    \left\|
        \sin\Theta\left(
            \hat a_1^{(0)},a_1
        \right)
    \right\|.
\]

Let $Y_1 = \MM_1(\Y)$ and $p_{-k} = \prod_{i \neq k}p_i$.
The appendix equation (1.16) in \cite{cai2018rate} yield
\beas
\PP\( \|a_{1,\perp}\T Y_1 P_{Y_1\T a_1}\| \geq x \) \leq C \exp \(Cp_1 -c\min\{x^2, x\sqrt{\lambda^2 +p_{-1}}\}\) + C
\exp\(-c(\lambda^2 +p_{-1})\),
\eeas
where $C$ and $c$ are absolute constants.
Taking $x = C\sqrt{p_1}$ gives
\beas
&&\PP\( \|a_{1,\perp}\T Y_1 P_{Y_1\T a_1}\| \geq C\sqrt{p_1} \) \\
&\leq& C \exp \(Cp_1 -c\min\{Cp_1, Cp_1^{1/2}\sqrt{\lambda^2 +p_{-1}}\}\) + C \exp\(-c(\lambda^2 +p_{-1})\) \\
&=& C \exp \(-Cp_1\) + C \exp\(-c(\lambda^2 +p_{-1})\) \\
&\leq& C \exp \(-Cp_1\),
\eeas
where the third line holds by $\lambda^2 + p_{-1} \geq C p_1^{1.5} \geq p_1$.

Equation (1.17) in the appendix of \cite{cai2018rate} yields
\[
    \PP\left(
        \left\|
            \sin\Theta\left(
                \hat a_1^{(0)},a_1
            \right)
        \right\|^2
        \leq
        \frac{
            C(\lambda^2+p_{-1})
            \left\|
                a_{1,\perp}\T
                Y_1
                P_{Y_1\T a_1}
            \right\|^2
        }{
            \lambda^4
        }
    \right)
    \geq
    1-
    C\exp\left(
        -c
        \frac{
            \lambda^4
        }{
            \lambda^2+p_{-1}
        }
    \right).
\]
Together with the preceding bound, this yields
\[
    \PP\left(
        \left\|
            \sin\Theta\left(
                \hat a_1^{(0)},a_1
            \right)
        \right\|^2
        \leq
        \frac{
            C(\lambda^2+p_{-1})p_1
        }{
            \lambda^4
        }
    \right)
    \geq
    1-
    C\exp\left(
        -c
        \frac{
            \lambda^4
        }{
            \lambda^2+p_{-1}
        }
    \right)
    -
    C\exp(-Cp_1).
\]
Note that
\[
    \frac{(\lambda^2+p_{-1})p_1}{\lambda^4} = \frac{p_1}{\lambda^2} + \frac{\prod_{j=1}^d p_j}{\lambda^4} \leq
    \frac{p_{\max}}{\lambda^2} + \frac{p_{\max}^d}{\lambda^4}.
\]
Moreover, by the inequalities at the beginning of this subsection,
\[
    \min\left\{ \|\hat a_1^{(0)}-a_1\|, \|\hat a_1^{(0)}+a_1\| \right\} \leq 2\left\| \sin\Theta\left(\hat a_1^{(0)},a_1\right)
    \right\|.
\]
Consequently,
\begin{align*}
     & \PP\left( \min\left\{ \|\hat a_1^{(0)}-a_1\|, \|\hat a_1^{(0)}+a_1\| \right\} \leq C \left( \frac{p_{\max}}{\lambda^2} +
    \frac{p_{\max}^d}{\lambda^4} \right)^{1/2} \right)                                                                          \\
     & \qquad\geq 1 - C\exp\left( -c\frac{\lambda^4}{\lambda^2+p_{-1}} \right) - C\exp(-cp_1).
\end{align*}

Under the signal-strength condition $|\lambda|\geq C p_{\max}^{d/4}$, we have, for every $k\in[d]$,
\[
    \frac{\lambda^4}{\lambda^2+p_{-k}} \gtrsim p_{\max} \geq p_{\min}.
\]
Applying the same argument to each mode and taking a union bound therefore gives
\[
    \begin{split}
         & \PP\left( \max_{k\in[d]} \min\left\{ \|\hat a_k^{(0)}-a_k\|, \|\hat a_k^{(0)}+a_k\| \right\} \leq C \left(
        \frac{p_{\max}}{\lambda^2} + \frac{p_{\max}^d}{\lambda^4} \right)^{1/2} \right)                               \\
         & \qquad\geq 1-Cd\exp(-cp_{\min}) \geq 1-C\exp(-c'p_{\min}),
    \end{split}
\]
where the last inequality follows from $p_{\min}\geq C\log d$.

Finally, restoring the general noise scale $\sigma$ by applying the preceding argument to $\Y/\sigma$ yields
\[
    \varepsilon_0 \leq C \left( \frac{\sigma^2p_{\max}}{\lambda^2} + \frac{\sigma^4p_{\max}^d}{\lambda^4} \right)^{1/2}
\]
with probability at least $1-C\exp(-c p_{\min})$.
After adjusting the absolute constants, this proves the initialization part of Theorem~\ref{thm_overall_R1}.

\subsection{Proof of Iteration}

We prove the iteration part of Theorem \ref{thm_overall_R1} by induction.
Consider the first mode for now.
Since our error metric does not account for the sign of the loading vectors, we assume $\lambda > 0$ without loss of generality.
In $(t+1)$th iteration of CP decomposition, we update via
$$
    \hat b_1^{(t+1)} = \MM_1(\Y)[(\hat a_2^{(t)}\otimes \cdots \otimes \hat a_d^{(t)})\T]^\dagger,\quad \hat \lambda^{(t+1)} =
    \|\hat b_1^{(t+1)}\|,\quad \hat a_1^{(t+1)} = (\hat \lambda^{(t+1)})^{-1} \hat b_1^{(t+1)}.
$$

We first derive a deterministic recurrence bound and then take the randomness into consideration.
\subsubsection{Find recurrence formula}

Denote $P = a_{1}a_{1}\T$ and $P_\perp = I - a_{1}a_{1}\T. $ Note that by Pythagorean theorem and Supplement Lemma \ref{lemma_vector_norm_quadratic_ine}, we have
\[
    \|\hat a_1^{(t+1)} - a_1\|^2 = \|P_\perp(\hat a_1^{(t+1)} - a_1)\|^2 + \|P(\hat a_1^{(t+1)} - a_1)\|^2 \Rightarrow \|\hat
    a_1^{(t+1)} - a_1\|^2 - \frac{1}{4} \|\hat a_1^{(t+1)} - a_1\|^4 = \|P_\perp\hat a_1^{(t+1)}\|^2.
\]
The above argument also holds if we flip the sign of $\hat a_1^{(t+1)}$.
Thus, we have
\be
\label{eq_recurrence_0}
\begin{split}
    \|\hat a_1^{(t+1)} \pm a_1\|^2 - \frac{1}{4} \|\hat a_1^{(t+1)} \pm a_1\|^4 = \frac{\|P_\perp(\hat \lambda^{(t+1)} \hat
    a_1^{(t+1)})\|^2}{(\hat \lambda^{(t+1)})^2} ,
\end{split}
\ee
for which we are going to derive the bound for the denominator and numerator

We introduce some additional notations to deal with the sign of $\hat a_i$, as $\hat a_i$ and $-\hat a_i$ should yield the same error.
Let $\beta_{t,j} \in \{-1, 1\}$ such that $\prod_{j = 2}^d \beta_{t,j} = 1$.
Then we can choose $\beta_{t,j}$ such that for $j = 2, \dots, d$, there is at most one $j_0$ among all $j$ such that $\beta_{t,j_0} \hat a_{j_0}^{(t)\top}  a_{j_0} \leq 0$.
Hence, we have for $j\neq j_0$,
\be
\label{eq_flip_label} \sqrt{2-2|\hat a_j^{(t)\top} a_j|} = \sqrt{2-2\beta_{t,j}\hat a_j^{(t)\top}  a_j} = \|\hat a_j^{(t)} -
\beta_{t,j}a_j\|.
\ee

\paragraph*{If there does not exist such $j_0$.}
We have
\begin{align}
        & \hat \lambda^{(t+1)} \hat a_1^{(t+1)}  - \lambda a_1 \notag                                                            \\
    =   & \MM_1(\Y) [(\hat a_2^{(t)}\otimes \cdots \otimes \hat a_d^{(t)})\T]^\dagger - \MM_1(\X) [(\beta_{t,2}a_2\otimes \cdots
    \otimes \beta_{t,d}a_d)\T]^\dagger \notag                                                                                    \\
    =   & (\MM_1(\Y) - \MM_1(\X)) [(\hat a_2^{(t)}\otimes \cdots \otimes \hat a_d^{(t)})\T]^\dagger\notag                        \\
        & + \MM_1(\X)\left( [(\hat a_2^{(t)}\otimes \cdots \otimes \hat a_d^{(t)})\T]^\dagger - [(\beta_{t,2}a_2\otimes \cdots
    \otimes \beta_{t,d}a_d)\T]^\dagger \right) \notag                                                                            \\
    {=} & \MM_1(\Z) (\hat a_2^{(t)}\otimes \cdots \otimes \hat a_d^{(t)}) + \lambda a_1 (a_2\otimes \cdots \otimes a_d)\T \left(
    (\hat a_2^{(t)}\otimes \cdots \otimes \hat a_d^{(t)}) - (\beta_{t,2}a_2\otimes \cdots \otimes \beta_{t,d}a_d)
    \right)\label{eq_bound_epsilon_tplus1_1_noj0}
\end{align}
By Supplement Lemmas~\ref{lemma_vector_norm_quadratic_ine} and~\ref{lemma_kronecker_bound} and equation \eqref{eq_flip_label}, we have
\begin{align}
         & \lambda \left| (a_2\otimes \cdots \otimes a_d)\T \left( (\hat a_2^{(t)}\otimes \cdots \otimes \hat a_d^{(t)}) -
    (\beta_{t,2}a_2\otimes \cdots \otimes \beta_{t,d}a_d) \right) \right|\notag                                                \\
    =    & \frac{\lambda }{2} \ln(\hat a_2^{(t)}\otimes \cdots \otimes \hat a_d^{(t)}) - (\beta_{t,2}a_2\otimes \cdots \otimes
    \beta_{t,d}a_d)\rn^2 \notag                                                                                                \\
    =    & \lambda-\lambda\prod_{j = 2}^d \( 1 - \frac{\|\beta_{t,j}a_j - \hat a_j^{(t)}\|^2}{2} \) \notag                     \\
    \leq & \lambda-\lambda\( 1 - \frac{\varepsilon_t^2}{2} \)^{d-1}.\label{eq_bound_epsilon_tplus1_11_noj0}
\end{align}
Combining \eqref{eq_bound_epsilon_tplus1_1_noj0} and  \eqref{eq_bound_epsilon_tplus1_11_noj0}, it yields that
\be
\label{eq_final_noj0} \ln\hat \lambda^{(t+1)} \hat a_1^{(t+1)}  - \lambda a_1 \rn \leq \lambda-\lambda\( 1 -
\frac{\varepsilon_t^2}{2} \)^{d-1} + {\delta_1^{(t+1)}},
\ee
where $\delta_1^{(t+1)} := \|\MM_1(\Z) (\hat a_2^{(t)}\otimes \cdots \otimes \hat a_d^{(t)})\|$ denotes the random error caused by noise.

\paragraph*{If there does exist such $j_0$.}
We have
\begin{align}
         & \hat \lambda^{(t+1)} \hat a_1^{(t+1)}  + \lambda a_1 \notag                                                            \\
    =    & \MM_1(\Y) [(\hat a_2^{(t)}\otimes \cdots \otimes \hat a_d^{(t)})\T]^\dagger - \MM_1(\X) [(\beta_{t,2}a_2\otimes \cdots
    \otimes (-\beta_{t,j_0})a_{j_0} \otimes \cdots \otimes \beta_{t,d}a_d)\T]^\dagger \notag                                      \\
    { =} & \MM_1(\Z) (\hat a_2^{(t)}\otimes \cdots \otimes \hat a_d^{(t)})\notag                                                  \\
         & + \lambda a_1 (a_2\otimes \cdots \otimes a_d)\T \left( (\hat a_2^{(t)}\otimes \cdots \otimes \hat a_d^{(t)}) -
    (\beta_{t,2}a_2\otimes \cdots \otimes (-\beta_{t,j_0})a_{j_0} \otimes \cdots \otimes \beta_{t,d}a_d) \right).
    \label{eq_bound_epsilon_tplus1_1_j0}
\end{align}
By Supplement Lemmas~\ref{lemma_vector_norm_quadratic_ine} and~\ref{lemma_kronecker_bound} and equation \eqref{eq_flip_label}, we have
\begin{align}
         & \left| (a_2\otimes \cdots \otimes a_d)\T \left( (\hat a_2^{(t)}\otimes \cdots \otimes \hat a_d^{(t)}) -
    (\beta_{t,2}a_2\otimes \cdots \otimes (-\beta_{t,j_0})a_{j_0} \otimes \cdots \otimes \beta_{t,d}a_d) \right)
    \right|\notag                                                                                                       \\
    =    & \frac{1}{2} \ln(\hat a_2^{(t)}\otimes \cdots \otimes \hat a_d^{(t)}) - (\beta_{t,2}a_2\otimes \cdots \otimes
    (-\beta_{t,j_0})a_{j_0} \otimes \cdots \otimes \beta_{t,d}a_d)\rn^2 \notag                                          \\
    =    & 1-\prod_{\substack{j = 2                                                                                     \\
            j \neq j_0}}^d \( 1 - \frac{\|\beta_{t,j}a_j - \hat a_j^{(t)}\|^2}{2} \) \( 1 - \frac{\|-\beta_{t,j_0}a_{j_0} - \hat
    a_{j_0}^{(t)}\|^2}{2} \)\notag                                                                                      \\
    \leq & 1-\( 1 - \frac{\varepsilon_t^2}{2} \)^{d-1}.\label{eq_bound_epsilon_tplus1_11_j0}
\end{align}
Combining \eqref{eq_bound_epsilon_tplus1_1_j0} and \eqref{eq_bound_epsilon_tplus1_11_j0}, it yields that
\be
\label{eq_final_j0} \ln\hat \lambda^{(t+1)} \hat a_1^{(t+1)}  + \lambda a_1 \rn \leq \lambda-\lambda\( 1 -
\frac{\varepsilon_t^2}{2} \)^{d-1} + {\delta_1^{(t+1)}}.
\ee

Hence, \eqref{eq_final_noj0} and \eqref{eq_final_j0} implies
\be
\notag \min\left\{\ln \hat\lambda^{(t+1)} \hat a_1^{(t+1)}  -  \lambda a_1\rn, \ln\hat\lambda^{(t+1)} \hat a_1^{(t+1)}  + \lambda a_1
\rn \right\} \leq \lambda-\lambda\( 1 - \frac{\varepsilon_t^2}{2} \)^{d-1} + {\delta_1^{(t+1)}}.
\ee
Note that
\be
\notag |\hat \lambda^{(t+1)} - \lambda | = |\|\hat \lambda^{(t+1)} \hat a_1^{(t+1)}\| - \|\lambda a_1\|| \leq \min_{s\in\{-1,1\}}
\left\|
    \hat \lambda^{(t+1)}
    \hat a_1^{(t+1)}
    -
    s\lambda a_1
\right\|.
\ee
Assuming $\lambda - \min_{s\in\{-1,1\}}
\left\|
    \hat \lambda^{(t+1)}
    \hat a_1^{(t+1)}
    -
    s\lambda a_1
\right\| \geq 0$, the above indicates
\be
\label{eq_denominator_0} (\hat \lambda^{(t+1)})^2 \geq \(\lambda - \min_{s\in\{-1,1\}}
\left\|
    \hat \lambda^{(t+1)}
    \hat a_1^{(t+1)}
    -
    s\lambda a_1
\right\|\)^2.
\ee

Furthermore, by \eqref{eq_bound_epsilon_tplus1_1_noj0} and \eqref{eq_bound_epsilon_tplus1_1_j0}, we have
\be
\label{eq_final_2} \|P_\perp \hat \lambda^{(t+1)} \hat a_1^{(t+1)} \| =\|P_\perp(\hat \lambda^{(t+1)} \hat a_1^{(t+1)} \pm
\lambda a_1)\| =  \|P_\perp \MM_1(\Z) (\hat a_2^{(t)}\otimes \cdots \otimes \hat a_d^{(t)})\| \leq \delta_1^{(t+1)}
\ee

For every $k\in[d]$, define
\[
    \delta_k^{(t+1)}
    :=
    \left\|
        \MM_k(\Z)
        \left(
            \bigotimes_{i\neq k}
            \hat a_i^{(t)}
        \right)
    \right\|,
    \qquad
    \delta^{(t+1)}
    :=
    \max_{k\in[d]}
    \delta_k^{(t+1)}.
\]
The preceding argument for mode $1$ applies identically to every mode $k\in[d]$.
Thus, taking the maximum over $k\in[d]$, by the mode-wise analogues of \eqref{eq_recurrence_0}, \eqref{eq_denominator_0}, \eqref{eq_final_2}, and $\varepsilon_t \leq \sqrt{2}$, we have
\be
\notag \frac{1}{2}\varepsilon_{t+1}^2 \leq \varepsilon_{t+1}^2 - \frac{1}{4} \varepsilon_{t+1}^4 \leq
\(\frac{\delta^{(t+1)}}{\lambda (1-\varepsilon_t^2 / 2)^{d-1} - \delta^{(t+1)}}\)^2 ,
\ee
i.e.,
\be
\label{eq_recurrence_formula} \varepsilon_{t+1} \leq \frac{\sqrt{2}\delta^{(t+1)}}{\lambda (1-\varepsilon_t^2 / 2)^{d-1} -
    \delta^{(t+1)}} = \frac{\sqrt{2}}{\frac{\lambda}{\delta^{(t+1)}} (1-\varepsilon_t^2 / 2)^{d-1} - 1}.
\ee

To make sure $\lambda - \min_{s\in\{-1,1\}}
\left\|
    \hat \lambda^{(t+1)}
    \hat a_1^{(t+1)}
    -
    s\lambda a_1
\right\| \geq 0$, we only need to check the denominator $\frac{\lambda}{\delta^{(t+1)}} (1-\varepsilon_t^2 / 2)^{d-1} - 1 \geq 0$.

\subsubsection{Bound $\delta_1^{(t+1)}$}\label{sec_random_noise_bound_alt}

Recall that
\[
    \delta_1^{(t+1)}
    :=
    \left\|
        \MM_1(\Z)
        \left(
            \hat a_2^{(t)}
            \otimes\cdots\otimes
            \hat a_d^{(t)}
        \right)
    \right\|.
\]
For every $k=2,\ldots,d$, choose
$s_{t,k}\in\{-1,1\}$ such that
\[
    \widetilde a_k^{(t)}
    :=
    s_{t,k}\hat a_k^{(t)}
\]
satisfies
\[
    \left\|
        \widetilde a_k^{(t)}-a_k
    \right\|
    \leq
    \varepsilon_t.
\]
Let
\[
    s_t
    :=
    \prod_{k=2}^d s_{t,k}.
\]
Then
\[
    \hat a_2^{(t)}
    \otimes\cdots\otimes
    \hat a_d^{(t)}
    =
    s_t
    \left(
        \widetilde a_2^{(t)}
        \otimes\cdots\otimes
        \widetilde a_d^{(t)}
    \right),
\]
so the overall sign does not affect the norm of the noise contraction.
Using the telescoping identity
\begin{align*}
    &
    \widetilde a_2^{(t)}
    \otimes\cdots\otimes
    \widetilde a_d^{(t)}
    -
    a_2\otimes\cdots\otimes a_d                                  \\
    &\quad=
    \sum_{h=2}^d
    \widetilde a_2^{(t)}
    \otimes\cdots\otimes
    \widetilde a_{h-1}^{(t)}
    \otimes
    \left(
        \widetilde a_h^{(t)}-a_h
    \right)
    \otimes
    a_{h+1}
    \otimes\cdots\otimes
    a_d,
\end{align*}
where empty products are omitted, we obtain
\be
\label{eq_noise_control_difference_decomp_r1_alt}
\begin{split}
    \delta_1^{(t+1)}
    &\leq
    \left\|
        \MM_1(\Z)
        \left(
            a_2\otimes\cdots\otimes a_d
        \right)
    \right\|                                                     \\
    &\quad+
    (d-1)\varepsilon_t
    \max_{\substack{\|b_k\|=1\\k=2,\ldots,d}}
    \left\|
        \MM_1(\Z)
        \left(
            b_2\otimes\cdots\otimes b_d
        \right)
    \right\|.
\end{split}
\ee

For deterministic $b_k$ with $\|b_k\|=1$, let
\[
    b
    :=
    b_2\otimes\cdots\otimes b_d.
\]
Then $\|b\|=1$. By independence of the entries of $\Z$ and the
sub-Gaussian moment-generating-function assumption following
\eqref{eq_main_model}, the coordinates of $\MM_1(\Z)b$ are independent
and satisfy
\[
    \EE\exp\left(
        t[\MM_1(\Z)b]_i
    \right)
    \leq
    \exp\left(
        \frac{\sigma^2t^2}{2}
    \right),
    \qquad
    i\in[p_1].
\]
Therefore, Supplement Lemma~\ref{lemma_subgaussian_vector_norm} gives, for every
$\rho\geq0$,
\be
\label{eq_subgaussian_tail_bound_r1_alt}
\PP\left(
    \left\|
        \MM_1(\Z)
        (b_2\otimes\cdots\otimes b_d)
    \right\|
    >
    \sigma(\sqrt{p_1}+\rho)
\right)
\leq
2\exp\left(
    -\frac{\rho^2}{2}
\right).
\ee

By Supplement Lemma \ref{lemma_e_net}, for any $b_k$ with $\|b_k\| = 1$, there exists $\tilde b_k \in \mathcal{N}_k$, which is the epsilon net in Supplement Lemma \ref{lemma_e_net}, such that $\|\tilde b_k - b_k\| \leq \varepsilon$.
Thus, using the telescoping identity
\[
\begin{split}
    &
    b_2\otimes\cdots\otimes b_d
    -
    \widetilde b_2\otimes\cdots\otimes\widetilde b_d
    \\
    &\quad=
    \sum_{h=2}^d
    b_2\otimes\cdots\otimes b_{h-1}
    \otimes
    \left(
        b_h-\widetilde b_h
    \right)
    \otimes
    \widetilde b_{h+1}\otimes\cdots\otimes\widetilde b_d,
\end{split}
\]
where empty products are omitted, we obtain
\[
\begin{split}
    &
    \left\|
        \MM_1(\Z)
        \left(
            b_2\otimes\cdots\otimes b_d
        \right)
    \right\|
    \\
    &\quad\leq
    \left\|
        \MM_1(\Z)
        \left(
            b_2\otimes\cdots\otimes b_d
            -
            \widetilde b_2\otimes\cdots\otimes\widetilde b_d
        \right)
    \right\|
    \\
    &\qquad+
    \left\|
        \MM_1(\Z)
        \left(
            \widetilde b_2\otimes\cdots\otimes\widetilde b_d
        \right)
    \right\|
    \\
    &\quad\leq
    \varepsilon(d-1)
    \max_{\substack{
        \|c_k\|=1\\
        k=2,\ldots,d
    }}
    \left\|
        \MM_1(\Z)
        \left(
            c_2\otimes\cdots\otimes c_d
        \right)
    \right\|
    \\
    &\qquad+
    \left\|
        \MM_1(\Z)
        \left(
            \widetilde b_2\otimes\cdots\otimes\widetilde b_d
        \right)
    \right\|.
\end{split}
\]
By taking the maximum, it yields
\[
    \max_{\|b_k\| = 1} \|\MM_1(\Z) (b_2 \otimes \cdots \otimes b_d)\| \leq \frac{1}{1-(d-1) \varepsilon }  \max_{b_k \in
        \mathcal{N}_k} \|\MM_1(\Z) (\tilde b_2 \otimes \cdots \otimes \tilde b_d)\|.
\]
Thus, taking $\varepsilon = \frac{1}{2(d-1)}$, it follows:
\be
\notag
\begin{split}
         & \PP\(
         \max_{\substack{
            \|b_k\|=1\\
            k=2,\ldots,d
        }}
        \left\|
            \MM_1(\Z)
            \left(
                b_2\otimes\cdots\otimes b_d
            \right)
        \right\|
         > 2\sigma(\sqrt{p_1} + \rho) \)               \\
    \leq & \PP\(
        \max_{\substack{
            \widetilde b_k\in\mathcal{N}_k\\
            k=2,\ldots,d
        }}
        \left\|
            \MM_1(\Z)
            \left(
                \widetilde b_2\otimes\cdots\otimes\widetilde b_d
            \right)
        \right\|
        > \sigma(\sqrt{p_1} +
    \rho) \)                                                                                                                    \\
    \leq & \sum_{b_k \in \mathcal{N}_k} \PP\(\|\MM_1(\Z) (\tilde b_2 \otimes \cdots \otimes \tilde b_d)\| > \sigma(\sqrt{p_1} +
    \rho) \)                                                                                                                    \\
        \leq & 2 \(1+\frac{2}{\varepsilon}\)^{\sum_{k>1}p_k} \exp\(-\frac{\rho^2}{2}\)                                              \\
    =    & 2 \exp\(\(\sum_{k>1}p_k\)\log\(4d - 3\)  -\frac{\rho^2}{2}\) .
\end{split}
\ee
{ We let $\rho = \sqrt{6(d-1)p_{\max}\log d}$.
    Since $\sum_{k>1}p_k\leq(d-1)p_{\max}$ and $4d-3\leq d^2$ for $d\geq3$, we further have
    \[
        \PP\(\max_{\|b_k\| = 1} \|\MM_1(\Z) (b_2 \otimes \cdots \otimes b_d)\| > 2\sigma(\sqrt{p_1} + \rho) \)
        \leq 2\exp\(-(d-1)p_{\max}\log 2\).
    \]
    Moreover, since $p_1\leq p_{\max}$ and $d\geq3$,
    \[
        2\sigma(\sqrt{p_1}+\rho)
        \leq
        7\sigma\sqrt{(d-1)p_{\max}\log d}.
    \]
    Thus, it holds with high probability that
    \be
    \label{eq_assumption_HP_1_r1_alt} \max_{\|b_k\| = 1} \|\MM_1(\Z) (b_2 \otimes \cdots \otimes b_d)\| \leq 7 \sigma
    \sqrt{(d-1)p_{\max}\log d}.
    \ee
}

Further note that by \eqref{eq_subgaussian_tail_bound_r1_alt}, we have
\be
\notag
\begin{split}
    \PP\(\ln \MM_1(\Z) (a_{2} \otimes \cdots \otimes  a_{d}) \rn > \sigma(\sqrt{p_1} + \rho)\) \leq & 2\exp\(-\frac{\rho^2}{2
    }\),
\end{split}
\ee
{
    So if we let $\rho = \sqrt{p_{\max}}$, then we have
    \[
        \PP\(\ln \MM_1(\Z) (a_{2} \otimes \cdots \otimes  a_{d}) \rn > \sigma(\sqrt{p_1} + \rho)\) \leq 2\exp\(-\frac{p_{\max}}{2 }\).
    \]
    Thus, it holds with high probability that
    \be
    \label{eq_assumption_HP_2_r1_alt} \ln \MM_1(\Z) (a_{2} \otimes \cdots \otimes  a_{d}) \rn \leq 2 \sigma \sqrt{p_{\max}},
    \ee
}

Then, \eqref{eq_noise_control_difference_decomp_r1_alt}, \eqref{eq_assumption_HP_1_r1_alt}, and \eqref{eq_assumption_HP_2_r1_alt} yield
\be
\label{eq_noise_random_term_r1_alt} \delta_1^{(t+1)} \leq 2 \sigma \sqrt{p_{\max}} + \varepsilon_t 7 (d-1) \sigma
\sqrt{(d-1)p_{\max}\log d}.
\ee

{
    If we further have $\varepsilon_t < \frac{1}{(d-1)^{3/2}\sqrt{\log d}}$, then we have
    \be
    \label{eq_noise_random_term_r1_alt_phase_final} \delta_1^{(t+1)} \leq 9 \sigma \sqrt{p_{\max}}.
    \ee
}

The same argument applies to every mode $k\in[d]$.
Therefore, on the corresponding common high-probability event,
\be
\label{eq_noise_random_term_r1_alt_all_modes}
\delta^{(t+1)}
\leq
2 \sigma \sqrt{p_{\max}}
+
7(d-1)\varepsilon_t\sigma
\sqrt{(d-1)p_{\max}\log d}.
\ee
In particular, if
\[
    \varepsilon_t
    \leq
    \frac{1}{(d-1)^{3/2}\sqrt{\log d}},
\]
then
\be
\label{eq_noise_random_term_r1_alt_phase_final_all_modes}
\delta^{(t+1)}
\leq
9\sigma\sqrt{p_{\max}}.
\ee

\subsection{Track the Error per Iteration}
Now, with recurrence bound \eqref{eq_recurrence_formula} and random error bounds \eqref{eq_noise_random_term_r1_alt_all_modes} and \eqref{eq_noise_random_term_r1_alt_phase_final_all_modes}, we are ready to track the error.
We consider the different phases of the algorithm according to when \eqref{eq_noise_random_term_r1_alt_phase_final_all_modes} is applicable.

\paragraph*{Phase 1: Initialization\\}

By the initialization bound in Theorem~\ref{thm_overall_R1}, and by taking the absolute constant $C_1$ in its signal-strength condition sufficiently large, we have $\varepsilon_0\leq1$.

\paragraph*{Phase 2\\}

In this phase we already have $\varepsilon_t \leq 1. $

By \eqref{eq_noise_random_term_r1_alt_all_modes}, we have
\be
\label{eq_noise_random_term_r1_alt_1} \delta^{(t+1)} \leq 8 (d-1) \sigma \sqrt{(d-1)p_{\max}\log d}.
\ee

Since
\[
    p_{\max}
    \geq
    p_{\min}
    \geq
    C_1\log d,
\]
by taking the absolute constant $C_1$ sufficiently large, we have
\[
    p_{\max}^{\frac{1}{4}-\frac{1}{2d}}
    >
    8.
\]
Raising this inequality to the $d$th power and using
\[
    8^d
    \geq
    14\cdot 2^d
    (d-1)^{3/2}
    \sqrt{\log d},
    \qquad
    d\geq3,
\]
we obtain
\[
    \frac{
        p_{\max}^{\frac{d}{4}-\frac{1}{2}}
    }{
        8(d-1)^{3/2}\sqrt{\log d}
    }
    \frac{1}{2^{d-1}}
    >
    \frac{7}{2}.
\]
Consequently, by
\eqref{eq_noise_random_term_r1_alt_1} and
$\lambda\geq\sigma p_{\max}^{d/4}$,
\[
\begin{split}
    \frac{\lambda}{\delta^{(t+1)}}
    \left(
        1-\frac{\varepsilon_t^2}{2}
    \right)^{d-1}
    &\geq
    \frac{
        p_{\max}^{\frac{d}{4}-\frac{1}{2}}
    }{
        8(d-1)^{3/2}\sqrt{\log d}
    }
    \left(
        1-\frac{\varepsilon_t^2}{2}
    \right)^{d-1}                                                \\
    &\geq
    \frac{
        p_{\max}^{\frac{d}{4}-\frac{1}{2}}
    }{
        8(d-1)^{3/2}\sqrt{\log d}
    }
    \frac{1}{2^{d-1}}                                            \\
    &>
    \frac{7}{2},
\end{split}
\]
where the second inequality follows from $\varepsilon_t\leq1$.

So the denominator of \eqref{eq_recurrence_formula} is greater than 0 in this case.
Furthermore, by \eqref{eq_recurrence_formula} and the fact that $\frac{1}{x-1} \leq \sqrt{2}/x$ for $x > 3.5$, we have
\be
\label{eq_recurrence_formula_1} \varepsilon_{t+1} \leq \frac{\sqrt{2}}{\frac{\lambda}{\delta^{(t+1)}} (1-\varepsilon_t^2 /
    2)^{d-1} - 1} \leq \frac{2\delta^{(t+1)}}{\lambda (1-\varepsilon_t^2 / 2)^{d-1}}.
\ee

Further note that as $d \geq 3$
\be
\label{eq_constant_bound1} (16 \log d)^{\frac{1}{d-1}} \leq 5, \quad (d-1)^{\frac{3}{d-1}} \leq 3.
\ee

Finally, since $p_{\max}$ is larger than a sufficiently large
absolute constant,
\[
    (p_{\max})^{\frac{2-d}{4(d-1)}}
    \leq
    \frac{1}{30}.
\]
Therefore,
\beas
& & \varepsilon_t^2 \leq 1 \\
&\Rightarrow&
1-\frac{\varepsilon_t^2}{2}
\geq
\frac{1}{2} \\
&\Rightarrow&
1-\frac{\varepsilon_t^2}{2}
\geq
15
(p_{\max})^{\frac{2-d}{4(d-1)}} \\
&\Rightarrow&
1-\frac{\varepsilon_t^2}{2}
\geq
(16\log d)^{\frac{1}{d-1}}
(p_{\max})^{\frac{2-d}{4(d-1)}}
(d-1)^{\frac{3}{d-1}}
\quad
\text{(by \eqref{eq_constant_bound1})} \\
&\Leftrightarrow&
16
(p_{\max})^{\frac{2-d}{4}}
(d-1)^3
\log d
\leq
\left(
    1-\frac{\varepsilon_t^2}{2}
\right)^{d-1} \\
&\Rightarrow &\frac{16 \sigma \sqrt{p_{\max}}}{\lambda }(d-1)^{3}\log d \leq (1-\varepsilon_t^2 / 2)^{d-1} \quad \text{(by
    $\lambda \geq \sigma p_{\max}^{d/4}$)} \\
&\Leftrightarrow &\frac{16 \sigma \sqrt{p_{\max}}}{\lambda (1-\varepsilon_t^2 / 2)^{d-1}} \leq \frac{1}{(d-1)^{3}\log d} \\
&\Leftrightarrow &\frac{16 (d-1) \sigma \sqrt{(d-1)p_{\max}\log d}}{\lambda (1-\varepsilon_t^2 / 2)^{d-1}} \leq
\frac{1}{(d-1)^{3/2}\sqrt{\log d}} \\
&\Rightarrow &\frac{2\delta^{(t+1)}}{\lambda (1-\varepsilon_t^2 / 2)^{d-1}} \leq \frac{1}{(d-1)^{3/2}\sqrt{\log d}} \quad
\text{(by \eqref{eq_noise_random_term_r1_alt_1})} \\
&\Rightarrow &\varepsilon_{t+1} \leq \frac{1}{(d-1)^{3/2}\sqrt{\log d}} \quad \text{(by \eqref{eq_recurrence_formula_1})}.
\eeas

\paragraph*{Phase 3: One-Step Convergence\\}

In this phase, suppose that $\varepsilon_t \leq \frac{1}{(d-1)^{3/2}\sqrt{\log d}}$.
Then, by \eqref{eq_recurrence_formula_1} and \eqref{eq_noise_random_term_r1_alt_phase_final_all_modes}, we have
\be
\notag \varepsilon_{t+1} \leq \frac{\sqrt{2}}{\frac{\lambda}{\delta^{(t+1)}} (1-\varepsilon_t^2 / 2)^{d-1} - 1} \leq
\frac{2\delta^{(t+1)}}{\lambda (1-\varepsilon_t^2 / 2)^{d-1}} \leq \frac{18\sigma\sqrt{p_{\max}}}{\lambda (1-\varepsilon_t^2
    / 2)^{d-1}} .
\ee
Since
\[
    \frac{\varepsilon_t^2}{2}
    \leq
    \frac{1}{
        2(d-1)^3\log d
    }
    \leq
    1,
\]
we have $-\varepsilon_t^2/2\geq-1$. Therefore, Bernoulli's
inequality, $(1+x)^k\geq1+kx$ for every integer $k\geq0$ and
$x\geq-1$, applied with $x=-\varepsilon_t^2/2$ and $k=d-1$, gives
\[
    \left(
        1-\frac{\varepsilon_t^2}{2}
    \right)^{d-1}
    \geq
    1-\frac{(d-1)\varepsilon_t^2}{2}.
\]
Thus, it yields
\be
\notag \varepsilon_{t+1} \leq \frac{18\sigma\sqrt{p_{\max}}}{\lambda \(1 - \frac{(d-1)\varepsilon_t^2}{2}\)} .
\ee
Further note that by {$\varepsilon_t^2 (d-1)/2 \leq \frac{1}{ (d-1)^2 \log d} \leq 1/2$} and $\frac{1}{1-x} \leq 2x+ 1$ for $0\leq x \leq 1/2$, we have
\be
\notag \varepsilon_{t+1} \leq \frac{18\sigma\sqrt{p_{\max}}}{\lambda } \(1 + (d-1)\varepsilon_t^2\).
\ee
This shows that
\[
    \varepsilon_{t+1}
    \lesssim
    \frac{\sigma\sqrt{p_{\max}}}{\lambda}.
\]
Moreover, since $\frac{1}{(d-1)^{3/2}\sqrt{\log d}}\leq1$, the Phase 2 argument implies that whenever $\varepsilon_t\leq\frac{1}{(d-1)^{3/2}\sqrt{\log d}}$, we also have $\varepsilon_{t+1}\leq\frac{1}{(d-1)^{3/2}\sqrt{\log d}}$.
By Phases 1 and 2, $\varepsilon_1\leq\frac{1}{(d-1)^{3/2}\sqrt{\log d}}$.
Therefore, induction shows that $\varepsilon_t\leq\frac{1}{(d-1)^{3/2}\sqrt{\log d}}$ for every $t\geq1$, and applying the Phase 3 bound at each $t\geq1$ proves \eqref{eq_bound_et_r1} for every $t\geq2$.

\paragraph*{Fixed-order case: one iteration suffices.}

It remains to prove the last assertion of
Theorem~\ref{thm_overall_R1}. Suppose that $d\leq C_4$. On the
initialization event established above, the signal-strength condition
$|\lambda|\geq C_1\sigma p_{\max}^{d/4}$ gives
\begin{align*}
    \varepsilon_0
    &\leq
    C
    \left(
        \frac{\sigma^2p_{\max}}{\lambda^2}
        +
        \frac{\sigma^4p_{\max}^d}{\lambda^4}
    \right)^{1/2}                                                    \\
    &\leq
    C\left(
        \frac{\sigma\sqrt{p_{\max}}}{|\lambda|}
        +
        \frac{\sigma^2p_{\max}^{d/2}}{\lambda^2}
    \right)                                                          \\
    &\leq
    C\left(
        C_1^{-1}p_{\max}^{1/2-d/4}
        +
        C_1^{-2}
    \right)                                                          \\
    &\leq
    C\left(
        C_1^{-1}+C_1^{-2}
    \right),
\end{align*}
where the last inequality follows from $d\geq3$ and
$p_{\max}\geq1$.

Since $C_4$ is an absolute constant, define
\[
    c_\star
    :=
    \frac{1}{
        (C_4-1)^{3/2}\sqrt{\log C_4}
    }.
\]
For every $3\leq d\leq C_4$, we have
\[
    \frac{1}{
        (d-1)^{3/2}\sqrt{\log d}
    }
    \geq
    c_\star.
\]
Therefore, by taking the absolute constant $C_1$ sufficiently large,
we obtain
\begin{equation}\label{eq_r1_fixed_order_initial_region}
    \varepsilon_0
    \leq
    \frac{c_\star}{2}
    <
    \frac{1}{
        (d-1)^{3/2}\sqrt{\log d}
    }.
\end{equation}

By \eqref{eq_noise_random_term_r1_alt_phase_final_all_modes} and
\eqref{eq_r1_fixed_order_initial_region}, we have
\[
    \max_{k\in[d]}
    \delta_k^{(1)}
    \leq
    9\sigma\sqrt{p_{\max}}.
\]
Moreover, Bernoulli's inequality gives
\begin{align*}
    \left(
        1-\frac{\varepsilon_0^2}{2}
    \right)^{d-1}
    &\geq
    1-\frac{(d-1)\varepsilon_0^2}{2}                                \\
    &\geq
    1-\frac{1}{
        2(d-1)^2\log d
    }
    \geq
    \frac{1}{2}.
\end{align*}
Since
\[
    \lambda
    \geq
    C_1\sigma p_{\max}^{d/4}
    \geq
    C_1\sigma\sqrt{p_{\max}},
\]
we may increase $C_1$ further, if necessary, so that
\[
    \lambda
    \left(
        1-\frac{\varepsilon_0^2}{2}
    \right)^{d-1}
    -
    \delta_k^{(1)}
    \geq
    \frac{\lambda}{4}
\]
simultaneously for every $k\in[d]$. Applying the mode-wise version of
\eqref{eq_recurrence_formula} therefore yields
\[
    \varepsilon_1
    \leq
    \frac{
        4\sqrt{2}\max_{k\in[d]}\delta_k^{(1)}
    }{
        \lambda
    }
    \leq
    36\sqrt{2}
    \frac{
        \sigma\sqrt{p_{\max}}
    }{
        \lambda
    }.
\]
Thus, after adjusting the absolute constant $C_3$,
\eqref{eq_bound_et_r1} holds at $t=1$.

Finally,
\[
    \frac{\sigma\sqrt{p_{\max}}}{\lambda}
    \leq
    C_1^{-1}p_{\max}^{1/2-d/4}
    \leq
    C_1^{-1}.
\]
By increasing $C_1$ once more, the preceding bound on
$\varepsilon_1$ is at most $c_\star/2$. Hence
$\varepsilon_1$ remains in the region
\eqref{eq_r1_fixed_order_initial_region}. Repeating the same argument
inductively gives
\[
    \varepsilon_t
    \lesssim
    \frac{\sigma\sqrt{p_{\max}}}{\lambda}
    \qquad
    \text{for every }t\geq1.
\]
The initialization and noise-control events hold simultaneously over
all modes with probability at least
$1-\exp(-C_2p_{\min})$, after adjusting the absolute constants and
using $p_{\min}\geq C_1\log d$. This proves the fixed-order assertion
of Theorem~\ref{thm_overall_R1}.

\section{Proof of Corollary \ref{corollary_lambda_R1}}\label{sec_proof_corollary_lambda_R1}

Work on the common high-probability event used in the proof of Theorem~\ref{thm_overall_R1}.
For $t\geq1$, define
\[
    u_t
    :=
    \hat a_2^{(t)}
    \otimes\cdots\otimes
    \hat a_d^{(t)},
    \qquad
    q_t
    :=
    \prod_{k=2}^d
    \left\langle
        a_k,\hat a_k^{(t)}
    \right\rangle.
\]
Since each $\hat a_k^{(t)}$ has unit norm, we have $\|u_t\|=1$ and hence $[u_t^\top]^\dagger=u_t$.
Therefore,
\[
    \hat b_1^{(t+1)}
    =
    \MM_1(\Y)u_t
    =
    \lambda q_t a_1
    +
    \MM_1(\Z)u_t.
\]
Define
\[
    \delta_1^{(t+1)}
    :=
    \left\|
        \MM_1(\Z)u_t
    \right\|.
\]
Since $\widehat{|\lambda|}^{(t)}=\|\hat b_1^{(t+1)}\|$ and $|q_t|\leq1$, the reverse triangle inequality gives
\begin{align*}
    \left|
        \widehat{|\lambda|}^{(t)}
        -
        |\lambda|
    \right|
    &\leq
    \left|
        \|\hat b_1^{(t+1)}\|
        -
        |\lambda q_t|
    \right|
    +
    \left|
        |\lambda q_t|
        -
        |\lambda|
    \right| \\
    &\leq
    \delta_1^{(t+1)}
    +
    |\lambda|
    \left(
        1-|q_t|
    \right).
\end{align*}
For every $k=2,\ldots,d$, the definition of $\varepsilon_t$ and the fact that both $a_k$ and $\hat a_k^{(t)}$ have unit norm imply
\[
    \left|
        \left\langle
            a_k,\hat a_k^{(t)}
        \right\rangle
    \right|
    =
    1
    -
    \frac{1}{2}
    \min_{s\in\{-1,1\}}
    \left\|
        \hat a_k^{(t)}
        -
        s a_k
    \right\|^2
    \geq
    1-\frac{\varepsilon_t^2}{2}.
\]
Consequently,
\[
    |q_t|
    \geq
    \left(
        1-\frac{\varepsilon_t^2}{2}
    \right)^{d-1}.
\]
Since $\varepsilon_t\leq\sqrt{2}$, Bernoulli's inequality yields
\[
    1-|q_t|
    \leq
    1-
    \left(
        1-\frac{\varepsilon_t^2}{2}
    \right)^{d-1}
    \leq
    \frac{d-1}{2}\varepsilon_t^2.
\]
It follows that
\[
    \left|
        \widehat{|\lambda|}^{(t)}
        -
        |\lambda|
    \right|
    \leq
    \delta_1^{(t+1)}
    +
    \frac{d-1}{2}
    |\lambda|
    \varepsilon_t^2.
\]
The proof of Theorem~\ref{thm_overall_R1} establishes, on the same high-probability event,
\[
    \varepsilon_t
    \leq
    \frac{1}{
        (d-1)^{3/2}\sqrt{\log d}
    }
    \qquad
    \text{for every }t\geq1.
\]
Therefore, by \eqref{eq_noise_random_term_r1_alt_phase_final_all_modes},
\[
    \delta_1^{(t+1)}
    \leq
    9\sigma\sqrt{p_{\max}}
    \qquad
    \text{for every }t\geq1.
\]
For every $t\geq2$, Theorem~\ref{thm_overall_R1} also gives
\[
    |\lambda|\varepsilon_t
    \leq
    C_3\sigma\sqrt{p_{\max}}.
\]
Combining the preceding two bounds, we obtain
\begin{align*}
    \frac{d-1}{2}
    |\lambda|
    \varepsilon_t^2
    &=
    \frac{d-1}{2}
    \varepsilon_t
    \left(
        |\lambda|\varepsilon_t
    \right) \\
    &\leq
    \frac{C_3}{
        2\sqrt{(d-1)\log d}
    }
    \sigma\sqrt{p_{\max}} \\
    &\lesssim
    \sigma\sqrt{p_{\max}}.
\end{align*}
Hence,
\[
    \left|
        \widehat{|\lambda|}^{(t)}
        -
        |\lambda|
    \right|
    \lesssim
    \sigma\sqrt{p_{\max}}
\]
for every $t\geq2$.
If $d\leq C_4$, Theorem~\ref{thm_overall_R1} gives the bound $|\lambda|\varepsilon_t\lesssim\sigma\sqrt{p_{\max}}$ for every $t\geq1$, so the same argument proves the conclusion for every $t\geq1$.
All the bounds above hold on the common event from the proof of Theorem~\ref{thm_overall_R1}, whose probability is at least $1-C\exp(-cp_{\min})$ after adjusting the absolute constants.

\section{Proof of Theorem \ref{thm_local}}\label{sec_proof_thm_local}

The proof extends the rank-one argument to a joint analysis of all $R$ components.
The main additional difficulties are to control the inverse Khatri--Rao Gram matrix, separate same-component perturbations from cross-component interactions, and establish noise bounds uniformly over the data-dependent ALS iterates.
We first derive the one-step recurrence and then analyze its quadratic and coherence-weighted linear regimes.

\subsection{Find recurrence formula}

Let
\[
    \mathfrak{D}_R
    :=
    \left\{
        \diag(\beta_1,\ldots,\beta_R)
        :
        \beta_r\in\{-1,1\},\ r\in[R]
    \right\}.
\]
For every $k\in[d]$, choose
\[
    \mathcal{B}_{t,k}
    \in
    \argmin_{\mathcal{B}\in\mathfrak{D}_R}
    \left\|
        \hat A_k^{(t)}-A_k\mathcal{B}
    \right\|_{1\rightarrow2}.
\]
Since $\mathcal{B}_{t,k}^2=I$, right multiplication by
$\mathcal{B}_{t,k}$ only changes the signs of the columns, and hence
\[
\begin{split}
    \varepsilon_t
    &=
    \max_{k\in[d]}
    \left\|
        \hat A_k^{(t)}-A_k\mathcal{B}_{t,k}
    \right\|_{1\rightarrow2}                                      \\
    &=
    \max_{k\in[d]}
    \left\|
        \hat A_k^{(t)}\mathcal{B}_{t,k}-A_k
    \right\|_{1\rightarrow2}.
\end{split}
\]
Write
\[
    \widetilde A_k^{(t)}
    :=
    \hat A_k^{(t)}\mathcal{B}_{t,k},
    \qquad
    E_k
    :=
    \widetilde A_k^{(t)}-A_k
    =
    [e_{k,1},\ldots,e_{k,R}].
\]
Equivalently, if
\[
    \mathcal{B}_{t,k}
    =
    \diag(\beta_{t,k,1},\ldots,\beta_{t,k,R}),
\]
then
\[
    \widetilde a_{k,r}^{(t)}
    =
    \beta_{t,k,r}\hat a_{k,r}^{(t)}
    =
    a_{k,r}+e_{k,r}.
\]
Therefore,
\be
\label{eq_vector_bound}
    \max_{r\in[R]}
    \|e_{k,r}\|
    =
    \|E_k\|_{1\rightarrow2}
    \leq
    \varepsilon_t,
    \qquad
    k\in[d].
\ee

By Supplement Lemmas~\ref{lemma_norm_transfer_inequality} and~\ref{lemma_generalized_inverse}, denote $\diag(|\lambda|) = \diag(|\lambda_1|, \ldots, |\lambda_R|)$ and we have
\be
\label{eq_update_main_inequality_general_d}
\begin{split}
         & \min_{\mathcal B\in\mathfrak D_R}
\left\|
    \hat B_1^{(t+1)}
    -
    A_1\diag(|\lambda|)\mathcal B
\right\|_{1\rightarrow2}                                                                                                         \\
=    & \min_{\mathcal B\in\mathfrak D_R}
\left\|
    \hat B_1^{(t+1)}
    -
    A_1\diag(\lambda)\mathcal B
\right\|_{1\rightarrow2}
\\                                                                                                                       \\
    =    & \min_{\beta_{t+1,1,r} \in \{-1, 1\}}\left\{  \|\hat B_1^{(t+1)} - A_1\diag(\lambda) \mathcal{B}_{t+1,1} \|_{1\rightarrow
    2} \right\}                                                                                                                       \\
    \leq & \ln A_1 \diag(\lambda) \(( A_d\odot \cdots \odot A_2) -  (\hat A_d^{(t)}\mathcal{B}_{t,d} \odot \cdots \odot \hat
    A_2^{(t)} \mathcal{B}_{t,2})\)\T \times [(\hat A_d^{(t)}\odot \cdots \odot \hat A_2^{(t)})\T]^\dagger \rn_{1\rightarrow
    2} +                                                                                                                              \\
         & \ln \MM_1(\Z) [(\hat A_d^{(t)}\odot \cdots \odot \hat A_2^{(t)})\T]^\dagger\rn_{1\rightarrow 2}                            \\
    \leq & \|A_1 \diag(|\lambda|)\|_{1\rightarrow 2} \ln \(( A_d\odot \cdots \odot A_2) -  (\hat A_d^{(t)}\mathcal{B}_{t,d}
    \odot \cdots \odot \hat A_2^{(t)} \mathcal{B}_{t,2})\)\T (\hat A_d^{(t)}\odot \cdots \odot \hat A_2^{(t)})
    \rn_{1\rightarrow 1} \times                                                                                                       \\
         & \ln \((\hat A_d^{(t)}\odot \cdots \odot \hat A_2^{(t)})\T (\hat A_d^{(t)}\odot \cdots \odot \hat A_2^{(t)})\)^\dagger
    \rn_{1\rightarrow 1} + \ln \MM_1(\Z) [(\hat A_d^{(t)}\odot \cdots \odot \hat A_2^{(t)})\T]^\dagger\rn_{1\rightarrow 2}.
\end{split}
\ee

The equality above holds because
\[
    \diag(\lambda)
    =
    \diag(|\lambda|)\mathcal S_{\lambda},
    \qquad
    \mathcal S_{\lambda}
    :=
    \diag\left(
        \operatorname{sign}(\lambda_1),
        \ldots,
        \operatorname{sign}(\lambda_R)
    \right)
    \in\mathfrak D_R,
\]
and the map
$\mathcal B\mapsto\mathcal S_{\lambda}\mathcal B$
is a bijection on $\mathfrak D_R$.
The inequality is obtained by evaluating the second minimum at
\[
    \mathcal B
    =
    \prod_{k=2}^d
    \mathcal B_{t,k}.
\]

\paragraph*{Bound the regular term. \\}
In this section we bound the first term on the right-hand side of \eqref{eq_update_main_inequality_general_d}.
We first consider the inversion term $\ln \((\hat A_d^{(t)}\odot \cdots \odot \hat A_2^{(t)})\T (\hat A_d^{(t)}\odot \cdots \odot \hat A_2^{(t)})\)^\dagger \rn_{1\rightarrow 1}$.
The key observation is that $(\hat A_d^{(t)}\odot \cdots \odot \hat A_2^{(t)})\T (\hat A_d^{(t)}\odot \cdots \odot \hat A_2^{(t)})$ should be close to an identity matrix.

We first consider the Gram matrix
\[
    M
    :=
    \left(
        \hat A_d^{(t)}
        \odot\cdots\odot
        \hat A_2^{(t)}
    \right)^\top
    \left(
        \hat A_d^{(t)}
        \odot\cdots\odot
        \hat A_2^{(t)}
    \right).
\]
Its diagonal entries satisfy $M_{ii}=1$. For $i\neq j$, the sign
alignment above gives
\begin{align*}
    |M_{ij}|
    &=
    \prod_{k=2}^d
    \left|
        \left\langle
            \hat a_{k,i}^{(t)},
            \hat a_{k,j}^{(t)}
        \right\rangle
    \right|                                                     \\
    &=
    \prod_{k=2}^d
    \left|
        \left\langle
            a_{k,i}+e_{k,i},
            a_{k,j}+e_{k,j}
        \right\rangle
    \right|                                                     \\
    &\leq
    \left(
        \xi+2\varepsilon_t+\varepsilon_t^2
    \right)^{d-1}.
\end{align*}
Consequently,
\be
\label{eq_bound_MI}
    \|M-I\|_{1\rightarrow1}
    \leq
    (R-1)
    \left(
        \xi+2\varepsilon_t+\varepsilon_t^2
    \right)^{d-1}.
\ee

Suppose that
\be
\label{eq_invertible_condition}
    (R-1)
    \left(
        \xi+2\varepsilon_t+\varepsilon_t^2
    \right)^{d-1}
    <1.
\ee
Then $\|I-M\|_{1\rightarrow1}<1$, so the Neumann series implies that
$M$ is invertible and
\[
    M^{-1}
    =
    \sum_{\ell=0}^{\infty}
    (I-M)^\ell.
\]
It follows that
\be
\label{eq_inverse_bound}
\begin{split}
    \|M^{-1}\|_{1\rightarrow1}
    &\leq
    \sum_{\ell=0}^{\infty}
    \|I-M\|_{1\rightarrow1}^{\ell}                              \\
    &=
    \frac{1}{
        1-\|M-I\|_{1\rightarrow1}
    }                                                           \\
    &\leq
    \frac{1}{
        1-
        (R-1)
        \left(
            \xi+2\varepsilon_t+\varepsilon_t^2
        \right)^{d-1}
    }.
\end{split}
\ee

Now we are going  to bound $\ln(\hat A_d^{(t)}\mathcal{B}_{t,d} \odot \cdots \odot \hat A_2^{(t)} \mathcal{B}_{t,2} -  A_d\odot \cdots \odot A_2)\T (\hat A_d^{(t)}\odot \cdots \odot \hat A_2^{(t)})\rn_{1\rightarrow 1}$.

Note that by Supplement Lemma \ref{lemma_vector_norm_quadratic_ine}, we have
$$
    |a_{k,i}\T \hat  a_{k,j}^{(t)}| = |\beta_{t,k,j}  a_{k,i}\T (a_{k,j} + e_{k,j})| \leq \begin{cases} \xi + \varepsilon_t,\ i\neq j \\ 1+0.5\varepsilon_t^2,\ i = j\end{cases},
$$
and
\[
    |e_{k,i}\T \hat  a_{k,j}^{(t)}| = |\beta_{t,k,j} e_{k,i}\T (a_{k,j} + e_{k,j})| \leq \begin{cases} \varepsilon_t^2 + \varepsilon_t,\ i\neq j \\ 1.5\varepsilon_t^2,\ i = j\end{cases}.
\]

For a subset $S\subseteq\{2,\ldots,d\}$, define
\[
    C_{k,S}
    :=
    \begin{cases}
        E_k, & k\in S,\\
        A_k, & k\notin S.
    \end{cases}
\]
If $|S|=h$, then the preceding bounds imply
\begin{align*}
    &
    \left\|
        \left(
            C_{d,S}
            \odot\cdots\odot
            C_{2,S}
        \right)^\top
        \left(
            \hat A_d^{(t)}
            \odot\cdots\odot
            \hat A_2^{(t)}
        \right)
    \right\|_{1\rightarrow1}                                   \\
    &\qquad\leq
    \left(
        1+\frac{\varepsilon_t^2}{2}
    \right)^{d-1-h}
    \left(
        \frac{3\varepsilon_t^2}{2}
    \right)^h
    +
    (R-1)
    \left(
        \varepsilon_t+\varepsilon_t^2
    \right)^h
    \left(
        \xi+\varepsilon_t
    \right)^{d-1-h}.
\end{align*}
Since
\[
\begin{split}
    &
    \hat A_d^{(t)}\mathcal{B}_{t,d}
    \odot\cdots\odot
    \hat A_2^{(t)}\mathcal{B}_{t,2}
    -
    A_d\odot\cdots\odot A_2                                    \\
    &\qquad=
    \sum_{\varnothing\neq S\subseteq\{2,\ldots,d\}}
    C_{d,S}\odot\cdots\odot C_{2,S},
\end{split}
\]
we obtain
\be
\label{eq_estimated_inverse_decomposition_t2}
\begin{split}
    &
    \left\|
        \left(
            \hat A_d^{(t)}\mathcal{B}_{t,d}
            \odot\cdots\odot
            \hat A_2^{(t)}\mathcal{B}_{t,2}
            -
            A_d\odot\cdots\odot A_2
        \right)^\top
        \left(
            \hat A_d^{(t)}
            \odot\cdots\odot
            \hat A_2^{(t)}
        \right)
    \right\|_{1\rightarrow1}                                   \\
    &\quad\leq
    \sum_{h=1}^{d-1}
    \binom{d-1}{h}
    \Bigg[
        \left(
            1+\frac{\varepsilon_t^2}{2}
        \right)^{d-1-h}
        \left(
            \frac{3\varepsilon_t^2}{2}
        \right)^h                                               \\
    &\hspace{4.7cm}
        +
        (R-1)
        \left(
            \varepsilon_t+\varepsilon_t^2
        \right)^h
        \left(
            \xi+\varepsilon_t
        \right)^{d-1-h}
    \Bigg]                                                       \\
    &\quad=
    \left(
        1+2\varepsilon_t^2
    \right)^{d-1}
    -
    \left(
        1+\frac{\varepsilon_t^2}{2}
    \right)^{d-1}                                               \\
    &\qquad+
    (R-1)
    \left[
        \left(
            \xi+2\varepsilon_t+\varepsilon_t^2
        \right)^{d-1}
        -
        \left(
            \xi+\varepsilon_t
        \right)^{d-1}
    \right]                                                     \\
    &\quad\leq
    \left(
        1+2\varepsilon_t^2
    \right)^{d-1}
    -1
    +
    (R-1)
    \left(
        \xi+2\varepsilon_t+\varepsilon_t^2
    \right)^{d-1}                                               \\
    &\qquad-
    (R-1)
    \left(
        \xi+\varepsilon_t
    \right)^{d-1}.
\end{split}
\ee

Combining \eqref{eq_inverse_bound} and \eqref{eq_estimated_inverse_decomposition_t2}, it yields
\be
\label{eq_B_hat_bound}
\begin{split}
         & \|A_1 \diag(|\lambda|)\|_{1\rightarrow 2} \ln \(( A_d\odot \cdots \odot A_2) -  (\hat A_d^{(t)} \mathcal{B}_{t,d} \odot
    \cdots \odot \hat A_2^{(t)} \mathcal{B}_{t,2})\)\T (\hat A_d^{(t)}\odot \cdots \odot \hat A_2^{(t)}) \rn_{1\rightarrow
    1} \times                                                                                                                      \\
         & \ln \((\hat A_d^{(t)}\odot \cdots \odot \hat A_2^{(t)})\T (\hat A_d^{(t)}\odot \cdots \odot \hat A_2^{(t)})\)\i
    \rn_{1\rightarrow 1}                                                                                                           \\
    \leq & \frac{\|A_1 \diag(|\lambda|)\|_{1\rightarrow 2} \(\(2\varepsilon_t^2 + 1\)^{d-1} - 1 + (R-1)(\varepsilon_t^2 +
    2\varepsilon_t + \xi)^{d-1} - (R-1)(\xi + \varepsilon_t)^{d-1}\)}{1 - (R-1)(\xi + 2 \varepsilon_t +
    \varepsilon_t^2)^{d-1}}                                                                                                        \\
    \leq & \frac{\lambda_{\max} \(\(2\varepsilon_t^2 + 1\)^{d-1} - 1 + (R-1)(\varepsilon_t^2 + 2\varepsilon_t + \xi)^{d-1} -
        (R-1)\xi^{d-1}\)}{1 - (R-1)(\xi + 2 \varepsilon_t + \varepsilon_t^2)^{d-1}}.
\end{split}
\ee

\paragraph*{Bound the noise term. \\}

Let
\[
    \hat K^{(t)}
    :=
    \hat A_d^{(t)}
    \odot\cdots\odot
    \hat A_2^{(t)}.
\]
Since $M=\hat K^{(t)\top}\hat K^{(t)}$ is invertible under
\eqref{eq_invertible_condition}, we have
\be
\label{eq_noise_term0}
    \left[
        \hat K^{(t)\top}
    \right]^\dagger
    =
    \hat K^{(t)}M^{-1}.
\ee
Therefore, by \eqref{eq_inverse_bound},
\be
\label{eq_noise_term1}
\begin{split}
    \left\|
        \MM_1(\Z)
        \left[
            \hat K^{(t)\top}
        \right]^\dagger
    \right\|_{1\rightarrow2}
    &\leq
    \left\|
        \MM_1(\Z)\hat K^{(t)}
    \right\|_{1\rightarrow2}
    \|M^{-1}\|_{1\rightarrow1}                                  \\
    &\leq
    \frac{
        \left\|
            \MM_1(\Z)\hat K^{(t)}
        \right\|_{1\rightarrow2}
    }{
        1-
        (R-1)
        \left(
            \xi+2\varepsilon_t+\varepsilon_t^2
        \right)^{d-1}
    }.
\end{split}
\ee

For every $j\in[R]$, define
\[
    s_{t,j}
    :=
    \prod_{k=2}^d
    \beta_{t,k,j}.
\]
Since
\[
    \hat a_{d,j}^{(t)}
    \otimes\cdots\otimes
    \hat a_{2,j}^{(t)}
    =
    s_{t,j}
    \left(
        \widetilde a_{d,j}^{(t)}
        \otimes\cdots\otimes
        \widetilde a_{2,j}^{(t)}
    \right),
\]
multiplication by the overall sign $s_{t,j}$ does not change the norm
of the noise contraction. Moreover,
\begin{align*}
    &
    \widetilde a_{d,j}^{(t)}
    \otimes\cdots\otimes
    \widetilde a_{2,j}^{(t)}
    -
    a_{d,j}\otimes\cdots\otimes a_{2,j}                         \\
    &\quad=
    \sum_{h=2}^d
    \widetilde a_{d,j}^{(t)}
    \otimes\cdots\otimes
    \widetilde a_{h+1,j}^{(t)}
    \otimes
    \left(
        \widetilde a_{h,j}^{(t)}-a_{h,j}
    \right)
    \otimes
    a_{h-1,j}
    \otimes\cdots\otimes
    a_{2,j},
\end{align*}
where empty products are omitted. Since all loading vectors have unit
norm and
\[
    \left\|
        \widetilde a_{h,j}^{(t)}-a_{h,j}
    \right\|
    \leq
    \varepsilon_t,
\]
we obtain
\be
\label{eq_noise_control_difference_decomp}
\begin{split}
    \left\|
        \MM_1(\Z)\hat K^{(t)}
    \right\|_{1\rightarrow2}
    &=
    \max_{j\in[R]}
    \left\|
        \MM_1(\Z)
        \left(
            \widetilde a_{d,j}^{(t)}
            \otimes\cdots\otimes
            \widetilde a_{2,j}^{(t)}
        \right)
    \right\|                                                     \\
    &\leq
    \max_{j\in[R]}
    \left\|
        \MM_1(\Z)
        \left(
            a_{d,j}\otimes\cdots\otimes a_{2,j}
        \right)
    \right\|                                                     \\
    &\quad+
    (d-1)\varepsilon_t
    \max_{\substack{\|b_k\|=1\\k=2,\ldots,d}}
    \left\|
        \MM_1(\Z)
        \left(
            b_d\otimes\cdots\otimes b_2
        \right)
    \right\|.
\end{split}
\ee

For deterministic $b_k$ with $\|b_k\|=1$, let
\[
    b
    :=
    b_d\otimes\cdots\otimes b_2.
\]
Then $\|b\|=1$. As above, the coordinates of $\MM_1(\Z)b$ are
independent and satisfy the sub-Gaussian moment-generating-function
condition with variance proxy $\sigma^2$. Therefore,
Supplement Lemma~\ref{lemma_subgaussian_vector_norm} gives, for every $\rho\geq0$,
\be
\label{eq_subgaussian_tail_bound}
\PP\left(
    \left\|
        \MM_1(\Z)
        (b_d\otimes\cdots\otimes b_2)
    \right\|
    >
    \sigma(\sqrt{p_1}+\rho)
\right)
\leq
2\exp\left(
    -\frac{\rho^2}{2}
\right).
\ee

By Supplement Lemma~\ref{lemma_e_net}, for every $k=2,\ldots,d$ and every
$b_k$ satisfying $\|b_k\|=1$, there exists
$\widetilde b_k\in\mathcal{N}_k$ such that
\[
    \|b_k-\widetilde b_k\|
    \leq
    \varepsilon.
\]
The telescoping identity
\begin{align*}
    &
    b_d\otimes\cdots\otimes b_2
    -
    \widetilde b_d\otimes\cdots\otimes\widetilde b_2             \\
    &\quad=
    \sum_{h=2}^d
    b_d\otimes\cdots\otimes b_{h+1}
    \otimes
    (b_h-\widetilde b_h)
    \otimes
    \widetilde b_{h-1}
    \otimes\cdots\otimes
    \widetilde b_2
\end{align*}
gives
\begin{align*}
    &
    \left\|
        \MM_1(\Z)
        \left(
            b_d\otimes\cdots\otimes b_2
        \right)
    \right\|                                                     \\
    &\quad\leq
    \varepsilon(d-1)
    \max_{\substack{\|c_k\|=1\\k=2,\ldots,d}}
    \left\|
        \MM_1(\Z)
        \left(
            c_d\otimes\cdots\otimes c_2
        \right)
    \right\|                                                     \\
    &\qquad+
    \left\|
        \MM_1(\Z)
        \left(
            \widetilde b_d
            \otimes\cdots\otimes
            \widetilde b_2
        \right)
    \right\|.
\end{align*}
Taking the maximum over the unit vectors $b_2,\ldots,b_d$ yields
\[
\begin{split}
    &
    \max_{\substack{\|b_k\|=1\\k=2,\ldots,d}}
    \left\|
        \MM_1(\Z)
        \left(
            b_d\otimes\cdots\otimes b_2
        \right)
    \right\|                                                     \\
    &\quad\leq
    \frac{1}{1-(d-1)\varepsilon}
    \max_{\substack{
        \widetilde b_k\in\mathcal{N}_k\\
        k=2,\ldots,d
    }}
    \left\|
        \MM_1(\Z)
        \left(
            \widetilde b_d
            \otimes\cdots\otimes
            \widetilde b_2
        \right)
    \right\|.
\end{split}
\]
Thus, taking $\varepsilon = \frac{1}{2(d-1)}$, it follows:
\be
\begin{split}
         & \PP\(\max_{\|b_k\| = 1} \|\MM_1(\Z) (b_d \otimes \cdots \otimes b_2)\|_F > 2\sigma(\sqrt{p_1} + \rho) \)             \\
    \leq & \PP\(\max_{b_k \in \mathcal{N}_k} \|\MM_1(\Z) (\tilde b_d \otimes \cdots \otimes \tilde b_2)\|_F > \sigma(\sqrt{p_1}
    + \rho) \)                                                                                                                  \\
    \leq & \sum_{b_k \in \mathcal{N}_k} \PP\(\|\MM_1(\Z) (\tilde b_d \otimes \cdots \otimes \tilde b_2)\|_F > \sigma(\sqrt{p_1}
    + \rho) \)                                                                                                                  \\
    \    \leq & 2 \(1+\frac{2}{\varepsilon}\)^{\sum_{k>1}p_k} \exp\(-\frac{\rho^2}{2}\)                                              \\
    =    & 2 \exp\(\(\sum_{k>1}p_k\)\log\(4d - 3\)  -\frac{\rho^2}{2}\) .
\end{split}
\ee
{Let $\rho = \sqrt{6(d-1)p_{\max}\log d}$.
    Since $\sum_{k>1}p_k\leq(d-1)p_{\max}$ and $4d-3\leq d^2$ for $d\geq3$, we further have
    \[
        \PP\(\max_{\|b_k\| = 1} \|\MM_1(\Z) (b_d \otimes \cdots \otimes b_2)\|_F > 2\sigma(\sqrt{p_1} + \rho) \)
        \leq 2\exp\(-(d-1)p_{\max}\log 2\).
    \]
    Moreover, since $p_1\leq p_{\max}$ and $d\geq3$,
    \[
        2\sigma(\sqrt{p_1}+\rho)
        \leq
        7\sigma\sqrt{(d-1)p_{\max}\log d}.
    \]
    Thus, it holds with high probability that
    \be
    \label{eq_assumption_HP_1} \max_{\|b_k\| = 1} \|\MM_1(\Z) (b_d \otimes \cdots \otimes b_2)\|_F \leq 7 \sigma
    \sqrt{(d-1)p_{\max}\log d}.
    \ee
}

Further note that by \eqref{eq_subgaussian_tail_bound}, we have
\be
\notag
\begin{split}
         & \PP\(\max_{j \in [R]} \ln \MM_1(\Z) (a_{d,j} \otimes \cdots \otimes  a_{2,j}) \rn > \sigma(\sqrt{p_1} + \rho)\) \\
    \leq & \sum_{j \in [R]} \PP\(\ln \MM_1(\Z) (a_{d,j} \otimes \cdots \otimes  a_{2,j}) \rn > \sigma(\sqrt{p_1} + \rho)\) \\
    \leq & 2 R\exp\(-\frac{\rho^2}{2 }\),
\end{split}
\ee
Hence, if we let $\rho = \sqrt{p_{\max}}$, then we have
\[
    \PP\(\max_{j \in [R]} \ln \MM_1(\Z) (a_{d,j} \otimes \cdots \otimes  a_{2,j}) \rn > \sigma(\sqrt{p_1} + \rho)\) \leq
    2\exp\(-\frac{p_{\max}}{2 } + \log R\).
\]
Thus, it holds with high probability that
\be
\label{eq_assumption_HP_2} \max_{j \in [R]} \ln \MM_1(\Z) (a_{d,j} \otimes \cdots \otimes  a_{2,j}) \rn \leq 2\sigma
\sqrt{p_{\max}},
\ee

Then, \eqref{eq_noise_control_difference_decomp}, \eqref{eq_assumption_HP_1}, and \eqref{eq_assumption_HP_2} yield
\be
\label{eq_noise_random_term_1} \ln\MM_1(\Z)  (\hat A_d^{(t)}\odot \cdots \odot \hat A_2^{(t)}) \rn_{1\rightarrow 2} \leq 2\sigma
\sqrt{p_{\max}} + \varepsilon_t 7 (d-1) \sigma \sqrt{(d-1)p_{\max}\log d}.
\ee

If we further have $\varepsilon_t < \frac{1}{(d-1)^{3/2}\sqrt{\log d}}$, then we have
\be
\label{eq_noise_random_term_2} \ln\MM_1(\Z)  (\hat A_d^{(t)}\odot \cdots \odot \hat A_2^{(t)}) \rn_{1\rightarrow 2} \leq 9
\sigma \sqrt{p_{\max}}.
\ee

Finally, by \eqref{eq_update_main_inequality_general_d}, \eqref{eq_B_hat_bound}, and \eqref{eq_noise_term1}, we have
\be
\label{eq_l1l2norm_bound}
\begin{split}
         & \min_{\mathcal B\in\mathfrak D_R}
\left\|
    \hat B_1^{(t+1)}
    -
    A_1\diag(|\lambda|)\mathcal B
\right\|_{1\rightarrow2}                                                                                                         \\                                                                                                                       \\
    \leq & \frac{\lambda_{\max} \(\(2\varepsilon_t^2 + 1\)^{d-1} - 1 + (R-1)(\varepsilon_t^2 + 2\varepsilon_t + \xi)^{d-1} -
        (R-1)\xi^{d-1}\) + \delta}{1 - (R-1)(\xi + 2 \varepsilon_t + \varepsilon_t^2)^{d-1}},
\end{split}
\ee
where $\delta = \ln \MM_1(\Z)  (\hat A_d^{(t)}\odot \cdots \odot \hat A_2^{(t)}) \rn_{1\rightarrow 2}$ and can be controlled by \eqref{eq_noise_random_term_1} or \eqref{eq_noise_random_term_2} in different cases.

For every $r\in[R]$, write
\[
    \hat\lambda_r^{(t+1)}
    :=
    \|\hat b_{1,r}^{(t+1)}\|,
    \qquad
    \hat a_{1,r}^{(t+1)}
    =
    \frac{
        \hat b_{1,r}^{(t+1)}
    }{
        \hat\lambda_r^{(t+1)}
    }.
\]
By the reverse triangle inequality, for every
$\beta\in\{-1,1\}$,
\[
    \left|
        \hat\lambda_r^{(t+1)}-|\lambda_r|
    \right|
    \leq
    \left\|
        \hat b_{1,r}^{(t+1)}
        -
        \beta|\lambda_r|a_{1,r}
    \right\|.
\]
Consequently,
\be
\label{eq_lambda_bound}
\begin{split}
    &
    \max_{r\in[R]}
    \left|
        \hat\lambda_r^{(t+1)}-|\lambda_r|
    \right|                                                     \\
    &\quad\leq
    \min_{\mathcal{B}\in\mathfrak{D}_R}
    \left\|
        \hat B_1^{(t+1)}
        -
        A_1\diag(|\lambda|)\mathcal{B}
    \right\|_{1\rightarrow2}.
\end{split}
\ee

We next use the elementary normalization inequality
\[
    a
    \left\|
        \frac{x}{\|x\|}-u
    \right\|
    \leq
    2\|x-au\|,
\]
which holds for every $x\neq0$, every unit vector $u$, and every
$a>0$. Indeed,
\[
\begin{split}
    a
    \left\|
        \frac{x}{\|x\|}-u
    \right\|
    &\leq
    |a-\|x\||
    +
    \|x-au\|                                                     \\
    &\leq
    2\|x-au\|.
\end{split}
\]
Applying this inequality columnwise with
$x=\hat b_{1,r}^{(t+1)}$, $a=|\lambda_r|$, and
$u=\beta a_{1,r}$ gives
\be
\label{eq_A_hat_bound}
\begin{split}
    &
    \lambda_{\min}
    \max_{r\in[R]}
    \min_{\beta\in\{-1,1\}}
    \left\|
        \hat a_{1,r}^{(t+1)}
        -
        \beta a_{1,r}
    \right\|                                                     \\
    &\quad\leq
    2
    \min_{\mathcal{B}\in\mathfrak{D}_R}
    \left\|
        \hat B_1^{(t+1)}
        -
        A_1\diag(|\lambda|)\mathcal{B}
    \right\|_{1\rightarrow2}                                    \\
    &\quad\leq
    \frac{
        2\lambda_{\max}
        \left[
            \left(
                1+2\varepsilon_t^2
            \right)^{d-1}
            -1
            +
            (R-1)
            \left(
                \varepsilon_t^2+2\varepsilon_t+\xi
            \right)^{d-1}
            -
            (R-1)\xi^{d-1}
        \right]
        +
        2\delta
    }{
        1-
        (R-1)
        \left(
            \xi+2\varepsilon_t+\varepsilon_t^2
        \right)^{d-1}
    },
\end{split}
\ee
where
\[
    \delta
    :=
    \left\|
        \MM_1(\Z)
        \left(
            \hat A_d^{(t)}
            \odot\cdots\odot
            \hat A_2^{(t)}
        \right)
    \right\|_{1\rightarrow2}.
\]
The same argument applies to every mode. Taking the maximum over
$k\in[d]$ and using the corresponding uniform noise bound therefore
yields
\be
\label{eq_recurrence_formula_general_R}
    \varepsilon_{t+1}
    \leq
    \frac{
        2\lambda_{\max}
        \left[
            \left(
                1+2\varepsilon_t^2
            \right)^{d-1}
            -1
            +
            (R-1)
            \left(
                \varepsilon_t^2+2\varepsilon_t+\xi
            \right)^{d-1}
            -
            (R-1)\xi^{d-1}
        \right]
        +
        2\delta
    }{
        \lambda_{\min}
        \left[
            1-
            (R-1)
            \left(
                \xi+2\varepsilon_t+\varepsilon_t^2
            \right)^{d-1}
        \right]
    }.
\ee

\subsection{Analysis of the recurrence formula}

We first verify that the bounds used below follow from the assumptions of Theorem~\ref{thm_local}. Since $\lambda_{\max}/\lambda_{\min}\geq1$, condition~\eqref{eq_assumption_1_general_R} implies $\varepsilon_0\lesssim C_1^{-1}(d+R)^{-1}$. If $d=3$, then $(d+R)^{-1}\leq R^{-1/2}$ because $R\geq2$. If $d>3$, the elementary bound $dR^{1/(d-3)}\leq4(d+R)$ gives $(d+R)^{-1}\leq4\{dR^{1/(d-3)}\}^{-1}$. Therefore, combining conditions~\eqref{eq_assumption_1_general_R} and~\eqref{eq_assumption_2_general_R}, and taking $C_1$ sufficiently large, yields \eqref{eq_assumption_general_R_0} and \eqref{eq_assumption_general_R_1} below. Condition~\eqref{eq_assumption_3_general_R} similarly yields \eqref{eq_assumption_general_R_2} and \eqref{eq_assumption_general_R_3}. Thus, we may use the following bounds in the proof:
{
\bea
0. & &
\varepsilon_0
\frac{\lambda_{\max}}{\lambda_{\min}}
(d+R)
\leq
c;
\label{eq_assumption_general_R_0}
\\
1. & &\text{If $d >3$, } \varepsilon_0 + \xi \leq \(C dR^{\frac{1}{d-3}}\)\i,\text{ if $d =3$, $\varepsilon_0 + \xi \leq
        (CR^{1/2})\i$; }\label{eq_assumption_general_R_1} \\
2. & &\lambda_{\min} \geq C \sigma p_{\max}^{1/2} d^2 \log d; \label{eq_assumption_general_R_2} \\
3. & &\lambda_{\min} \geq C \frac{\lambda_{\max}}{\lambda_{\min}}\sigma \sqrt{dp_{\max}\log d} (d+R);
\label{eq_assumption_general_R_3} \\
4. & &\frac{\lambda_{\max}}{\lambda_{\min}} R d \xi^{d-2} \leq c. \label{eq_assumption_general_R_4}
\eea
}
Here, $C$ and $c$ are absolute constants. Condition~\eqref{eq_assumption_general_R_4} is needed only when $\sigma\sqrt{p_{\max}}/\lambda_{\min}<\tau_{\mathrm{ALS}}$, corresponding to the second conclusion of Theorem~\ref{thm_local}.

\subsubsection{Simplify the recurrence formula}

The simplifications below are derived under the local bound $\varepsilon_t\leq\varepsilon_0$.
In the convergence analysis below, we verify by induction that, until the target accuracy is reached,
\[
    \varepsilon_{t+1}
    \leq
    \varepsilon_t
    \leq
    \varepsilon_0.
\]
Thus, the use of the local bound in deriving the recurrence is self-consistent.

The denominator of \eqref{eq_recurrence_formula_general_R} is lower bounded as
\be
\label{eq_denominator_bound_1} 1 - (R-1)(\xi + 2 \varepsilon_t + \varepsilon_t^2)^{d-1} \geq 1 - (R-1)(\xi + 3
\varepsilon_t)^{d-1} > c
\ee
where $c$ is some absolute constant and the final inequality holds by \eqref{eq_assumption_general_R_1}.

Note the fact that for $x >0$ and integer $k$ such that $kx \leq 1/2$, we have
\be
\label{eq_exp_bound} (1+x)^k = \exp(k\log (1+x)) \leq \exp(kx) \leq 1.5 kx +1.
\ee
Then by \eqref{eq_assumption_general_R_1}, we have $\varepsilon_0 \leq (4d)^{-1/2}$, which by \eqref{eq_exp_bound} further yields
\be
\label{eq_bound_first_term_formal}
\begin{split}
    \(2\varepsilon_t^2 + 1\)^{d-1} - 1 \leq & 3(d-1) \varepsilon_t^2.
\end{split}
\ee

For brevity, define
\[
    \mathcal{N}_d(e) := (1+2e^2)^{d-1}-1 +(R-1)(e^2+2e+\xi)^{d-1} -(R-1)\xi^{d-1},
\]
and recall the transition scale
\[
    \tau_{\mathrm{ALS}} =
    \begin{cases}
        \xi,            & d=3, \\[2mm]
        (R-1)\xi^{d-2}, & d>3.
    \end{cases}
\]

We first consider $d=3$.
A direct expansion gives
\begin{align*}
    \mathcal{N}_3(e) ={} & 4(R-1)\xi e + \left\{ 4R+2(R-1)\xi \right\}e^2 \\
                         & + 4(R-1)e^3 + (R+3)e^4.
\end{align*}
By \eqref{eq_assumption_general_R_1} and $e=\varepsilon_t\leq\varepsilon_0$, both $e$ and $\xi$ are bounded by sufficiently small absolute constants.
Consequently,
\[
    \mathcal{N}_3(e) \lesssim R\left(e^2+\xi e\right).
\]
If $e>\xi=\tau_{\mathrm{ALS}}$, then
\[
    \mathcal{N}_3(e) \lesssim R e^2 \lesssim (d+R)e^2.
\]
If $e\leq\xi=\tau_{\mathrm{ALS}}$, then
\[
    \mathcal{N}_3(e) \lesssim R\xi e \lesssim Rd\xi^{d-2}e.
\]

We next consider $d>3$.
Suppose first that
\[
    e>\frac{\xi}{2d}.
\]
Since $e\leq\varepsilon_0\leq1$ and $\xi<2de$, we have
\[
    e^2+2e+\xi \leq 3de.
\]
Furthermore, after increasing the absolute constant in \eqref{eq_assumption_general_R_1}, we have
\[
    R(3de)^{d-3} \leq R(3d\varepsilon_0)^{d-3} \leq \left(\frac{3}{C}\right)^{d-3} \leq \frac{1}{d}.
\]
It follows from \eqref{eq_bound_first_term_formal} that
\begin{align*}
    \mathcal{N}_d(e) & \leq 3(d-1)e^2 +(R-1)(3de)^{d-1}  \\
                     & \lesssim d e^2 \lesssim (d+R)e^2.
\end{align*}

Suppose now that
\[
    e\leq\frac{\xi}{2d}.
\]
Since $e\leq1$, the mean value theorem gives
\begin{align*}
     & (e^2+2e+\xi)^{d-1}-\xi^{d-1}           \\
     & \qquad\leq (d-1)(\xi+3e)^{d-2}(e^2+2e) \\
     & \qquad\leq C d\xi^{d-2}e,
\end{align*}
where we used
\[
    \xi+3e \leq \xi\left(1+\frac{3}{2d}\right)
\]
and
\[
    \left(1+\frac{3}{2d}\right)^{d-2} \leq C.
\]
Together with \eqref{eq_bound_first_term_formal}, this yields
\[
    \mathcal{N}_d(e) \lesssim d e^2 +(R-1)d\xi^{d-2}e.
\]

Moreover, \eqref{eq_assumption_general_R_1} implies
\[
    (R-1)\xi^{d-3} \leq \frac{1}{C^{d-3}d^{d-3}} \leq \frac{1}{2d}
\]
after increasing the absolute constant $C$.
Therefore,
\[
    \tau_{\mathrm{ALS}} = (R-1)\xi^{d-2} \leq \frac{\xi}{2d}.
\]
If $e>\tau_{\mathrm{ALS}}$, then either $e>\xi/(2d)$, in which case the preceding large-error bound applies, or
\[
    \tau_{\mathrm{ALS}} < e \leq \frac{\xi}{2d},
\]
in which case
\[
    (R-1)d\xi^{d-2}e = d\tau_{\mathrm{ALS}}e \leq de^2.
\]
Thus,
\[
    \mathcal{N}_d(e) \lesssim (d+R)e^2 \qquad \text{when }e>\tau_{\mathrm{ALS}}.
\]
If $e\leq\tau_{\mathrm{ALS}}$, then necessarily $e\leq\xi/(2d)$ and
\[
    e^2 \leq \tau_{\mathrm{ALS}}e.
\]
Consequently,
\[
    \mathcal{N}_d(e) \lesssim Rd\xi^{d-2}e \qquad \text{when }e\leq\tau_{\mathrm{ALS}}.
\]

Combining the cases $d=3$ and $d>3$ with \eqref{eq_recurrence_formula_general_R} and \eqref{eq_denominator_bound_1}, we obtain
\be
\label{eq_recurrence_order2} \varepsilon_{t+1} \lesssim \frac{\lambda_{\max}}{\lambda_{\min}} (d+R)\varepsilon_t^2 +
\frac{\delta}{\lambda_{\min}}, \qquad \text{if }\varepsilon_t>\tau_{\mathrm{ALS}},
\ee
and
\be
\label{eq_recurrence_order1} \varepsilon_{t+1} \lesssim \frac{\lambda_{\max}}{\lambda_{\min}} Rd\xi^{d-2}\varepsilon_t +
\frac{\delta}{\lambda_{\min}}, \qquad \text{if }\varepsilon_t\leq\tau_{\mathrm{ALS}}.
\ee

\subsubsection{Convergence analysis}

We are going to calculate how many steps it will need to get the error level of $\sigma p_{\max}^{1/2} / \lambda_{\min}$.
Note that \eqref {eq_noise_random_term_1} and \eqref{eq_assumption_general_R_1}, we have
\[
    \delta \lesssim \delta_1 := \sigma \sqrt{dp_{\max}\log d},
\]
and if we further have $\varepsilon_t < \frac{1}{(d-1)^{3/2}\sqrt{\log d}}$, then
\[
    \delta \lesssim \delta_2 := \sigma \sqrt{p_{\max}}.
\]
We will first prove that $\varepsilon_t$ will reach the order of $\delta_1/ \lambda_{\min}$, and then by condition \eqref{eq_assumption_general_R_2}, we have $\varepsilon_t < \frac{1}{(d-1)^{3/2}\sqrt{\log d}}$ thereafter, and finally prove $\varepsilon_t$ converge to $\delta_2 /\lambda_{\min} = \sigma p_{\max}^{1/2} / \lambda_{\min}$.

By \eqref{eq_recurrence_order2} and \eqref{eq_recurrence_order1}, the quadratic recurrence applies while $\varepsilon_t>\tau_{\mathrm{ALS}}$, and the linear recurrence applies once $\varepsilon_t\leq\tau_{\mathrm{ALS}}$.
We therefore distinguish two cases according to whether the final statistical error lies above or below the transition scale.

\paragraph*{When
    $\delta_2/\lambda_{\min}\geq\tau_{\mathrm{ALS}}$: \\}

Choose an absolute constant $C\geq1$ sufficiently large so that,
whenever the quadratic recurrence
\eqref{eq_recurrence_order2} applies, it can be written as
\begin{equation}\label{eq_quadratic_recurrence_concrete}
    \varepsilon_{t+1}
    \leq
    A_{\mathrm q}\varepsilon_t^2+b_j,
\end{equation}
where
\[
    A_{\mathrm q}
    :=
    C
    \frac{\lambda_{\max}}{\lambda_{\min}}
    (d+R),
    \qquad
    b_j
    :=
    C\frac{\delta_j}{\lambda_{\min}},
    \qquad
    j\in\{1,2\}.
\]
By \eqref{eq_assumption_general_R_3}, after increasing the absolute
constant in that assumption if necessary,
\[
    4A_{\mathrm q}b_1
    \leq
    \frac12.
\]
Since $\delta_2\leq\delta_1$, the same inequality also holds with
$b_1$ replaced by $b_2$.

For $j\in\{1,2\}$, define the smaller fixed point
\begin{equation}\label{eq_quadratic_fixed_point}
    \varepsilon_j^\star
    :=
    \frac{
        1-\sqrt{1-4A_{\mathrm q}b_j}
    }{
        2A_{\mathrm q}
    }.
\end{equation}
Then
\[
    A_{\mathrm q}(\varepsilon_j^\star)^2
    -
    \varepsilon_j^\star
    +
    b_j
    =
    0,
\]
and, equivalently,
\[
    b_j
    =
    \varepsilon_j^\star
    -
    A_{\mathrm q}(\varepsilon_j^\star)^2.
\]
Moreover,
\[
    b_j
    \leq
    \varepsilon_j^\star
    =
    \frac{
        2b_j
    }{
        1+\sqrt{1-4A_{\mathrm q}b_j}
    }
    \leq
    2b_j.
\]

We first use the coarser noise level $\delta_1$. Since
\[
    \frac{\delta_2}{\lambda_{\min}}
    \geq
    \tau_{\mathrm{ALS}}
\]
and $\delta_1\geq\delta_2$, we have
\[
    \varepsilon_1^\star
    \geq
    b_1
    \geq
    \tau_{\mathrm{ALS}},
\]
after adjusting the absolute constant $C$ if necessary. Hence, as long
as
\[
    \varepsilon_t
    >
    3\varepsilon_1^\star,
\]
we also have $\varepsilon_t>\tau_{\mathrm{ALS}}$, so the quadratic
recurrence \eqref{eq_quadratic_recurrence_concrete} applies.

Define the first hitting time
\[
    T_1
    :=
    \inf\left\{
        t\geq0:
        \varepsilon_t
        \leq
        3\varepsilon_1^\star
    \right\}.
\]
For every $t<T_1$, set
\[
    x_t
    :=
    \varepsilon_t-\varepsilon_1^\star.
\]
Since $\varepsilon_t>3\varepsilon_1^\star$, we have
\[
    \varepsilon_t+\varepsilon_1^\star
    \leq
    2\left(
        \varepsilon_t-\varepsilon_1^\star
    \right)
    =
    2x_t.
\]
Using \eqref{eq_quadratic_recurrence_concrete} and the fixed-point
identity, for every $t$ such that $t+1<T_1$,
\begin{align*}
    x_{t+1}
    &=
    \varepsilon_{t+1}-\varepsilon_1^\star                                      \\
    &\leq
    A_{\mathrm q}
    \left(
        \varepsilon_t^2
        -
        (\varepsilon_1^\star)^2
    \right)                                                                     \\
    &=
    A_{\mathrm q}
    \left(
        \varepsilon_t-\varepsilon_1^\star
    \right)
    \left(
        \varepsilon_t+\varepsilon_1^\star
    \right)                                                                     \\
    &\leq
    2A_{\mathrm q}x_t^2.
\end{align*}

Define
\[
    q_t
    :=
    2A_{\mathrm q}x_t.
\]
By \eqref{eq_assumption_general_R_0}, taking the sufficiently small
absolute constant $c$ therein gives
\[
    q_0
    \leq
    2A_{\mathrm q}\varepsilon_0
    \leq
    \frac12.
\]
The preceding recurrence gives
\[
    q_{t+1}
    \leq
    q_t^2.
\]
Consequently, a direct induction yields
\begin{equation}\label{eq_quadratic_double_exponential_phase1}
    q_t
    \leq
    q_0^{2^t}
    \leq
    2^{-2^t},
    \qquad
    t<T_1.
\end{equation}
Indeed, if the claim holds at $t$, then
\[
    q_{t+1}
    \leq
    q_t^2
    \leq
    \left(
        q_0^{2^t}
    \right)^2
    =
    q_0^{2^{t+1}}.
\]

This induction also verifies monotonicity. As long as $t+1<T_1$,
\[
    x_{t+1}
    \leq
    q_t x_t
    \leq
    \frac12x_t,
\]
and therefore
\[
    \varepsilon_{t+1}
    \leq
    \varepsilon_t.
\]
It follows that
\[
    \varepsilon_t
    \leq
    \varepsilon_0
\]
throughout the first quadratic phase, which verifies the local
invariant assumed when deriving
\eqref{eq_recurrence_order2}.

Let $m_1$ be the smallest nonnegative integer satisfying
\[
    2^{-2^{m_1}}
    \leq
    4A_{\mathrm q}\varepsilon_1^\star.
\]
If $T_1>m_1$, then
\eqref{eq_quadratic_double_exponential_phase1} gives
\[
    x_{m_1}
    =
    \frac{q_{m_1}}{2A_{\mathrm q}}
    \leq
    2\varepsilon_1^\star,
\]
and hence
\[
    \varepsilon_{m_1}
    \leq
    3\varepsilon_1^\star,
\]
contradicting the definition of $T_1$. Therefore,
\[
    T_1
    \leq
    m_1
    \lesssim
    1+
    \log\log\left(
        \frac{1}{
            A_{\mathrm q}\varepsilon_1^\star
        }
    \right)
    \lesssim
    1+
    \log\log\left(
        \frac{\lambda_{\min}}{\delta_1}
    \right).
\]
At this time,
\begin{equation}\label{eq_reach_delta1_region}
    \varepsilon_{T_1}
    \leq
    3\varepsilon_1^\star
    \lesssim
    \frac{\delta_1}{\lambda_{\min}}.
\end{equation}

By \eqref{eq_assumption_general_R_2}, after increasing the absolute
constant therein if necessary,
\[
    \varepsilon_{T_1}
    <
    \frac{1}{
        (d-1)^{3/2}\sqrt{\log d}
    }.
\]
Therefore, the refined noise bound
\eqref{eq_noise_random_term_2} applies from this point onward.

We next use the refined noise level $\delta_2$. Define
$\varepsilon_2^\star$ by
\eqref{eq_quadratic_fixed_point} with $j=2$. As above,
\[
    \frac{\delta_2}{\lambda_{\min}}
    \lesssim
    \varepsilon_2^\star
    \lesssim
    \frac{\delta_2}{\lambda_{\min}},
\]
and, in the present case,
\[
    \varepsilon_2^\star
    \geq
    \tau_{\mathrm{ALS}}
\]
after adjusting absolute constants.

By \eqref{eq_reach_delta1_region} and
\eqref{eq_assumption_general_R_3},
\[
    2A_{\mathrm q}
    \varepsilon_{T_1}
    \lesssim
    A_{\mathrm q}
    \frac{\delta_1}{\lambda_{\min}}
    \leq
    \frac12.
\]
Define
\[
    T_2
    :=
    \inf\left\{
        t\geq T_1:
        \varepsilon_t
        \leq
        3\varepsilon_2^\star
    \right\}.
\]
Repeating the preceding induction with
$\varepsilon_1^\star$ replaced by $\varepsilon_2^\star$ and with
initial index $T_1$, we obtain, for every $m\geq0$ such that
$T_1+m<T_2$,
\begin{equation}\label{eq_quadratic_double_exponential_phase2}
    2A_{\mathrm q}
    \left(
        \varepsilon_{T_1+m}
        -
        \varepsilon_2^\star
    \right)
    \leq
    \left[
        2A_{\mathrm q}
        \left(
            \varepsilon_{T_1}
            -
            \varepsilon_2^\star
        \right)
    \right]^{2^m}
    \leq
    2^{-2^m}.
\end{equation}
In particular,
\[
    T_2-T_1
    \lesssim
    1+
    \log\log\left(
        \frac{\lambda_{\min}}{\delta_2}
    \right),
\]
and
\[
    \varepsilon_{T_2}
    \leq
    3\varepsilon_2^\star
    \lesssim
    \frac{\delta_2}{\lambda_{\min}}.
\]
The same induction shows that the iterates are nonincreasing before
$T_2$, so they remain in the local region in which the recurrence was
derived.

Since $\delta_1\geq\delta_2$, combining the two phases gives
\begin{align*}
    T_2
    &\lesssim
    1+
    \log\log\left(
        \frac{\lambda_{\min}}{\delta_1}
    \right)
    +
    1+
    \log\log\left(
        \frac{\lambda_{\min}}{\delta_2}
    \right)                                                              \\
    &\lesssim
    \log\log\left(
        \frac{\lambda_{\min}}{\delta_2}
    \right),
\end{align*}
after adjusting the absolute constants. Since
\[
    \delta_2
    =
    \sigma\sqrt{p_{\max}},
\]
we conclude that
\[
    \varepsilon_{T_2}
    \lesssim
    \frac{
        \sigma\sqrt{p_{\max}}
    }{
        \lambda_{\min}
    }
\]
within
\[
    T_2
    \lesssim
    \log\log\left(
        \frac{
            \lambda_{\min}
        }{
            \sigma\sqrt{p_{\max}}
        }
    \right)
\]
iterations.

\paragraph*{When
    $\delta_2/\lambda_{\min}<\tau_{\mathrm{ALS}}$: \\}

We first record a variant of the preceding stopping-time argument.
For $j\in\{1,2\}$, define
\[
    r_j
    :=
    \tau_{\mathrm{ALS}}
    \lor
    3\varepsilon_j^\star.
\]
Suppose that, from some initial index $t_{\mathrm{in}}$, the quadratic
recurrence with noise level $b_j$ applies whenever
$\varepsilon_t>\tau_{\mathrm{ALS}}$, and that
\[
    2A_{\mathrm q}\varepsilon_{t_{\mathrm{in}}}
    \leq
    \frac12.
\]
If $\varepsilon_{t_{\mathrm{in}}}\leq r_j$, no further argument is
needed. Otherwise, define
\[
    T_j
    :=
    \inf\left\{
        m\geq0:
        \varepsilon_{t_{\mathrm{in}}+m}
        \leq
        r_j
    \right\}.
\]
For every $m<T_j$, we have
\[
    \varepsilon_{t_{\mathrm{in}}+m}
    >
    r_j
    \geq
    \tau_{\mathrm{ALS}}
    \lor
    3\varepsilon_j^\star.
\]
Hence, the quadratic recurrence applies, and the same calculation as
in \eqref{eq_quadratic_double_exponential_phase1} gives
\[
    2A_{\mathrm q}
    \left(
        \varepsilon_{t_{\mathrm{in}}+m}
        -
        \varepsilon_j^\star
    \right)
    \leq
    2^{-2^m},
    \qquad
    m<T_j.
\]
Since
\[
    \varepsilon_j^\star
    \leq
    \frac{r_j}{3},
\]
taking $m$ sufficiently large that
\[
    2^{-2^m}
    \leq
    \frac{4}{3}A_{\mathrm q}r_j
\]
would imply
\[
    \varepsilon_{t_{\mathrm{in}}+m}
    \leq
    r_j.
\]
Consequently,
\[
    T_j
    \lesssim
    1+
    \log\log\left(
        \frac{1}{A_{\mathrm q}r_j}
    \right)
    \lesssim
    1+
    \log\log\left(
        \frac{\lambda_{\min}}{\delta_j}
    \right).
\]
As in the preceding argument, the iterates are nonincreasing before
the corresponding stopping time.

We first apply this argument with $j=1$ and
$t_{\mathrm{in}}=0$. Condition
\eqref{eq_assumption_general_R_0} gives
\[
    2A_{\mathrm q}\varepsilon_0
    \leq
    \frac12
\]
after adjusting the absolute constants. Thus, there exists an index
$T_{\mathrm c}$ satisfying
\[
    T_{\mathrm c}
    \lesssim
    1+
    \log\log\left(
        \frac{\lambda_{\min}}{\delta_1}
    \right)
\]
such that
\[
    \varepsilon_{T_{\mathrm c}}
    \leq
    r_1
    =
    \tau_{\mathrm{ALS}}
    \lor
    3\varepsilon_1^\star.
\]
Moreover, since
\[
    \varepsilon_1^\star
    \leq
    2b_1
    \lesssim
    \frac{\delta_1}{\lambda_{\min}},
\]
conditions \eqref{eq_assumption_general_R_1} and
\eqref{eq_assumption_general_R_2}, after increasing their absolute
constants if necessary, imply
\[
    r_1
    <
    \frac{1}{
        (d-1)^{3/2}\sqrt{\log d}
    }.
\]
Therefore, the refined noise bound
\eqref{eq_noise_random_term_2} applies from this point onward.

Set
\[
    r_2
    :=
    \tau_{\mathrm{ALS}}
    \lor
    3\varepsilon_2^\star.
\]
If
\[
    \varepsilon_{T_{\mathrm c}}
    \leq
    r_2,
\]
let $T_{\mathrm q}:=T_{\mathrm c}$. Otherwise,
$\varepsilon_{T_{\mathrm c}}>r_2\geq\tau_{\mathrm{ALS}}$.
In this case, necessarily
\[
    r_1
    =
    3\varepsilon_1^\star,
\]
and hence
\[
    \varepsilon_{T_{\mathrm c}}
    \leq
    3\varepsilon_1^\star
    \lesssim
    \frac{\delta_1}{\lambda_{\min}}.
\]
By \eqref{eq_assumption_general_R_3}, after increasing the absolute
constant therein if necessary,
\[
    2A_{\mathrm q}\varepsilon_{T_{\mathrm c}}
    \leq
    \frac12.
\]
We may therefore apply the preceding stopping-time argument with
$j=2$ and $t_{\mathrm{in}}=T_{\mathrm c}$. It follows that there
exists an index $T_{\mathrm q}\geq T_{\mathrm c}$ such that
\[
    \varepsilon_{T_{\mathrm q}}
    \leq
    r_2
\]
and
\[
    T_{\mathrm q}-T_{\mathrm c}
    \lesssim
    1+
    \log\log\left(
        \frac{\lambda_{\min}}{\delta_2}
    \right).
\]
Since $\delta_1\geq\delta_2$, we have
\[
    T_{\mathrm q}
    \lesssim
    \log\log\left(
        \frac{\lambda_{\min}}{\delta_2}
    \right).
\]

If
\[
    3\varepsilon_2^\star
    \geq
    \tau_{\mathrm{ALS}},
\]
then
\[
    \varepsilon_{T_{\mathrm q}}
    \leq
    3\varepsilon_2^\star
    \lesssim
    \frac{\delta_2}{\lambda_{\min}},
\]
and the desired statistical accuracy has already been reached.

It remains to consider the case
\[
    3\varepsilon_2^\star
    <
    \tau_{\mathrm{ALS}}.
\]
Then
\[
    \varepsilon_{T_{\mathrm q}}
    \leq
    \tau_{\mathrm{ALS}}.
\]
Choose a sufficiently large absolute constant $C_{\mathrm l}$ and
define
\[
    A_{\mathrm l}
    :=
    C_{\mathrm l}
    \frac{\lambda_{\max}}{\lambda_{\min}}
    Rd\xi^{d-2},
    \qquad
    b_{\mathrm l}
    :=
    C_{\mathrm l}
    \frac{\delta_2}{\lambda_{\min}},
\]
so that, whenever
$\varepsilon_t\leq\tau_{\mathrm{ALS}}$, the refined linear recurrence
can be written as
\[
    \varepsilon_{t+1}
    \leq
    A_{\mathrm l}\varepsilon_t
    +
    b_{\mathrm l}.
\]
By condition \eqref{eq_assumption_general_R_4}, after taking the
absolute constant therein sufficiently small,
\[
    A_{\mathrm l}
    \leq
    \frac12.
\]

If
\[
    \tau_{\mathrm{ALS}}
    \leq
    4b_{\mathrm l},
\]
then
\[
    \varepsilon_{T_{\mathrm q}}
    \leq
    \tau_{\mathrm{ALS}}
    \lesssim
    \frac{\delta_2}{\lambda_{\min}},
\]
so again the desired accuracy has already been reached.

Otherwise,
\[
    b_{\mathrm l}
    <
    \frac{\tau_{\mathrm{ALS}}}{4}.
\]
In this case, whenever
$\varepsilon_t\leq\tau_{\mathrm{ALS}}$, we have
\[
    \varepsilon_{t+1}
    \leq
    \frac12\tau_{\mathrm{ALS}}
    +
    \frac14\tau_{\mathrm{ALS}}
    <
    \tau_{\mathrm{ALS}}.
\]
Thus, the iterates remain in the linear region. Iterating the linear
recurrence gives, for every $m\geq0$,
\[
    \varepsilon_{T_{\mathrm q}+m}
    \leq
    2^{-m}\tau_{\mathrm{ALS}}
    +
    2b_{\mathrm l}.
\]
Therefore,
\[
    \varepsilon_{T_{\mathrm q}+m}
    \lesssim
    \frac{\delta_2}{\lambda_{\min}}
\]
once
\[
    m
    \lesssim
    \log\left(
        \frac{
            \tau_{\mathrm{ALS}}\lambda_{\min}
        }{
            \delta_2
        }
    \right).
\]
Combining the quadratic and linear phases yields
\[
    T
    \lesssim
    \log\log\left(
        \frac{\lambda_{\min}}{\delta_2}
    \right)
    +
    \log\left(
        \frac{
            \tau_{\mathrm{ALS}}\lambda_{\min}
        }{
            \delta_2
        }
    \right).
\]
Since
\[
    \delta_2
    =
    \sigma\sqrt{p_{\max}},
\]
this is the second iteration bound in
Theorem~\ref{thm_local}.

It remains to verify that, once the statistical-accuracy region is
reached, all subsequent iterates remain in that region. Set
\[
    s
    :=
    \frac{\delta_2}{\lambda_{\min}}
    =
    \frac{\sigma\sqrt{p_{\max}}}{\lambda_{\min}},
\]
and let $K\geq1$ be a sufficiently large absolute constant. By
\eqref{eq_assumption_general_R_2} and
\eqref{eq_assumption_general_R_3}, after increasing the absolute
constants in these assumptions if necessary, we have
\[
    Ks
    <
    \frac{1}{
        (d-1)^{3/2}\sqrt{\log d}
    }
\]
and
\[
    \frac{\lambda_{\max}}{\lambda_{\min}}
    (d+R)Ks
    \leq
    c_1,
\]
where $c_1>0$ is a sufficiently small absolute constant. The same
choice ensures that $Ks+\xi$ lies in the local region used to derive
\eqref{eq_recurrence_order2} and \eqref{eq_recurrence_order1}.

Suppose that
\[
    \varepsilon_t
    \leq
    Ks.
\]
The preceding bound on $Ks$ implies that the refined noise estimate
\eqref{eq_noise_random_term_2} applies, and hence
\[
    \delta
    \lesssim
    \delta_2.
\]

If
\[
    \varepsilon_t
    >
    \tau_{\mathrm{ALS}},
\]
then \eqref{eq_recurrence_order2} gives
\[
\begin{split}
    \varepsilon_{t+1}
    &\lesssim
    \frac{\lambda_{\max}}{\lambda_{\min}}
    (d+R)K^2s^2
    +
    s                                                     \\
    &\leq
    Ks,
\end{split}
\]
where the last inequality follows by taking $c_1$ sufficiently small
and $K$ sufficiently large.

Suppose next that
\[
    \varepsilon_t
    \leq
    \tau_{\mathrm{ALS}}.
\]
If $s<\tau_{\mathrm{ALS}}$, then
\eqref{eq_recurrence_order1} and
\eqref{eq_assumption_general_R_4} yield
\[
\begin{split}
    \varepsilon_{t+1}
    &\lesssim
    \frac{\lambda_{\max}}{\lambda_{\min}}
    Rd\xi^{d-2}Ks
    +
    s                                                     \\
    &\leq
    Ks,
\end{split}
\]
after taking the absolute constant in
\eqref{eq_assumption_general_R_4} sufficiently small and $K$
sufficiently large.

It remains to consider the case
\[
    s
    \geq
    \tau_{\mathrm{ALS}}
    \qquad\text{and}\qquad
    \varepsilon_t
    \leq
    \tau_{\mathrm{ALS}}.
\]
When $d=3$, the bound derived above for $\mathcal{N}_3$ gives
\[
    \mathcal{N}_3(\varepsilon_t)
    \lesssim
    R\left(
        \varepsilon_t^2
        +
        \xi\varepsilon_t
    \right)
    \lesssim
    (d+R)s^2.
\]
When $d>3$, we similarly have
\[
\begin{split}
    \mathcal{N}_d(\varepsilon_t)
    &\lesssim
    Rd\xi^{d-2}\varepsilon_t                              \\
    &=
    \frac{R}{R-1}
    d\tau_{\mathrm{ALS}}\varepsilon_t                     \\
    &\lesssim
    (d+R)s^2.
\end{split}
\]
Substituting these bounds into
\eqref{eq_recurrence_formula_general_R} gives
\[
\begin{split}
    \varepsilon_{t+1}
    &\lesssim
    \frac{\lambda_{\max}}{\lambda_{\min}}
    (d+R)s^2
    +
    s                                                     \\
    &\leq
    Ks,
\end{split}
\]
again by the choices of $c_1$ and $K$.

Thus, in either regime,
\[
    \varepsilon_t
    \leq
    Ks
    \quad\Longrightarrow\quad
    \varepsilon_{t+1}
    \leq
    Ks.
\]
The preceding convergence arguments show that
$\varepsilon_T\leq Ks$ for some $T$ satisfying the corresponding
iteration bound in Theorem~\ref{thm_local}. Induction therefore gives
\[
    \varepsilon_t
    \leq
    K
    \frac{\sigma\sqrt{p_{\max}}}{\lambda_{\min}}
    \qquad
    \text{for every }t\geq T.
\]
Absorbing $K$ into the absolute constant $C_2$ proves the persistence
claim in Theorem~\ref{thm_local}.

We finally record the probability of the event on which the preceding
bounds hold. For any fixed mode, the uniform noise event
\eqref{eq_assumption_HP_1} has failure probability at most
\[
    C\exp(-cp_{\max}),
\]
whereas the fixed-component noise event
\eqref{eq_assumption_HP_2} has failure probability at most
\[
    CR\exp(-cp_{\max}).
\]
The event in \eqref{eq_assumption_HP_1} is uniform over all choices of
the unit loading vectors, and hence over all ALS iterates, so no union
bound over the iteration index is required. Taking a union bound over
all $d$ modes therefore shows that all the bounds used above hold
simultaneously with probability at least
\[
    1-CdR\exp(-cp_{\max}).
\]
After adjusting the absolute constants, this completes the proof of
Theorem~\ref{thm_local}.

\section{Proof of Theorem \ref{thm_lower_bound}}

By symmetry, it suffices to prove the result for $k=1$. For two
vectors $u,v\in\RR^{p_1}$, define the sign-invariant distance
\[
    d_{\pm}(u,v)
    :=
    \min_{s\in\{-1,1\}}
    \|u-sv\|.
\]

By a standard volumetric packing argument, there exist absolute
constants $c_0,c_1>0$ and a set
\[
    \mathcal{P}
    =
    \{v_1,\ldots,v_M\}
    \subset
    \mathbb{S}^{p_1-2}
    \subset
    \RR^{p_1-1}
\]
such that
\[
    \log M
    \geq
    c_0(p_1-1)
\]
and, for every $i\neq j$,
\[
    \|v_i-v_j\|
    \geq
    \frac12.
\]

Let $\delta\in(0,1/64]$, to be chosen later, and write
\[
    c_\delta
    :=
    \sqrt{1-\delta},
    \qquad
    s_\delta
    :=
    \sqrt{\delta}.
\]
For every $i\in[M]$, define
\[
    u_i
    :=
    \begin{pmatrix}
        c_\delta\\
        s_\delta v_i
    \end{pmatrix}
    \in
    \mathbb{S}^{p_1-1}.
\]
For $i\neq j$,
\[
    \|u_i-u_j\|
    =
    \sqrt{\delta}\|v_i-v_j\|
    \in
    \left[
        \frac{\sqrt{\delta}}{2},
        2\sqrt{\delta}
    \right].
\]
Moreover,
\[
    \|u_i+u_j\|^2
    =
    4(1-\delta)
    +
    \delta\|v_i+v_j\|^2
    \geq
    4(1-\delta)
    >
    3.
\]
Since $\|u_i-u_j\|\leq2\sqrt{\delta}<1$, it follows that
\begin{equation}\label{eq_lower_bound_packing_distance}
    d_{\pm}(u_i,u_j)
    =
    \|u_i-u_j\|
    \geq
    \frac{\sqrt{\delta}}{2}.
\end{equation}

For every $i\in[M]$, define the $p_1$-by-$p_1$ matrix
\[
    Q_i
    :=
    \begin{pmatrix}
        c_\delta
        &
        -s_\delta v_i^\top
        \\
        s_\delta v_i
        &
        I_{p_1-1}
        -
        (1-c_\delta)v_iv_i^\top
    \end{pmatrix}.
\]
The matrix $Q_i$ acts as a rotation on the subspace generated by
$e_{1,1}$ and $(0,v_i^\top)^\top$, and as the identity on its
orthogonal complement. In particular,
\begin{equation}\label{eq_lower_bound_Q_orthogonal}
    Q_i^\top Q_i
    =
    I_{p_1},
    \qquad
    Q_i e_{1,1}
    =
    u_i.
\end{equation}

For later use, observe that
\begin{align}
    \|Q_i-Q_j\|_F^2
    ={}&
    2\delta\|v_i-v_j\|^2
    \notag\\
    &+
    (1-c_\delta)^2
    \left\|
        v_iv_i^\top-v_jv_j^\top
    \right\|_F^2.
    \label{eq_lower_bound_Q_difference_exact}
\end{align}
For unit vectors $v_i$ and $v_j$,
\[
    \left\|
        v_iv_i^\top-v_jv_j^\top
    \right\|_F^2
    \leq
    2\|v_i-v_j\|^2.
\]
Since $1-c_\delta\leq\delta$, we obtain
\begin{equation}\label{eq_lower_bound_Q_difference}
    \|Q_i-Q_j\|_F^2
    \leq
    3\delta\|v_i-v_j\|^2.
\end{equation}

For $h\in[d]$, let $e_{h,r}$ denote the $r$th standard basis vector
in $\RR^{p_h}$, and let
\[
    E_h
    :=
    [e_{h,1},\ldots,e_{h,R}]
    \in
    \RR^{p_h\times R}.
\]
For every $i\in[M]$, set
\[
    A_1^{(i)}
    :=
    Q_iE_1,
    \qquad
    A_h^{(i)}
    :=
    E_h,
    \quad
    h=2,\ldots,d.
\]
By \eqref{eq_lower_bound_Q_orthogonal},
\[
    A_h^{(i)\top}A_h^{(i)}
    =
    I_R
\]
for every $h\in[d]$. We construct
\begin{equation}\label{eq_lower_bound_tensor_construction}
    \X_i
    :=
    \lambda
    \sum_{r=1}^R
    a_{1,r}^{(i)}
    \circ
    e_{2,r}
    \circ\cdots\circ
    e_{d,r}.
\end{equation}
Thus, every $\X_i$ belongs to
$\mathcal{F}_{R,\lambda}^{\mathrm{orth}}$.

For completeness, every $\X_i$ has CP rank exactly $R$. Indeed, its
mode-$2$ unfolding satisfies
\[
    \MM_2(\X_i)
    =
    \lambda
    E_2
    \left(
        A_1^{(i)}
        \odot
        E_3
        \odot\cdots\odot
        E_d
    \right)^\top.
\]
Both matrices on the right-hand side have rank $R$, and hence
\[
    \operatorname{rank}
    \left(
        \MM_2(\X_i)
    \right)
    =
    R.
\]
It follows that
\[
    \operatorname{rank}_{\mathrm{CP}}(\X_i)
    =
    R.
\]

Let $\PP_i$ denote the distribution of
\[
    \Y_i
    =
    \X_i+\Z,
\]
where the entries of $\Z$ are independently distributed as
$\mathcal{N}(0,\sigma^2)$. For $i\neq j$, orthogonality of the
standard-basis factors in modes $2,\ldots,d$ gives
\begin{align}
    \|\X_i-\X_j\|_F^2
    &=
    \lambda^2
    \sum_{r=1}^R
    \left\|
        (Q_i-Q_j)e_{1,r}
    \right\|^2
    \notag\\
    &=
    \lambda^2
    \left\|
        (Q_i-Q_j)E_1
    \right\|_F^2
    \notag\\
    &\leq
    \lambda^2
    \|Q_i-Q_j\|_F^2
    \notag\\
    &\leq
    3\lambda^2\delta
    \|v_i-v_j\|^2
    \notag\\
    &\leq
    12\lambda^2\delta.
    \label{eq_lower_bound_tensor_difference}
\end{align}
Therefore, the Gaussian Kullback--Leibler divergence satisfies
\begin{equation}\label{eq_lower_bound_KL}
    D_{\mathrm{KL}}(\PP_i\|\PP_j)
    =
    \frac{
        \|\X_i-\X_j\|_F^2
    }{
        2\sigma^2
    }
    \leq
    \frac{
        6\lambda^2\delta
    }{
        \sigma^2
    }.
\end{equation}

Choose
\begin{equation}\label{eq_lower_bound_delta}
    \delta
    =
    c_2
    \left(
        1
        \land
        \frac{
            \sigma^2(p_1-1)
        }{
            \lambda^2
        }
    \right),
\end{equation}
where $c_2>0$ is a sufficiently small absolute constant. We take
$c_2\leq1/64$, so that all preceding bounds apply. By
\eqref{eq_lower_bound_KL} and
$\log M\geq c_0(p_1-1)$, $c_2$ may be chosen so that
\[
    \max_{i\neq j}
    D_{\mathrm{KL}}(\PP_i\|\PP_j)
    \leq
    \frac18\log M.
\]
The generalized Fano inequality and
\eqref{eq_lower_bound_packing_distance} therefore imply
\begin{equation}\label{eq_lower_bound_vector}
    \inf_{\hat u}
    \max_{i\in[M]}
    \EE_i
    d_{\pm}(\hat u,u_i)
    \gtrsim
    \sqrt{\delta}
    \gtrsim
    1
    \land
    \frac{
        \sigma\sqrt{p_1}
    }{
        \lambda
    },
\end{equation}
where the infimum is over all measurable estimators of $u_i$ based
on $\Y_i$.

It remains to transfer \eqref{eq_lower_bound_vector} to estimation of
the unordered columns of $A_1^{(i)}$. Let
\[
    \hat A_1
    =
    [\hat a_{1,1},\ldots,\hat a_{1,R}]
\]
be an arbitrary estimator. We may project every estimated column onto
the closed Euclidean unit ball, since this cannot increase its
distance to any unit vector.

From the projected columns, construct $\hat u$ by selecting the
column having the largest absolute first coordinate and multiplying
it by a sign so that its first coordinate is nonnegative. Define
\[
    L_i
    :=
    \mathcal{L}_1
    \left(
        \hat A_1,A_1^{(i)}
    \right).
\]
The first column of $A_1^{(i)}$ is $u_i$ and has first coordinate
$c_\delta$. From the explicit form of $Q_i$, all its other columns
satisfy
\[
    \left|
        e_{1,1}^\top Q_i e_{1,r}
    \right|
    \leq
    s_\delta,
    \qquad
    r=2,\ldots,R.
\]
If $L_i<1/4$, the estimated column matched to $u_i$ has absolute first
coordinate at least
\[
    c_\delta-L_i
    \geq
    \sqrt{\frac{63}{64}}-\frac14,
\]
whereas every estimated column matched to another column of
$A_1^{(i)}$ has absolute first coordinate at most
\[
    s_\delta+L_i
    \leq
    \frac18+\frac14.
\]
Thus, the selected column is the one matched to $u_i$, and
\[
    d_{\pm}(\hat u,u_i)
    \leq
    L_i.
\]
If $L_i\geq1/4$, then both $\hat u$ and $u_i$ belong to the closed
unit ball, so
\[
    d_{\pm}(\hat u,u_i)
    \leq
    2
    \leq
    8L_i.
\]
Consequently,
\[
    d_{\pm}(\hat u,u_i)
    \leq
    8L_i
\]
in all cases. Combining this inequality with
\eqref{eq_lower_bound_vector} gives
\[
    \inf_{\hat A_1}
    \max_{i\in[M]}
    \EE_i
    \mathcal{L}_1
    \left(
        \hat A_1,A_1^{(i)}
    \right)
    \gtrsim
    1
    \land
    \frac{
        \sigma\sqrt{p_1}
    }{
        \lambda
    }.
\]
Since
\[
    \{\X_1,\ldots,\X_M\}
    \subset
    \mathcal{F}_{R,\lambda}^{\mathrm{orth}},
\]
we conclude that
\[
    \inf_{\hat A_1}
    \sup_{\X\in\mathcal{F}_{R,\lambda}^{\mathrm{orth}}}
    \EE_{\X}
    \mathcal{L}_1(\hat A_1,A_1)
    \gtrsim
    1
    \land
    \frac{
        \sigma\sqrt{p_1}
    }{
        \lambda
    }.
\]
The same construction applies to any mode $k\in[d]$, which completes
the proof.

\section{Proof of Theorem \ref{thm_general_tucker}}\label{sec_proof_thm_general_tucker}

By applying the following argument to $\Y/\sigma$, it suffices to prove
the result for $\sigma=1$. Under this normalization, the assumptions of
Theorem~\ref{thm_general_tucker} give
\bea
1.&&
\lambda_{\operatorname{Tucker}}
\geq
Cd p_{\max}^{d/4};
\label{eq_tucker_assumption_1}
\\
2.&&
\lambda_{\operatorname{Tucker}}
\geq
dC^d r_{\max}^{\frac{d-1}{2}}
\lor
Cd\sqrt{p_{\max}r_{\max}};
\label{eq_tucker_assumption_2}
\\
3.&&
p_{\min}>C\log d.
\notag
\eea

For brevity, define
\[
    q_0
    :=
    \sqrt{p_{\max}}
    +
    \sqrt{r_{\max}^{d-1}},
\]
and
\[
    q_1
    :=
    q_0
    +
    \sqrt{
        \log(d)
        \sum_{k=1}^d p_kr_k
    }.
\]
We next verify that the preceding assumptions control the uniform
projected-noise term appearing below. Since $r_k\leq p_k$ for every
$k\in[d]$, we have $r_{\max}\leq p_{\max}$ and
\[
    \sqrt{
        \log(d)
        \sum_{k=1}^d p_kr_k
    }
    \leq
    \sqrt{d\log d}\,
    \sqrt{p_{\max}r_{\max}}.
\]
When $d=3$, the factor $\sqrt{d\log d}$ is an absolute constant, so the
mixed term is controlled by
$C\sqrt{p_{\max}r_{\max}}$. When $d=4$, we have
\[
    \sqrt{p_{\max}r_{\max}}
    \leq
    p_{\max}
    =
    p_{\max}^{d/4}.
\]
When $d\geq5$, the conditions
$p_{\max}\geq p_{\min}>C\log d$ and
$r_{\max}\leq p_{\max}$ imply, after increasing the absolute constant
$C$ if necessary, that
\[
    d
    \sqrt{
        \log(d)
        \sum_{k=1}^d p_kr_k
    }
    \lesssim
    p_{\max}^{d/4}.
\]
Consequently, after increasing the absolute constants in
\eqref{eq_tucker_assumption_1}--\eqref{eq_tucker_assumption_2} if
necessary,
\begin{equation}
\label{eq_tucker_uniform_absorption}
    dq_1
    \lesssim
    \lambda_{\operatorname{Tucker}}.
\end{equation}
Moreover, the implied constant in
\eqref{eq_tucker_uniform_absorption} can be made arbitrarily small by
increasing the constants in the signal-strength assumptions.

Let $Y_1 = \MM_1(\Y)$, $p_{-k} = \prod_{i \neq k}p_i$, and $L_t = \max_k \ln\sin\Theta\(\hat U_k^{(t)}, U_k\)\rn$.
The appendix equation (1.16) in \cite{cai2018rate} yield
\beas
    \PP\(
        \|U_{1,\perp}\T Y_1 P_{Y_1\T U_1}\| \geq x
    \)
    &\leq&
    C \exp\left(
        Cp_1
        -c\min\left\{
            x^2,
            x\sqrt{\lambda_{\operatorname{Tucker}}^2+p_{-1}}
        \right\}
    \right)
    \\
    &&\quad
    + C \exp\left(
        -c(\lambda_{\operatorname{Tucker}}^2+p_{-1})
    \right),
\eeas
where $C$ and $c$ are absolute constants.
Let $x = C\sqrt{p_1}$, it yields
\bea
&&\PP\( \|U_{1,\perp}\T Y_1 P_{Y_1\T U_1}\| \geq C\sqrt{p_1} \)\notag \\
&\leq& C \exp \(Cp_1 -c\min\{Cp_1, Cp_1^{1/2}\sqrt{\lambda_{\operatorname{Tucker}}^2 +p_{-1}}\}\) + C
\exp\(-c(\lambda_{\operatorname{Tucker}}^2 +p_{-1})\) \notag \\
&=& C \exp \(-Cp_1\) + C \exp\(-c(\lambda_{\operatorname{Tucker}}^2 +p_{-1})\) \notag \\
&\leq& C \exp \(-Cp_1\), \label{eq_tucker1}
\eea
where the third line holds by \eqref{eq_tucker_assumption_1}, $\lambda_{\operatorname{Tucker}} \gtrsim  p_{\max}^{d/4} \geq  p_{1}^{1/2}$, and $\lambda_{\operatorname{Tucker}}^2 + p_{-1} \geq C p_1^{1.5} \geq p_1$.

The appendix equation (1.17) in \cite{cai2018rate} yields
\beas
\PP\( \ln\sin\Theta\(\hat U_1^{(0)}, U_1\)\rn^2 \leq \frac{C(\lambda_{\operatorname{Tucker}}^2 + p_{-1})\|U_{1,\perp}\T Y_1
P_{Y_1\T U_1}\|^2}{\lambda_{\operatorname{Tucker}}^4} \) \geq 1- C \exp
\(-c\frac{\lambda_{\operatorname{Tucker}}^4}{\lambda_{\operatorname{Tucker}}^2 + p_{-1}}\),
\eeas
which with \eqref{eq_tucker1} further yields
\beas
\PP\( \ln\sin\Theta\(\hat U_1^{(0)}, U_1\)\rn^2 \leq \frac{C(\lambda_{\operatorname{Tucker}}^2 +
    p_{-1})p_1}{\lambda_{\operatorname{Tucker}}^4} \) &\geq& 1- C \exp
\(-c\frac{\lambda_{\operatorname{Tucker}}^4}{\lambda_{\operatorname{Tucker}}^2 + p_{-1}}\)-C \exp \(-Cp_1\).
\eeas

Note that
\[
    \frac{ (\lambda_{\operatorname{Tucker}}^2+p_{-1})p_1 }{ \lambda_{\operatorname{Tucker}}^4 } \leq
    \frac{p_{\max}}{\lambda_{\operatorname{Tucker}}^2} + \frac{p_{\max}^d}{\lambda_{\operatorname{Tucker}}^4}.
\]
Taking square roots in the preceding sine-angle bound and applying the
same argument to every mode yield
\begin{equation}
\label{eq_hosvd_bound}
    \PP\left(
        L_0
        \leq
        \widetilde\varepsilon
    \right)
    \geq
    1-Cd\exp(-cp_{\min}),
\end{equation}
where
\[
    \widetilde\varepsilon
    :=
    C
    \left(
        \frac{
            \sqrt{p_{\max}}
        }{
            \lambda_{\operatorname{Tucker}}
        }
        +
        \frac{
            p_{\max}^{d/2}
        }{
            \lambda_{\operatorname{Tucker}}^2
        }
    \right).
\]
By \eqref{eq_tucker_assumption_1},
\begin{align*}
    \widetilde\varepsilon
    &\leq
    C
    \left(
        \frac{
            \sqrt{p_{\max}}
        }{
            Cd p_{\max}^{d/4}
        }
        +
        \frac{
            p_{\max}^{d/2}
        }{
            C^2d^2p_{\max}^{d/2}
        }
    \right)                                                     \\
    &\leq
    C
    \left(
        \frac{1}{Cd}
        +
        \frac{1}{C^2d^2}
    \right).
\end{align*}
Therefore, for any sufficiently small absolute constant $c_0>0$, the
absolute constant in the signal-strength assumption can be chosen so
that
\begin{equation}
\label{eq_hooi_initial_small_region}
    L_0
    \leq
    \frac{c_0}{d}
\end{equation}
on the event in \eqref{eq_hosvd_bound}.

Let
\[
    P_h=U_hU_h^\top, \qquad P_{h\perp}=U_{h\perp}U_{h\perp}^\top,
\]
and define
\[
    X_1^{(t)} = \MM_1\left( \X\times_{k=2}^d(\hat U_k^{(t)})^\top \right), \qquad Z_1^{(t)} = \MM_1\left( \Z\times_{k=2}^d(\hat
    U_k^{(t)})^\top \right).
\]
Expanding
\[
    \bigotimes_{h=2}^d \left(P_h+P_{h\perp}\right)\hat U_h^{(t)}
\]
and grouping the terms according to the number of factors of the form $P_{h\perp}\hat U_h^{(t)}$, Supplement Lemma~\ref{lemma_maxZ_bound} and its proof give
\begin{align*}
    \|Z_1^{(t)}\| \lesssim{} & \sqrt{p_1} + \sqrt{\prod_{k=2}^d r_k}                                                                                 \\
                             & + \left( \sqrt{p_1} + \sqrt{\prod_{k=2}^d r_k} + \sqrt{ \log(d) \sum_{k=2}^d p_kr_k } \right) \sum_{j=1}^{d-1}
    \binom{d-1}{j}L_t^j                                                                                                                              \\
    ={}                      & \sqrt{p_1} + \sqrt{\prod_{k=2}^d r_k}                                                                                 \\
                             & + \left( \sqrt{p_1} + \sqrt{\prod_{k=2}^d r_k} + \sqrt{ \log(d) \sum_{k=2}^d p_kr_k } \right) \left\{ (1+L_t)^{d-1}-1
    \right\}.
\end{align*}
Meanwhile,
\beas
\sigma_{r_1}\(X_1^{(t)}\) &=& \sigma_{r_1}\( \MM_1(\X) \( \bigotimes_{k = 2}^d\(U_kU_k\T\) \) \( \bigotimes_{k = 2}^d \hat
U_k^{(t)} \) \) \\
&=& \sigma_{r_1}\( \MM_1(\X) \( \bigotimes_{k = 2}^d U_k \) \( \bigotimes_{k = 2}^d \( U_k\T \hat U_k^{(t)}\) \) \) \\
&\geq & \sigma_{r_1}\( \MM_1(\X) \( \bigotimes_{k = 2}^d U_k \)\) \prod_{k = 2}^d\sigma_{\min}\( U_k\T \hat U_k^{(t)} \) \\
&\geq & \lambda_{\operatorname{Tucker}} \(1 - L_t\)^{(d-1)/2},
\eeas
where the last line is by Lemma 1 in \cite{cai2018rate}.
Note that $\operatorname{span}(U_1)$ is the rank-$r_1$ left singular subspace of $X_1^{(t)}$, while $\hat U_1^{(t+1)}$ is the leading rank-$r_1$ left singular subspace of
\[
    X_1^{(t)}+Z_1^{(t)}
    =
    \MM_1\left(
        \Y\times_{k=2}^d
        (\hat U_k^{(t)})^\top
    \right).
\]
Suppose that $L_t\leq c_0/d$, as will be verified inductively below. Then
\[
    (1+L_t)^{d-1}-1
    \lesssim
    dL_t,
\]
and hence the preceding bound on $Z_1^{(t)}$ gives
\[
    \|Z_1^{(t)}\|
    \lesssim
    q_0+q_1dL_t
    \lesssim
    q_1.
\]
On the other hand,
\[
    \sigma_{r_1}\left(X_1^{(t)}\right)
    \geq
    \lambda_{\operatorname{Tucker}}
    (1-L_t)^{(d-1)/2}
    \gtrsim
    \lambda_{\operatorname{Tucker}}.
\]
By \eqref{eq_tucker_uniform_absorption}, after increasing the absolute constants in the signal-strength assumptions if necessary, we therefore have
\[
    \|Z_1^{(t)}\|
    \leq
    \frac{1}{2}
    \sigma_{r_1}\left(X_1^{(t)}\right).
\]
Consequently, Wedin's theorem together with Weyl's inequality gives
\beas
&&
\ln
\sin\Theta\left(
    \hat U_1^{(t+1)},U_1
\right)
\rn
\leq
\frac{
    \|Z_1^{(t)}\|
}{
    \sigma_{r_1}\left(X_1^{(t)}\right)
    -
    \|Z_1^{(t)}\|
}
\lesssim
\frac{
    \|Z_1^{(t)}\|
}{
    \sigma_{r_1}\left(X_1^{(t)}\right)
}
\\
&\lesssim&
\frac{
    \sqrt{p_1}
    +
    \sqrt{\prod_{k=2}^d r_k}
}{
    \lambda_{\operatorname{Tucker}}
    (1-L_t)^{(d-1)/2}
}
+
\frac{
    \sqrt{p_1}
    +
    \sqrt{\prod_{k=2}^d r_k}
    +
    \sqrt{
        \log(d)
        \sum_{k=2}^d p_kr_k
    }
}{
    \lambda_{\operatorname{Tucker}}
}
\frac{
    (1+L_t)^{d-1}-1
}{
    (1-L_t)^{(d-1)/2}
}.
\eeas

We now work on the event in \eqref{eq_hosvd_bound}. Suppose that
\[
    L_t
    \leq
    \frac{c_0}{d},
\]
where $c_0>0$ is a sufficiently small absolute constant. Since
$L_t\leq1/2$, we have
\[
    (1-L_t)^{-\frac{d-1}{2}}
    \leq
    \exp\left(
        C(d-1)L_t
    \right)
    \leq
    C,
\]
and
\[
    (1+L_t)^{d-1}-1
    \leq
    \exp\left(
        (d-1)L_t
    \right)-1
    \lesssim
    dL_t.
\]
Consequently, the preceding recurrence can be written as
\begin{equation}
\label{eq_hooi_affine_recurrence}
    L_{t+1}
    \leq
    a+bL_t,
\end{equation}
where
\[
    a
    :=
    C
    \frac{
        q_0
    }{
        \lambda_{\operatorname{Tucker}}
    }
\]
and
\[
    b
    :=
    C d
    \frac{
        q_1
    }{
        \lambda_{\operatorname{Tucker}}
    }.
\]
By \eqref{eq_tucker_assumption_1},
\eqref{eq_tucker_assumption_2}, and
\eqref{eq_tucker_uniform_absorption}, the absolute constants can be
chosen so that
\[
    a
    \leq
    \frac{c_0}{2d}
    \qquad\text{and}\qquad
    b
    \leq
    \frac12.
\]
Therefore, if $L_t\leq c_0/d$, then
\[
    L_{t+1}
    \leq
    a+bL_t
    \leq
    \frac{c_0}{2d}
    +
    \frac12\frac{c_0}{d}
    =
    \frac{c_0}{d}.
\]
Since \eqref{eq_hooi_initial_small_region} gives
$L_0\leq c_0/d$, induction shows that
\begin{equation}
\label{eq_hooi_small_region_all_t}
    L_t
    \leq
    \frac{c_0}{d}
    \qquad
    \text{for every }t\geq0.
\end{equation}

Iterating \eqref{eq_hooi_affine_recurrence} gives
\[
    L_t
    \leq
    \frac{a}{1-b}
    +
    b^tL_0
    \leq
    2a
    +
    2^{-t}L_0.
\]
Furthermore,
\begin{align*}
    \frac{
        L_0
    }{
        q_0/\lambda_{\operatorname{Tucker}}
    }
    &\lesssim
    \frac{
        \sqrt{p_{\max}}/
        \lambda_{\operatorname{Tucker}}
        +
        p_{\max}^{d/2}/
        \lambda_{\operatorname{Tucker}}^2
    }{
        q_0/
        \lambda_{\operatorname{Tucker}}
    }                                                        \\
    &\lesssim
    1+
    \frac{
        p_{\max}^{d/2}
    }{
        \lambda_{\operatorname{Tucker}}q_0
    }.
\end{align*}
It follows that
\[
    L_T
    \lesssim
    \frac{
        q_0
    }{
        \lambda_{\operatorname{Tucker}}
    }
\]
after
\[
    T
    \lesssim
    1+
    \log\left(
        1+
        \frac{
            p_{\max}^{d/2}
        }{
            \lambda_{\operatorname{Tucker}}q_0
        }
    \right)
\]
iterations.

Finally, restoring the general noise scale $\sigma$ by applying the
preceding argument to $\Y/\sigma$ gives
\[
    T
    \lesssim
    1+
    \log\left(
        1+
        \frac{
            \sigma p_{\max}^{d/2}
        }{
            \lambda_{\operatorname{Tucker}}
            \left(
                \sqrt{p_{\max}}
                +
                \sqrt{r_{\max}^{d-1}}
            \right)
        }
    \right).
\]

Once we have $L_t \lesssim \frac{\sqrt{p_{\max}} + \sqrt{r_{\max}^{d-1}}}{\lambda_{\operatorname{Tucker}}}$, it yields
\beas
\|Z_1^{(t)}\| &\lesssim& \sqrt{p_1} + \sqrt{\prod r_k} + \( \sqrt{p_1} + \sqrt{\prod r_k} + \sqrt{\log(d)\sum p_k r_k } \)\( (1+
L_t)^{d-1} - 1\) \\
&\lesssim&  \sqrt{p_1} + \sqrt{\prod r_k} +  d L_t \( \sqrt{p_1} + \sqrt{\prod r_k} + \sqrt{\log(d)\sum p_k r_k } \) \\
&\lesssim&  \sqrt{p_{\max}} + \sqrt{r_{\max}^{d-1}}.
\eeas

The signal-strength assumptions used above are exactly \eqref{eq_tucker_assumption_1} and \eqref{eq_tucker_assumption_2}.
As shown at the beginning of the proof, these assumptions absorb the uniform projected-noise term without requiring an additional logarithmic signal-strength condition.
Finally, restoring the general noise scale $\sigma$ by applying the preceding argument to $\Y/\sigma$ proves
\[
    \max_{k\in[d]} \left\| U_kU_k^\top-\hat U_k\hat U_k^\top \right\| \lesssim \sigma \frac{ \sqrt{p_{\max}} + \sqrt{r_{\max}^{d-1}}
    }{ \lambda_{\operatorname{Tucker}} }.
\]
The corresponding projected-noise bound follows from the preceding display in the same way.
This completes the proof of Theorem~\ref{thm_general_tucker}.

\section{Proof of Proposition \ref{prop_tasd_noiseless}}\label{sec_proof_prop_tasd_noiseless}

\begin{proof}
Fix a target mode $h\in[d]$, and let $\ell\neq h$ denote the second mode left uncontracted in Algorithm~\ref{algorithm_tasd}.
For each $k\in[d]$, the mode-$k$ unfolding of the noiseless tensor can be written as
\[
    \MM_k(\X)
    =
    A_k
    \diag(\lambda_1,\ldots,\lambda_R)
    \left(
        \odot_{j\neq k} A_j
    \right)^\top.
\]
Since $\X$ has CP rank $R$, every $\lambda_r$ is nonzero.
Moreover, because every $A_j$ has full column rank, the Khatri--Rao product $\odot_{j\neq k}A_j$ also has full column rank.
Indeed, its Gram matrix is the Hadamard product of the positive-definite matrices $A_j^\top A_j$, $j\neq k$, and is therefore positive definite by the Schur product theorem.
Hence
\[
    \operatorname{rank}\left(\MM_k(\X)\right)=R,
    \qquad
    \operatorname{col}\left(\MM_k(\X)\right)
    =
    \operatorname{col}(A_k).
\]
Therefore, an exact rank-$(R,\ldots,R)$ Tucker decomposition of $\X$ has factor matrices $U_k$ satisfying $\operatorname{col}(U_k)=\operatorname{col}(A_k)$.
In particular, the HOSVD initialization has these exact mode subspaces, and HOOI preserves them in the noiseless setting.

Since $U_k$ has orthonormal columns and spans the column space of $A_k$, there exists an invertible matrix $V_k\in\RR^{R\times R}$ such that
\[
    A_k=U_kV_k.
\]
Writing $V_k=[v_{k,1},\ldots,v_{k,R}]$, the Tucker core therefore satisfies
\[
    \S
    =
    \sum_{r=1}^R
    \lambda_r
    v_{1,r}\circ\cdots\circ v_{d,r}.
\]

Let $J=[d]\setminus\{h,\ell\}$.
For $i\in\{1,2\}$, define
\[
    M_i
    :=
    \MM_h\left(
        \S\times_{k\in J}w_{ik}\T
    \right).
\]
Then
\[
    M_i
    =
    V_hD_iV_\ell^\top,
\]
where $D_i$ is diagonal with
\[
    (D_i)_{rr}
    =
    \lambda_r
    \prod_{k\in J}
    \langle w_{ik},v_{k,r}\rangle.
\]
Because $V_k$ is invertible, every $v_{k,r}$ is nonzero.
Thus, for Gaussian $w_{ik}$, each $\langle w_{ik},v_{k,r}\rangle$ is a nondegenerate Gaussian random variable and is nonzero with probability one.
Since there are only finitely many such inner products, $D_1$ and $D_2$ are invertible simultaneously with probability one.
On this event,
\[
    M_1M_2^\dagger
    =
    V_hD_1D_2^{-1}V_h^{-1}.
\]
Its eigenvalues are the diagonal entries
\[
    \mu_r
    =
    \prod_{k\in J}
    \frac{
        \langle w_{1k},v_{k,r}\rangle
    }{
        \langle w_{2k},v_{k,r}\rangle
    },
    \qquad
    r\in[R].
\]

We next show that these eigenvalues are pairwise distinct almost surely.
The case $R=1$ is immediate.
For $R\geq2$, fix $r\neq s$ and choose any $k_0\in J$.
Conditional on all Gaussian contraction vectors except $w_{1k_0}$, write
\[
    c_q
    :=
    \frac{
        \displaystyle
        \prod_{k\in J\setminus\{k_0\}}
        \langle w_{1k},v_{k,q}\rangle
    }{
        \displaystyle
        \prod_{k\in J}
        \langle w_{2k},v_{k,q}\rangle
    },
    \qquad
    q\in\{r,s\}.
\]
On the preceding probability-one nonvanishing event, $c_r$ and $c_s$ are nonzero, and
\[
    \mu_r-\mu_s
    =
    \left\langle
        w_{1k_0},
        c_rv_{k_0,r}-c_sv_{k_0,s}
    \right\rangle.
\]
Since $V_{k_0}$ is invertible, $v_{k_0,r}$ and $v_{k_0,s}$ are linearly independent, so
\[
    c_rv_{k_0,r}-c_sv_{k_0,s}\neq0.
\]
Conditional on the remaining contraction vectors, $\mu_r-\mu_s$ is therefore a nondegenerate Gaussian random variable.
Hence $\PP(\mu_r=\mu_s)=0$.
A finite union over all $r\neq s$ shows that $\mu_1,\ldots,\mu_R$ are pairwise distinct with probability one.

Consequently, the columns of $V_h$ are uniquely determined as the eigenvectors of $M_1M_2^\dagger$, up to column scaling and permutation.
Thus, the eigenvector matrix returned by SimDiag can be written as
\[
    \hat V_h
    =
    V_h\Pi C,
\]
where $\Pi$ is a permutation matrix and $C$ is an invertible diagonal matrix.
Since $A_h=U_hV_h$,
\[
    U_h\hat V_h
    =
    A_h\Pi C.
\]
After normalizing every column and recalling that the columns of $A_h$ have unit norm, the diagonal scaling is reduced to a sign ambiguity.
Therefore, for some diagonal sign matrix $\mathcal S$,
\[
    \hat A_h
    =
    A_h\Pi\mathcal S.
\]
This proves exact recovery up to permutation and columnwise sign changes, with probability one over the Gaussian contraction vectors generated by TASD.
\end{proof}

\section{Proof of Theorem \ref{thm_tasd}}\label{sec_proof_thm_tasd}

We consider the estimation for the first mode.
Other modes can be proved similarly.
Since the rank-one case has been studied separately in Theorem \ref{thm_overall_R1}, we assume $R\geq 2$ throughout this proof.

We restate the TASD procedure as follows.
First, perform $\operatorname{TuckerDecomp}(\Y)$ with Tucker rank $(R,\ldots,R)$.
Denote the output of HOSVD/HOOI as $\hat U_k$ and $\hat \S$.
Second, do $\operatorname{SimDiag}(\hat \S)$.
Specifically, let $w_{1k},w_{2k}\in\RR^R$ be independent standard Gaussian vectors for $k=3,\ldots,d$, independently of $\Y$.
Let $\hat V_1$ be obtained from the eigen-decomposition of
\[
    \MM_1(\hat\S \times_{k=3}^d w_{1k}\T)
    \left\{
        \MM_1(\hat\S \times_{k=3}^d w_{2k}\T)
    \right\}^{\dagger}.
\]
Finally, we let $\hat A_1=\hat U_1\operatorname{Re}(\hat V_1)$ and normalize each column to have unit Euclidean norm.

\subsection{HOOI in this context and possible orthogonal transformation}

Denote the output of HOSVD/HOOI as $\hat U_k$ and $\hat \S$.
Note that the Tucker decomposition can only estimate the span of $U_k$ instead of $U_k$ itself.
However, this will not be an issue by the following arguments.

First, by Lemma 2.6 in \cite{chen2021spectral}, there exist orthogonal
matrices $O_k\in\mathbb{O}_{R,R}$ such that
\[
    \|\hat U_k-U_kO_k\|\leq \sqrt{2}\|\hat U_k\hat U_k\T-U_kU_k\T\|.
\]
By replacing $U_k$ with $U_kO_k$ and rotating the Tucker core accordingly, we may assume without loss of generality that
\be
\label{eq_tucker_specialU} \|\hat U_k-U_k\|\leq \sqrt{2}\|\hat U_k\hat U_k\T-U_kU_k\T\|.
\ee

Recall that
\[
    \X
    =
    \Lambda
    \times_{k=1}^d A_k.
\]
Since $A_k$ and $U_k$ have the same column space and $A_k$ has full column rank, there exists an invertible matrix $V_k\in\RR^{R\times R}$ such that
\[
    A_k=U_kV_k.
\]
Let $v_{r,k}$ denote the $r$th column of $V_k$.
Then
\[
    \S=\Lambda\times_{k=1}^d V_k, \qquad \X=\S\times_{k=1}^d U_k.
\]
Let
\[
    \hat\S=\S+\E.
\]
For $i=1,2$, define
\begin{align*}
    S_i      & =\MM_1(\S\times_{k=3}^d w_{ik}\T) =\sum_{r=1}^R\lambda_r \left(\prod_{k=3}^d\langle w_{ik},v_{r,k}\rangle\right)
    v_{r,1}v_{r,2}\T,                                                                                                         \\
    \hat S_i & =\MM_1(\hat\S\times_{k=3}^d w_{ik}\T) =S_i+E_i,
\end{align*}
where
\[
    E_i=\MM_1(\E\times_{k=3}^d w_{ik}\T).
\]
Let $\hat V_1$ be obtained from the eigen-decomposition of $\hat S_1\hat S_2^\dagger$.
Finally, we have $\hat A_1=\hat U_1\operatorname{Re}(\hat V_1)$ after column normalization.

\subsection{Bound on the estimated Tucker core}

We first bound the perturbation of the estimated core.
For $i=1,2$, we have
\begin{align*}
    \|E_i\| & =\|\MM_1((\hat\S-\S)\times_{k=3}^d w_{ik}\T)\|                      \\
            & =\|\MM_1(\hat\S-\S) [w_{id}\otimes\cdots\otimes w_{i3}\otimes I_R]\| \\
            & \leq \|\MM_1(\hat\S-\S)\|\prod_{k=3}^d\|w_{ik}\|.
\end{align*}

Note that
\begin{align*}
    \|\MM_1(\hat\S-\S)\| & =\left\|\MM_1\left((\X+\Z)\times_{k=1}^d\hat U_k\T -\X\times_{k=1}^dU_k\T\right)\right\|                              \\
                         & \leq \left\|\MM_1\left(\X\times_{k=1}^d\hat U_k\T -\X\times_{k=1}^dU_k\T\right)\right\| +\|\MM_1(\Z\times_{k=1}^d\hat
    U_k\T)\|.
\end{align*}
By a telescoping argument,
\begin{align*}
     & \left\|\MM_1\left(\X\times_{k=1}^d\hat U_k\T -\X\times_{k=1}^dU_k\T\right)\right\| \\
     & \qquad\leq \bar\lambda_{\operatorname{Tucker}} \sum_{k=1}^d\|\hat U_k-U_k\|.
\end{align*}
Therefore, by Theorem \ref{thm_general_tucker}, \eqref{eq_tucker_specialU}, and the definition $\kappa_T=\bar\lambda_{\operatorname{Tucker}}/\lambda_{\operatorname{Tucker}}$, we have, on the high-probability event in Theorem \ref{thm_general_tucker},
\be
\label{eq_bound_Shat_general_d} \max_{i\in\{1,2\}}\|E_i\| \lesssim d\kappa_T\sigma \left(\sqrt{p_{\max}}+\sqrt{R^{d-1}}\right)
\max_{i\in\{1,2\}}\prod_{k=3}^d\|w_{ik}\|.
\ee

\subsection{SimDiag}

We next analyze simultaneous diagonalization.
For $i=1,2$, let
\[
    D_i =\diag\left( \lambda_r\prod_{k=3}^d\langle w_{ik},v_{r,k}\rangle: r\in[R] \right).
\]
Then
\[
    S_i=V_1D_iV_2\T.
\]
Since $V_1$ and $V_2$ are invertible and all diagonal entries of $D_2$ are nonzero with probability one,
\[
    S_1S_2^\dagger =V_1D_1D_2^\dagger V_1^\dagger.
\]
The diagonal entries of $D_1D_2^\dagger$ are
\[
    \mu_r =\prod_{k=3}^d \frac{\langle w_{1k},v_{r,k}\rangle} {\langle w_{2k},v_{r,k}\rangle}, \qquad r\in[R].
\]
Thus, the columns of $V_1$ are the right eigenvectors of $S_1S_2^\dagger$.

\paragraph*{Generalized inverse perturbation. \\}

Let
\[
    \Delta_S =\|\hat S_1\hat S_2^\dagger-S_1S_2^\dagger\|.
\]
Note that
\[
    \hat S_1\hat S_2^\dagger-S_1S_2^\dagger =S_1(\hat S_2^\dagger-S_2^\dagger) +(\hat S_1-S_1)\hat S_2^\dagger.
\]
By \cite{xu2020perturbation},
\[
    \|\hat S_2^\dagger-S_2^\dagger\| \leq 3\|S_2^\dagger\|\|\hat S_2^\dagger\|\|\hat S_2-S_2\|,
\]
which implies
\be
\label{eq_bound_SShat_general_d} \Delta_S \leq 3\|S_2^\dagger\|\|\hat S_2^\dagger\| \|\hat S_2-S_2\|\|S_1\| +\|\hat
S_2^\dagger\|\|\hat S_1-S_1\|.
\ee

Define
\begin{align*}
    m_2 & =\min_{r\in[R]} \left|\prod_{k=3}^d\langle w_{2k},v_{r,k}\rangle\right|, \\
    M_1 & =\max_{r\in[R]} \left|\prod_{k=3}^d\langle w_{1k},v_{r,k}\rangle\right|, \\
    W   & =\max_{i\in\{1,2\}}\prod_{k=3}^d\|w_{ik}\|.
\end{align*}
Then
\[
    \|S_1\| \leq \lambda_{\max}\|V_1\|\|V_2\|M_1,
\]
and
\[
    \|S_2^\dagger\| \leq \frac{1}{\sigma_R(V_1)\sigma_R(V_2)\lambda_{\min}m_2}.
\]
Moreover, by Weyl's inequality,
\[
    \|\hat S_2^\dagger\| \leq \left\{ \lambda_{\min}\sigma_R(V_1)\sigma_R(V_2)m_2 -\|\hat S_2-S_2\| \right\}^{-1},
\]
provided that the quantity in braces is positive.
Combining these bounds with \eqref{eq_bound_SShat_general_d} and \eqref{eq_bound_Shat_general_d} yields
\bea
\Delta_S &\lesssim& \frac{WM_1}{m_2} \frac{\kappa_T\kappa_{\max}^2\lambda_{\max}d\sigma
    \left(\sqrt{p_{\max}}+\sqrt{R^{d-1}}\right)} {\lambda_{\min} \left\{ \lambda_{\min}m_2\sigma_{\min}^2 -Cd\kappa_T\sigma
    \left(\sqrt{p_{\max}}+\sqrt{R^{d-1}}\right)W \right\}} \notag \\
&&+ \frac{d\kappa_T\sigma \left(\sqrt{p_{\max}}+\sqrt{R^{d-1}}\right)W} {\lambda_{\min}m_2\sigma_{\min}^2 -Cd\kappa_T\sigma
    \left(\sqrt{p_{\max}}+\sqrt{R^{d-1}}\right)W}. \label{eq_bound_SS_general_d}
\eea
where
\[
    \kappa_{\max}=\max_{k\in[d]}\kappa(A_k), \qquad \sigma_{\min}=\min_{k\in[d]}\sigma_R(A_k).
\]

\paragraph*{Eigen-decomposition perturbation. \\}

Let $C_{\operatorname{gap}}>0$ be the deterministic eigengap lower bound appearing in Supplement Assumption~\ref{assumption_eigengap_general_d}.
By that assumption,
\[
    \min_{\substack{r_1,r_2\in[R] \\
            r_1\neq r_2}}
    |\mu_{r_1}-\mu_{r_2}|
    \geq
    C_{\operatorname{gap}},
\]
so $S_1S_2^\dagger$ has distinct eigenvalues.

In the context of Theorem 3.11 in \cite{stewart_matrix_2001}, if
\be
\label{eq_eigenbound_condition_general_d} (\operatorname{Sep}(A)-2\kappa\|A-\tilde A\|)^2 >4\kappa^2\|A-\tilde A\|^2,
\ee
then the corresponding eigenvector perturbation satisfies
\be
\label{eq_bound_p_general_d} \|p\| \leq \frac{2\kappa\|A-\tilde A\|} {\operatorname{Sep}(A)-2\kappa\|A-\tilde A\|}.
\ee
We let $A=S_1S_2^\dagger$ and $\tilde A=\hat S_1\hat S_2^\dagger$.
Since
\[
    A=V_1\diag(\mu_1,\ldots,\mu_R)V_1^\dagger,
\]
we can take $\kappa=\kappa(V_1)=\kappa(A_1)$, while
\[
    \operatorname{Sep}(A)
    =
    \min_{\substack{r_1,r_2\in[R] \\
            r_1\neq r_2}}
    |\mu_{r_1}-\mu_{r_2}|
    \geq
    C_{\operatorname{gap}}.
\]
Hence,
\be
\label{eq_bound_p_1_general_d}
\|p\|
\leq
\frac{2\kappa(A_1)\Delta_S}
{C_{\operatorname{gap}}-2\kappa(A_1)\Delta_S}.
\ee
It is therefore sufficient to verify
\begin{equation}
\label{eq_tasd_eigen_side_conditions}
    C_{\operatorname{gap}}
    >
    4\kappa(A_1)\Delta_S,
    \qquad
    \|p\|\|A_1\|
    <
    1.
\end{equation}
Indeed, the first inequality implies
\[
    \operatorname{Sep}(A)
    -
    2\kappa(A_1)\Delta_S
    \geq
    C_{\operatorname{gap}}
    -
    2\kappa(A_1)\Delta_S
    >
    2\kappa(A_1)\Delta_S,
\]
and hence \eqref{eq_eigenbound_condition_general_d} holds.
We verify both inequalities in
\eqref{eq_tasd_eigen_side_conditions} below, after deriving the
simplified bound on $\Delta_S$.

After a suitable permutation of the eigenvectors, let
$\bar v_{1,r}$ denote the eigenvector of
$\hat S_1\hat S_2^\dagger$ matched with $v_{r,1}$ and scaled as in
Theorem~3.11 of \cite{stewart_matrix_2001}. Define
\[
    \bar a_{1,r}
    :=
    \hat U_1\bar v_{1,r}.
\]
Provided that
\eqref{eq_tasd_eigen_side_conditions} holds, Theorem~3.11 of
\cite{stewart_matrix_2001} gives
\[
    \min_{s\in\{-1,1\}}
    \left\|
        \bar a_{1,r}
        -
        s a_{1,r}
    \right\|
    \leq
    \|\hat U_1-U_1\|
    +
    C\|A_1\|\|p\|.
\]

Moreover, $\hat S_1\hat S_2^\dagger$ is a real matrix. Under the
separation condition in
\eqref{eq_tasd_eigen_side_conditions}, each perturbed eigenvalue is
uniquely matched with one of the distinct real eigenvalues
$\mu_1,\ldots,\mu_R$. Since non-real eigenvalues of a real matrix occur
in conjugate pairs, each matched perturbed eigenvalue is real.
Consequently, $\bar v_{1,r}$ can be chosen real, and the real-part
operation in Algorithm~\ref{algorithm_tasd} does not change this
choice.

The normalized TASD output is therefore
\[
    \hat a_{1,r}
    =
    \frac{
        \bar a_{1,r}
    }{
        \|\bar a_{1,r}\|
    }.
\]
For every nonzero vector $x$ and every unit vector $u$,
\[
\begin{split}
    \left\|
        \frac{x}{\|x\|}
        -
        u
    \right\|
    &\leq
    \left\|
        \frac{x}{\|x\|}
        -
        x
    \right\|
    +
    \|x-u\|                                                   \\
    &=
    |1-\|x\||
    +
    \|x-u\|                                                   \\
    &\leq
    2\|x-u\|.
\end{split}
\]
Applying this inequality with
$x=\bar a_{1,r}$ and $u=s a_{1,r}$ gives
\bea
\label{eq_bound_ahat_a_general_d}
    \min_{s\in\{-1,1\}}
    \|\hat a_{1,r}-s a_{1,r}\|
    &\lesssim&
    \|\hat U_1-U_1\|
    +
    \|A_1\|\|p\|.
\eea

\subsection{Related quantities}

Recall that each column of $A_k$ has unit Euclidean norm and that the mutual coherence is bounded by $\xi$.
By the Gershgorin circle theorem,
\bea
\|A_k\|^2 &=&\|A_k\T A_k\| \leq 1+(R-1)\xi,\notag \\
\sigma_R^2(A_k) &=&\sigma_R(A_k\T A_k) \geq 1-(R-1)\xi,\notag \\
\kappa(A_k) &\leq& \sqrt{\frac{1+(R-1)\xi}{1-(R-1)\xi}}. \label{eq_loading_condition_general_d}
\eea
Moreover, since $V_k=U_k\T A_k$, the matrices $V_k$ and $A_k$ have the same singular values, and
\be
\label{eq_V_condition_number_general_d} \kappa(V_k)=\kappa(A_k).
\ee
Also,
\be
\label{eq_V_column_cor_general_d} v_{r_1,k}\T v_{r_2,k} =a_{k,r_1}\T a_{k,r_2},
\ee
so every column of $V_k$ has unit norm and
\[
    |v_{r_1,k}\T v_{r_2,k}|\leq\xi \qquad \text{for }r_1\neq r_2.
\]
Finally,
\[
    \lambda_{\operatorname{Tucker}} \geq \lambda_{\min} \left\{1-(R-1)\xi\right\}^{d/2}.
\]

\begin{Assumption}\label{assumption_xiR_general_d}
    Assume that
    \[
        (R-1)\xi\leq 1-c_0
    \]
    for some fixed absolute constant $c_0\in(0,1)$.
\end{Assumption}

\begin{Remark}\label{remark_xiR_general_d}
    Under Supplement Assumption \ref{assumption_xiR_general_d}, $\|A_k\|^2\leq 2-c_0$, $\sigma_R^2(A_k)\geq c_0$, and $\kappa(A_k)\leq\sqrt{(2-c_0)/c_0}$ uniformly over $k\in[d]$.
    The same conclusion holds for $V_k$.
\end{Remark}

Fix the confidence level $\delta\in(0,1/2)$ appearing in Theorem~\ref{thm_tasd}.
Throughout the remainder of this proof, let $c_\delta\in(0,1)$ and $C_\delta\geq1$ denote sufficiently small and sufficiently large constants, respectively, depending only on $\delta$.
Their values may change from line to line.

We next introduce the random events used in the proof.

\begin{Assumption}\label{assumption_eigengap_general_d}
    Assume that
    \[
        \min_{\substack{r_1,r_2\in[R] \\
                r_1\neq r_2}} \left| \prod_{k=3}^d \frac{\langle w_{1k},v_{r_1,k}\rangle} {\langle w_{2k},v_{r_1,k}\rangle} - \prod_{k=3}^d
        \frac{\langle w_{1k},v_{r_2,k}\rangle} {\langle w_{2k},v_{r_2,k}\rangle} \right| \geq C_{\operatorname{gap}},
    \]
        where
    \[
    \begin{split}
        C_{\operatorname{gap}}
        >
        C_{\operatorname{eig}}
        \kappa_T
        \frac{\lambda_{\max}}{\lambda_{\min}}
        \frac{
            \sigma
            C_\delta^d
            R^{\frac{3(d-2)}{2}}
            (\log R)^{\frac{d-2}{2}}
            d^{2d-3}
            (R\lor\log d)^{d-2}
        }{
            \lambda_{\min}
        }
        \left(
            \sqrt{p_{\max}}
            +
            \sqrt{R^{d-1}}
        \right),
    \end{split}
    \]
    and $C_{\operatorname{eig}}>0$ is a sufficiently large constant
    depending only on the fixed constant $c_0$ in
    Supplement Assumption~\ref{assumption_xiR_general_d}.
\end{Assumption}

\begin{Assumption}\label{assumption_inner_general_d}
    Assume that
    \[
        \min_{r\in[R]}
        \left|
            \prod_{k=3}^d
            \langle w_{2k},v_{r,k}\rangle
        \right|
        >
        \left(
            \frac{c_\delta}{dR}
        \right)^{d-2},
    \]
    and
    \[
        \max_{r\in[R]}
        \left|
            \prod_{k=3}^d
            \langle w_{1k},v_{r,k}\rangle
        \right|
        <
        \left(
            C_\delta\sqrt{\log R}
        \right)^{d-2}.
    \]
\end{Assumption}

\begin{Assumption}\label{assumption_w_norm_general_d}
    Assume that
    \[
        \max_{i\in\{1,2\}} \prod_{k=3}^d \|w_{ik}\| < \left( C_\delta(R\lor\log d) \right)^{\frac{d-2}{2}}.
    \]
\end{Assumption}

\begin{Assumption}\label{assumption_lambda_pR_general_d}
    Assume that
    \[
        \lambda_{\min} > C_\delta^d \kappa_T d^{d-1} R^{d-2} \sigma \left( \sqrt{p_{\max}} + \sqrt{R^{d-1}} \right) (R\lor\log
        d)^{\frac{d-2}{2}}.
    \]
\end{Assumption}

By Supplement Assumptions~\ref{assumption_inner_general_d}--\ref{assumption_lambda_pR_general_d} and Supplement Remark~\ref{remark_xiR_general_d}, after enlarging $C_\delta$ if necessary, we have
\[
    Cd\kappa_T\sigma
    \left(
        \sqrt{p_{\max}}
        +
        \sqrt{R^{d-1}}
    \right)
    W
    \leq
    \frac{1}{2}
    \lambda_{\min}m_2\sigma_{\min}^2.
\]
Consequently, \eqref{eq_bound_SS_general_d} and Supplement Remark~\ref{remark_xiR_general_d} imply
\[
\begin{split}
    \Delta_S
    \lesssim{}&
    \kappa_T
    \frac{\lambda_{\max}}{\lambda_{\min}}
    \frac{
        d\sigma
        \left(
            \sqrt{p_{\max}}
            +
            \sqrt{R^{d-1}}
        \right)
    }{
        \lambda_{\min}
    }
    \frac{WM_1}{m_2^2}
    \\
    &+
    \kappa_T
    \frac{
        d\sigma
        \left(
            \sqrt{p_{\max}}
            +
            \sqrt{R^{d-1}}
        \right)
    }{
        \lambda_{\min}
    }
    \frac{W}{m_2}.
\end{split}
\]
Under Supplement Assumptions~\ref{assumption_inner_general_d} and~\ref{assumption_w_norm_general_d}, we have
\[
\begin{aligned}
    \frac{WM_1}{m_2^2}
    &\lesssim
    C_\delta^d
    d^{2d-4}
    R^{2d-4}
    (\log R)^{\frac{d-2}{2}}
    (R\lor\log d)^{\frac{d-2}{2}}
    \\
    &\leq
    C_\delta^d
    d^{2d-4}
    R^{\frac{3(d-2)}{2}}
    (\log R)^{\frac{d-2}{2}}
    (R\lor\log d)^{d-2},
\end{aligned}
\]
where the last inequality follows from $R\leq R\lor\log d$. Moreover,
\[
    \frac{W}{m_2}
    \lesssim
    C_\delta^d
    d^{d-2}
    R^{d-2}
    (R\lor\log d)^{\frac{d-2}{2}}.
\]
Since $R\geq2$, $d\geq3$, $\log R\geq\log 2$, and $\lambda_{\max}/\lambda_{\min}\geq1$, the second term is also dominated by the right-hand side below after enlarging $C_\delta$ if necessary. Therefore,
\be
\label{eq_bound_SS_1_general_d}
\Delta_S
\lesssim
\kappa_T
\frac{\lambda_{\max}}{\lambda_{\min}}
\frac{
    \sigma
    C_\delta^d
    R^{\frac{3(d-2)}{2}}
    (\log R)^{\frac{d-2}{2}}
    d^{2d-3}
    (R\lor\log d)^{d-2}
    \left(
        \sqrt{p_{\max}}
        +
        \sqrt{R^{d-1}}
    \right)
}{
    \lambda_{\min}
}.
\ee

For brevity, define
\[
\begin{split}
    B_\delta
    &:=
    \kappa_T
    \frac{\lambda_{\max}}{\lambda_{\min}}
    \frac{
        \sigma
        C_\delta^d
        R^{\frac{3(d-2)}{2}}
        (\log R)^{\frac{d-2}{2}}
        d^{2d-3}
        (R\lor\log d)^{d-2}
    }{
        \lambda_{\min}
    }
    \left(
        \sqrt{p_{\max}}
        +
        \sqrt{R^{d-1}}
    \right).
\end{split}
\]
Then \eqref{eq_bound_SS_1_general_d} implies
\[
    \Delta_S
    \leq
    C B_\delta
\]
for some absolute constant $C>0$.

By Supplement Remark~\ref{remark_xiR_general_d}, there exists a constant
$C_0>0$, depending only on the fixed constant $c_0$, such that
\[
    \kappa(A_1)
    \leq
    C_0,
    \qquad
    \|A_1\|
    \leq
    C_0.
\]
Moreover, since every column of $A_1$ has unit Euclidean norm,
\[
    \|A_1\|
    \geq
    1.
\]
Supplement Assumption~\ref{assumption_eigengap_general_d} gives
\[
    C_{\operatorname{gap}}
    >
    C_{\operatorname{eig}}B_\delta.
\]
By taking $C_{\operatorname{eig}}$ sufficiently large, depending only
on $c_0$, we therefore have
\begin{equation}
\label{eq_gap_dominates_DeltaS}
    \kappa(A_1)\|A_1\|\Delta_S
    \leq
    \frac{
        C_{\operatorname{gap}}
    }{
        8
    }.
\end{equation}
Since $\|A_1\|\geq1$, it follows that
\[
    \kappa(A_1)\Delta_S
    \leq
    \frac{
        C_{\operatorname{gap}}
    }{
        8
    },
\]
and hence
\[
    C_{\operatorname{gap}}
    >
    4\kappa(A_1)\Delta_S.
\]
Thus, the first condition in
\eqref{eq_tasd_eigen_side_conditions} holds.

Moreover,
\[
    C_{\operatorname{gap}}
    -
    2\kappa(A_1)\Delta_S
    \geq
    \frac{
        3C_{\operatorname{gap}}
    }{
        4
    }.
\]
Therefore, by \eqref{eq_bound_p_1_general_d} and
\eqref{eq_gap_dominates_DeltaS},
\[
\begin{split}
    \|p\|\|A_1\|
    &\leq
    \frac{
        2\kappa(A_1)\|A_1\|\Delta_S
    }{
        C_{\operatorname{gap}}
        -
        2\kappa(A_1)\Delta_S
    }                                                        \\
    &\leq
    \frac{
        C_{\operatorname{gap}}/4
    }{
        3C_{\operatorname{gap}}/4
    }
    =
    \frac13
    <
    1.
\end{split}
\]
This verifies the second condition in
\eqref{eq_tasd_eigen_side_conditions}.

Finally, \eqref{eq_bound_p_1_general_d},
\eqref{eq_bound_SS_1_general_d}, and
\eqref{eq_gap_dominates_DeltaS} give
\be
\label{eq_bound_p_2_general_d}
\begin{split}
    \|p\|
    \lesssim{}&
    \kappa_T
    \frac{\lambda_{\max}}{\lambda_{\min}}
    \frac{
        \sigma
        C_\delta^d
        R^{\frac{3(d-2)}{2}}
        (\log R)^{\frac{d-2}{2}}
        d^{2d-3}
        (R\lor\log d)^{d-2}
    }{
        C_{\operatorname{gap}}
        \lambda_{\min}
    }                                                        \\
    &\qquad\qquad\times
    \left(
        \sqrt{p_{\max}}
        +
        \sqrt{R^{d-1}}
    \right).
\end{split}
\ee

Combining \eqref{eq_bound_ahat_a_general_d}, \eqref{eq_bound_p_2_general_d}, \eqref{eq_bound_U_hat_general_d}, \eqref{eq_tucker_specialU}, and Supplement Remark \ref{remark_xiR_general_d}, we obtain
\bea
\min_{s\in\{-1,1\}} \|\hat a_{1,r}-s a_{1,r}\| &\lesssim& \sigma\frac{\sqrt{p_{\max}}+\sqrt{R^{d-1}}} {\lambda_{\operatorname{Tucker}}}
+\notag \\
&& \kappa_T\frac{\lambda_{\max}}{\lambda_{\min}} \frac{ \sigma C_\delta^dR^{\frac{3(d-2)}{2}}(\log R)^{\frac{d-2}{2}}
d^{2d-3}(R\lor\log d)^{d-2} \left(\sqrt{p_{\max}}+\sqrt{R^{d-1}}\right) }{C_{\operatorname{gap}}\lambda_{\min}}.
\label{eq_bound_ahat_a_1_general_d}
\eea

\subsection{Probability of assumptions to hold}

\paragraph*{Supplement Assumption \ref{assumption_eigengap_general_d}: Eigengap\\}

For $k=3,\ldots,d$ and $r\in[R]$, define
\[
    q_{k,r} = \frac{\langle w_{1k},v_{r,k}\rangle} {\langle w_{2k},v_{r,k}\rangle}, \qquad \mu_r=\prod_{k=3}^dq_{k,r}.
\]
Fix $r_1\neq r_2$, and define
\[
    \gamma_r=\prod_{k=4}^dq_{k,r},
\]
where the empty product is understood as one when $d=3$.
Let $\mathcal{F}_{4:d}$ be the sigma-field generated by $\{w_{1k},w_{2k}:k=4,\ldots,d\}$, with $\mathcal{F}_{4:d}$ taken to be the trivial sigma-field when $d=3$.

Conditional on $\mathcal{F}_{4:d}$, the quantities $\gamma_{r_1}$ and $\gamma_{r_2}$ are fixed, while $w_{13}$ and $w_{23}$ remain independent standard Gaussian vectors.
For simplicity, write
\[
    x_r=\langle w_{13},v_{r,3}\rangle, \qquad y_r=\langle w_{23},v_{r,3}\rangle.
\]
Then
\[
    \mu_{r_1}-\mu_{r_2} = \gamma_{r_1}\frac{x_{r_1}}{y_{r_1}} - \gamma_{r_2}\frac{x_{r_2}}{y_{r_2}} = \frac{
    \gamma_{r_1}x_{r_1}y_{r_2} - \gamma_{r_2}x_{r_2}y_{r_1} }{y_{r_1}y_{r_2}}.
\]
Conditional additionally on $w_{23}$, the numerator is a centered Gaussian random variable with conditional variance
\begin{align*}
     & \operatorname{Var}\left( \gamma_{r_1}x_{r_1}y_{r_2} - \gamma_{r_2}x_{r_2}y_{r_1} \mid\mathcal{F}_{4:d},w_{23} \right) \\
     & \qquad= \left\| \gamma_{r_1}y_{r_2}v_{r_1,3} - \gamma_{r_2}y_{r_1}v_{r_2,3} \right\|^2                                \\
     & \qquad\geq (1-\xi) \left( \gamma_{r_1}^2y_{r_2}^2 + \gamma_{r_2}^2y_{r_1}^2 \right),
\end{align*}
where the last inequality follows from \eqref{eq_V_column_cor_general_d}.
The Gaussian small-ball inequality therefore yields, for every $t>0$,
\begin{align*}
     & \PP\left( |\mu_{r_1}-\mu_{r_2}|\leq t \mid\mathcal{F}_{4:d},w_{23} \right)                                        \\
     & \qquad\leq \frac{ Ct|y_{r_1}y_{r_2}| }{ \sqrt{1-\xi} \sqrt{ \gamma_{r_1}^2y_{r_2}^2 + \gamma_{r_2}^2y_{r_1}^2 } } \\
     & \qquad\leq \frac{ Ct\sqrt{y_{r_1}^2+y_{r_2}^2} }{ \sqrt{1-\xi} \min\{|\gamma_{r_1}|,|\gamma_{r_2}|\} }.
\end{align*}
Taking the conditional expectation with respect to $w_{23}$ and using
\[
    \EE\sqrt{y_{r_1}^2+y_{r_2}^2} \leq \sqrt{\EE(y_{r_1}^2+y_{r_2}^2)} =\sqrt{2},
\]
we obtain
\be
\label{eq_conditional_product_small_ball} \PP\left( |\mu_{r_1}-\mu_{r_2}|\leq t \mid\mathcal{F}_{4:d} \right) \leq \frac{Ct}
{\sqrt{1-\xi}\min\{|\gamma_{r_1}|,|\gamma_{r_2}|\}}.
\ee

For $0<\tau\leq1$, define
\[
    \mathcal{E}_{\tau} = \left\{ \min_{\substack{k=4,\ldots,d \\
            r\in[R]}} |q_{k,r}|\geq\tau \right\},
\]
where $\mathcal{E}_{\tau}$ is the entire sample space when $d=3$.
For every fixed $k$ and $r$, $q_{k,r}$ has the standard Cauchy distribution.
Hence,
\[
    \PP(|q_{k,r}|<\tau) =\frac{2}{\pi}\arctan(\tau) \leq C\tau,
\]
and a union bound gives
\be
\label{eq_probability_E_tau} \PP(\mathcal{E}_{\tau}^c) \leq C(d-3)R\tau.
\ee
On $\mathcal{E}_{\tau}$,
\[
    \min_{r\in[R]}|\gamma_r|\geq\tau^{d-3}.
\]
Combining \eqref{eq_conditional_product_small_ball} and \eqref{eq_probability_E_tau}, and taking a union bound over all pairs $r_1\neq r_2$, yields
\be
\label{eq_product_eigengap_probability} \PP\left( \min_{\substack{r_1,r_2\in[R] \\
        r_1\neq r_2}} |\mu_{r_1}-\mu_{r_2}|\leq t \right) \leq C(d-3)R\tau + \frac{CR^2t}{\sqrt{1-\xi}\tau^{d-3}}.
\ee
By Supplement Assumption~\ref{assumption_xiR_general_d}, $1-\xi$ is bounded below by an absolute positive constant.
Taking
\[
    \tau = \frac{c_\delta}{dR}, \qquad t = \frac{c_\delta}{dR^2} \left( \frac{c_\delta}{dR} \right)^{d-3},
\]
where $c_\delta>0$ is sufficiently small depending only on $\delta$, Equation~\eqref{eq_product_eigengap_probability} gives
\begin{equation}\label{eq_product_eigengap_final}
    \PP\left( \min_{\substack{r_1,r_2\in[R] \\
            r_1\neq r_2}} |\mu_{r_1}-\mu_{r_2}| \geq \frac{c_\delta}{dR^2} \left( \frac{c_\delta}{dR} \right)^{d-3} \right) \geq
    1-\frac{\delta}{4}.
\end{equation}
Thus, we can choose
\[
    C_{\operatorname{gap}} = \frac{c_\delta}{dR^2} \left( \frac{c_\delta}{dR} \right)^{d-3}.
\]
After absorbing powers of $c_\delta^{-1}$ and the sufficiently large
constant $C_{\operatorname{eig}}$ into $C_\delta^d$, a sufficient
condition for Supplement Assumption~\ref{assumption_eigengap_general_d}, and
hence for the two perturbation conditions in
\eqref{eq_tasd_eigen_side_conditions}, is
\[
    \lambda_{\min} \gtrsim \kappa_T\sigma \frac{\lambda_{\max}}{\lambda_{\min}} C_\delta^d d^{3d-5} R^{\frac{5d}{2}-4} (\log
    R)^{\frac{d-2}{2}} (R\lor\log d)^{d-2} \left( \sqrt{p_{\max}} + \sqrt{R^{d-1}} \right).
\]

\paragraph*{Supplement Assumption \ref{assumption_inner_general_d}: Inner product\\}

Conditional on $V_3,\ldots,V_d$, for every $i\in\{1,2\}$, $k=3,\ldots,d$, and $r\in[R]$,
\[
    \langle w_{ik},v_{r,k}\rangle
    \sim
    N(0,1).
\]
Moreover, for every fixed $i\in\{1,2\}$ and $r\in[R]$, the random variables
\[
    \left\{
        \langle w_{ik},v_{r,k}\rangle:
        k=3,\ldots,d
    \right\}
\]
are conditionally independent. Independence across $r$ is not required.

The Gaussian small-ball inequality and a union bound yield
\[
    \PP\left(
        \min_{\substack{
            k=3,\ldots,d\\
            r\in[R]
        }}
        \left|
            \langle w_{2k},v_{r,k}\rangle
        \right|
        \leq
        t
    \right)
    \leq
    C(d-2)Rt.
\]
Taking
\[
    t
    =
    \frac{c_\delta}{dR}
\]
for a sufficiently small constant $c_\delta>0$ depending only on $\delta$, we obtain
\begin{equation}\label{eq_bound_P_min}
    \PP\left(
        \min_{r\in[R]}
        \left|
            \prod_{k=3}^d
            \langle w_{2k},v_{r,k}\rangle
        \right|
        >
        \left(
            \frac{c_\delta}{dR}
        \right)^{d-2}
    \right)
    \geq
    1-\frac{\delta}{8}.
\end{equation}

For each $r\in[R]$, define
\[
    G_r
    :=
    \prod_{k=3}^d
    \left|
        \langle w_{1k},v_{r,k}\rangle
    \right|.
\]
For every $q\geq2$, conditional independence across $k$ and the standard Gaussian moment bound give
\[
\begin{aligned}
    \left\{
        \EE\left(
            G_r^q
            \mid
            V_3,\ldots,V_d
        \right)
    \right\}^{1/q}
    &=
    \prod_{k=3}^d
    \left\{
        \EE\left(
            \left|
                \langle w_{1k},v_{r,k}\rangle
            \right|^q
            \mid
            V_3,\ldots,V_d
        \right)
    \right\}^{1/q}
    \\
    &\leq
    \left(
        C\sqrt q
    \right)^{d-2}.
\end{aligned}
\]
Therefore, Markov's inequality gives
\[
    \PP\left(
        G_r
        >
        \left(
            eC\sqrt q
        \right)^{d-2}
        \,\middle|\,
        V_3,\ldots,V_d
    \right)
    \leq
    \exp\left(
        -q(d-2)
    \right)
    \leq
    \exp(-q).
\]
Taking a union bound over $r\in[R]$ and then removing the conditioning yields
\[
    \PP\left(
        \max_{r\in[R]}G_r
        >
        \left(
            eC\sqrt q
        \right)^{d-2}
    \right)
    \leq
    R\exp(-q).
\]
Set
\[
    q
    :=
    \log\left(
        \frac{8R}{\delta}
    \right).
\]
Since $R\geq2$ and $\delta\in(0,1/2)$, we have $q\geq2$ and
\[
    q
    \leq
    C_\delta\log R
\]
for a sufficiently large constant $C_\delta$ depending only on $\delta$. Absorbing the absolute constants into $C_\delta$, we conclude that
\begin{equation}\label{eq_bound_P_max}
    \PP\left(
        \max_{r\in[R]}
        \left|
            \prod_{k=3}^d
            \langle w_{1k},v_{r,k}\rangle
        \right|
        <
        \left(
            C_\delta\sqrt{\log R}
        \right)^{d-2}
    \right)
    \geq
    1-\frac{\delta}{8}.
\end{equation}
Therefore, Supplement Assumption~\ref{assumption_inner_general_d} holds with probability at least $1-\delta/4$.

\paragraph*{Supplement Assumption \ref{assumption_w_norm_general_d}: Vector norm\\}

For every $i\in\{1,2\}$ and $k=3,\ldots,d$, $w_{ik}\sim N(0,I_R)$, and hence
\[
    \PP(\|w_{ik}\|\geq\sqrt{R}+t) \leq\exp(-ct^2).
\]
Taking
\[
    t = C_\delta\sqrt{R\lor\log d}
\]
and applying a union bound over $i$ and $k$, we obtain, for a sufficiently large constant $C_\delta$ depending only on $\delta$,
\begin{equation}\label{eq_bound_w_norm_probability}
    \PP\left( \max_{i\in\{1,2\}} \prod_{k=3}^d \|w_{ik}\| < \left( C_\delta(R\lor\log d) \right)^{\frac{d-2}{2}} \right) \geq
    1-\frac{\delta}{4}.
\end{equation}
Thus, Supplement Assumption~\ref{assumption_w_norm_general_d} holds with probability at least $1-\delta/4$.

\subsection{Summary}

Supplement Assumption~\ref{assumption_lambda_pR_general_d} is implied by the third condition of Theorem~\ref{thm_tasd}.
By \eqref{eq_product_eigengap_final}, \eqref{eq_bound_P_min}, \eqref{eq_bound_P_max}, and \eqref{eq_bound_w_norm_probability}, the eigengap event, the inner-product event, and the Gaussian-vector norm event hold simultaneously with probability at least $1-3\delta/4$.

Moreover, by the first two conditions of Theorem~\ref{thm_tasd} and Theorem~\ref{thm_general_tucker}, the HOOI bounds \eqref{eq_bound_U_hat_general_d} and \eqref{eq_bound_UZ_hat_general_d} hold with probability at least
\[
    1-Cd\exp(-cp_{\min}).
\]
The condition
\[
    p_{\min} \geq C\log\left(\frac{d}{\delta}\right)
\]
ensures that
\[
    Cd\exp(-cp_{\min}) \leq \frac{\delta}{4}.
\]
A union bound therefore shows that all required events hold simultaneously with probability at least $1-\delta$.

On this event, substituting
\[
    C_{\operatorname{gap}} = \frac{c_\delta}{dR^2} \left( \frac{c_\delta}{dR} \right)^{d-3}
\]
into \eqref{eq_bound_ahat_a_1_general_d}, and absorbing all powers of $c_\delta^{-1}$ into $C_\delta^d$, yields
\begin{align*}
    \min_{s\in\{-1,1\}} \|\hat a_{1,r}-s a_{1,r}\| \lesssim{} & \sigma \frac{ \sqrt{p_{\max}}+\sqrt{R^{d-1}} }{
    \lambda_{\operatorname{Tucker}} }                                                                                                                           \\
                                                   & + \sigma\kappa_T \frac{\lambda_{\max}}{\lambda_{\min}} \frac{ C_\delta^d d^{3d-5} R^{\frac{5d}{2}-4} (\log
        R)^{\frac{d-2}{2}} (R\lor\log d)^{d-2} \left( \sqrt{p_{\max}}+\sqrt{R^{d-1}} \right) }{ \lambda_{\min} }.
\end{align*}
Since
\[
    \sqrt{p_{\max}}+\sqrt{R^{d-1}} \lesssim \sqrt{p_{\max}\lor R^{d-1}},
\]
the desired result follows for the first mode.
The other modes follow by the same argument.
The eigenvalue matching induces a permutation of the $R$ components, and the sign ambiguity is handled by taking the minimum over the two possible signs.
This completes the proof.

\subsection{Proof of Remark~\ref{remark_tasd_confidence}}\label{sec_proof_remark_tasd_confidence}

We verify the explicit choice of $C_\delta$ asserted in Remark~\ref{remark_tasd_confidence} by returning to the bounds before the confidence-dependent constants were absorbed.
The argument has three steps: obtaining explicit bounds from the Gaussian small-ball, moment, and norm inequalities; multiplying the resulting factors in the inverse and eigenvector perturbation bounds; and verifying the perturbation conditions under the third signal-strength condition of Theorem~\ref{thm_tasd}.
We use the notation $m_2$, $M_1$, $W$, $\Delta_S$, and $C_{\operatorname{gap}}$ from Section~\ref{sec_proof_thm_tasd}.
Throughout this proof, $C,c>0$ denote constants depending only on the fixed coherence margin $c_0$.

\paragraph*{Explicit probability bounds.}
Conditional on the observed tensor, the contraction vectors retain their independent standard Gaussian distributions.
Take $c_\delta=c\delta$ for a sufficiently small constant $c>0$.
The Gaussian small-ball bound preceding \eqref{eq_bound_P_min} gives
\[
    m_2\geq\left(\frac{c\delta}{dR}\right)^{d-2}
\]
with failure probability at most $\delta/8$.
For the eigengap, set
\[
    \tau=\frac{c\delta}{dR},
    \qquad
    C_{\operatorname{gap}}
    :=
    \frac{(c\delta)^{d-2}}{d^{d-2}R^{d-1}}.
\]
Substituting $t=C_{\operatorname{gap}}$ into \eqref{eq_product_eigengap_probability} bounds its two terms by $Cc\delta(d-3)/d$ and $Cc\delta/(d\sqrt{1-\xi})$, respectively.
Since the coherence assumption gives $1-\xi\geq c_0$, choosing $c$ sufficiently small yields
\[
    \min_{r_1\neq r_2}|\mu_{r_1}-\mu_{r_2}|
    \geq C_{\operatorname{gap}}
\]
with failure probability at most $\delta/4$.

For the upper bound on $M_1$, the moment calculation preceding \eqref{eq_bound_P_max}, with $q=\log(8R/\delta)$, gives
\[
    \PP\left(
        M_1>
        \left\{C\log\left(\frac{8R}{\delta}\right)\right\}^{(d-2)/2}
    \right)
    \leq R\exp(-q)
    =\frac{\delta}{8}.
\]
For $W$, the Gaussian norm inequality used in the proof of \eqref{eq_bound_w_norm_probability} states that $\PP(\|w_{ik}\|\geq\sqrt R+t)\leq\exp(-ct^2)$.
Taking $t=C\sqrt{\log(8d/\delta)}$ and applying a union bound over the $2(d-2)$ contraction vectors gives
\[
    W
    \leq
    \left\{C\left[R+\log\left(\frac{8d}{\delta}\right)\right]\right\}^{(d-2)/2}
\]
with failure probability at most $\delta/4$.
Consequently, these four bounds hold simultaneously with probability at least $1-3\delta/4$.
The first two conditions of Theorem~\ref{thm_tasd} ensure that the HOOI event has failure probability at most $\delta/4$, as established in Section~\ref{sec_proof_thm_tasd}.
We work on the intersection of these events, which has probability at least $1-\delta$.

\paragraph*{Tracking the perturbation factors.}
Since $R\geq2$, $d\geq3$, and $\delta\in(0,1/2)$,
\[
\begin{aligned}
    \log\left(\frac{8R}{\delta}\right)
    &\lesssim
    (\log R)\log(e/\delta),
    \\
    R+\log\left(\frac{8d}{\delta}\right)
    &\lesssim
    (R\lor\log d)\log(e/\delta).
\end{aligned}
\]
Multiplying the explicit upper and lower bounds therefore yields
\[
\begin{split}
    \frac{dWM_1}{m_2^2C_{\operatorname{gap}}}
    \lesssim{}&
    C^d\delta^{-3(d-2)}
    \{\log(e/\delta)\}^{d-2}
    d^{3d-5}R^{3d-5}
    \\
    &\times
    (\log R)^{(d-2)/2}
    (R\lor\log d)^{(d-2)/2}.
\end{split}
\]
Here, $m_2^{-2}$ contributes $\delta^{-2(d-2)}$, $C_{\operatorname{gap}}^{-1}$ contributes $\delta^{-(d-2)}$, and the product of the upper bounds for $W$ and $M_1$ contributes $\{\log(e/\delta)\}^{d-2}$.
The deterministic lower bounds used for $m_2$ and $C_{\operatorname{gap}}$ are at most one, and the upper bound used for $M_1$ is at least one.
Thus, the same right-hand side also bounds $dW/m_2$ and $dW/(m_2C_{\operatorname{gap}})$.
Using $R^{(d-2)/2}\leq(R\lor\log d)^{(d-2)/2}$ gives
\[
\begin{split}
    &\frac{dW}{m_2}
    +
    \frac{d}{C_{\operatorname{gap}}}
    \left(
        \frac{WM_1}{m_2^2}
        +
        \frac{W}{m_2}
    \right)
    \\
    &\qquad\lesssim
    C^d\delta^{-3(d-2)}
    \{\log(e/\delta)\}^{d-2}
    d^{3d-5}R^{\frac{5d}{2}-4}
    \\
    &\qquad\quad\times
    (\log R)^{\frac{d-2}{2}}
    (R\lor\log d)^{d-2}.
\end{split}
\]
For a sufficiently large constant $C_*$ depending only on $c_0$,
\[
    C^d\delta^{-3(d-2)}\{\log(e/\delta)\}^{d-2}
    \leq C_\delta^d,
    \qquad
    C_\delta=C_*\delta^{-3}\log(e/\delta).
\]

\paragraph*{Verification of the perturbation conditions.}
We now check that this choice also suffices for the signal-strength condition and the perturbation argument.
Since $\lambda_{\max}/\lambda_{\min}\geq1$ and $\sigma_{\min}^2\geq c_0$, the third condition of Theorem~\ref{thm_tasd}, together with the preceding bound on $dW/m_2$, implies
\[
    Cd\kappa_T\sigma
    \left(\sqrt{p_{\max}}+\sqrt{R^{d-1}}\right)W
    \leq
    \frac12\lambda_{\min}m_2\sigma_{\min}^2,
\]
after choosing $C_*$ sufficiently large.
The denominator in \eqref{eq_bound_SS_general_d} is therefore positive and bounded below by $\lambda_{\min}m_2\sigma_{\min}^2/2$.
That equation consequently gives
\[
\begin{split}
    \Delta_S
    \lesssim{}&
    \frac{\kappa_T\sigma}{\lambda_{\min}}
    \frac{\lambda_{\max}}{\lambda_{\min}}
    \left(\sqrt{p_{\max}}+\sqrt{R^{d-1}}\right)
    \\
    &\times
    d\left(
        \frac{WM_1}{m_2^2}
        +
        \frac{W}{m_2}
    \right).
\end{split}
\]
Combining this inequality with the bound involving $C_{\operatorname{gap}}^{-1}$ and the same signal-strength condition yields
\[
    \kappa(A_1)\|A_1\|\Delta_S
    \leq
    \frac{C_{\operatorname{gap}}}{8},
\]
with $C_*$ sufficiently large.
As established in Section~\ref{sec_proof_thm_tasd}, this implies both conditions in \eqref{eq_tasd_eigen_side_conditions}.
Equations~\eqref{eq_bound_p_1_general_d} and \eqref{eq_bound_ahat_a_general_d}, together with the HOOI bounds, then yield the error bound of Theorem~\ref{thm_tasd} with this explicit choice of $C_\delta$.
The argument applies to any fixed target mode by permuting the mode indices.
This proves the confidence dependence asserted in Remark~\ref{remark_tasd_confidence}.

\section{Proof of Corollaries}

\subsection{Corollary \ref{corollary_lambda_X}}

We separate the Corollary \ref{corollary_lambda_X} into following two corollaries and prove them respectively.

\begin{Corollary}\label{corollary_lambda}
    In the same setting as Theorem \ref{thm_local}, we also have the following bound: for $t > T$,
    \[
        \max_{r\in[R]} \left| \widehat{|\lambda_r|}-|\lambda_r| \right| \lesssim \sigma \sqrt{p_{\max}}.
    \]
\end{Corollary}

\begin{Corollary}\label{corollary_signal_tensor}
    In the setting of Theorem~\ref{thm_local}, assume that ALS has been performed for $t \geq T+1$ iterations.
    Let $\hat B_1^{(t+1)}$ be calculated by \eqref{eq_update_formula}, and define
    \[
        \hat \X = \sum_{r=1}^R \hat b_{1,r}^{(t+1)} \circ \hat a_{2,r}^{(t)} \circ \cdots \circ \hat a_{d,r}^{(t)},
    \]
    where $\hat b_{1,r}^{(t+1)}$ and $\hat a_{k,r}^{(t)}$ are the $r$th columns of $\hat B_1^{(t+1)}$ and $\hat A_k^{(t)}$, respectively.
    Then, on the same high-probability event as in Theorem~\ref{thm_local},
    \[
        \|\hat \X-\X\|_F \lesssim d\sigma\sqrt{R p_{\max}}.
    \]
\end{Corollary}

\subsubsection{Supplement Corollary \ref{corollary_lambda}}\label{sec_proof_corollary_lambda}

This corollary holds by \eqref{eq_lambda_bound} and \eqref{eq_A_hat_bound}, we have
\[
    \max_{i\in [R]} |\hat \lambda^{(t+1)}_i - |\lambda_i|| \leq \frac{2\lambda_{\max} \left((2\varepsilon_t^2 + 1)^{d-1} - 1 +
        (R-1)(\varepsilon_t^2 + 2\varepsilon_t + \xi)^{d-1} - (R-1)\xi^{d-1}\right) + 2\delta}{\(1 - (R-1)(\xi + 2 \varepsilon_t +
        \varepsilon_t^2)^{d-1}\)}.
\]
Set
\[
    s := \frac{\sigma\sqrt{p_{\max}}}{\lambda_{\min}}, \qquad \eta := \sigma\sqrt{p_{\max}}.
\]
For $t>T$, Theorem~\ref{thm_local} gives $\varepsilon_t\lesssim s$, and \eqref{eq_noise_random_term_2} gives $\delta\lesssim\eta$.

If $\varepsilon_t>\tau_{\mathrm{ALS}}$, the quadratic numerator bound gives
\[
    \max_{i\in[R]} \left| \hat\lambda_i^{(t+1)}-|\lambda_i| \right| \lesssim \lambda_{\max}(d+R)\varepsilon_t^2+\delta.
\]
If $\varepsilon_t\leq\tau_{\mathrm{ALS}}$, the linear numerator bound gives
\[
    \max_{i\in[R]} \left| \hat\lambda_i^{(t+1)}-|\lambda_i| \right| \lesssim \lambda_{\max}Rd\xi^{d-2}\varepsilon_t+\delta.
\]

Suppose first that $s\geq\tau_{\mathrm{ALS}}$.
If $\varepsilon_t\leq\tau_{\mathrm{ALS}}$, then, when $d=3$,
\[
    Rd\xi^{d-2}\varepsilon_t = 3R\xi\varepsilon_t \lesssim R s^2 \lesssim (d+R)s^2.
\]
When $d>3$, we have
\[
    Rd\xi^{d-2}\varepsilon_t = \frac{R}{R-1} d\tau_{\mathrm{ALS}}\varepsilon_t \lesssim d s^2 \lesssim (d+R)s^2.
\]
Therefore, regardless of which recurrence regime contains the actual iterate,
\[
    \max_{i\in[R]} \left| \hat\lambda_i^{(t+1)}-|\lambda_i| \right| \lesssim \lambda_{\max}(d+R)s^2+\eta.
\]
Condition~\eqref{eq_assumption_3_general_R} implies
\[
    \lambda_{\max}(d+R)s^2 \lesssim \eta.
\]

Suppose next that $s<\tau_{\mathrm{ALS}}$.
If $\varepsilon_t>\tau_{\mathrm{ALS}}$, the same quadratic argument yields a bound of order $\eta$.
If $\varepsilon_t\leq\tau_{\mathrm{ALS}}$, then condition~\eqref{eq_assumption_general_R_4} gives
\begin{align*}
    \lambda_{\max}Rd\xi^{d-2}\varepsilon_t & = \lambda_{\min} \left( \frac{\lambda_{\max}}{\lambda_{\min}} Rd\xi^{d-2} \right)
    \varepsilon_t                                                                                                              \\
                                           & \lesssim \lambda_{\min}s = \eta.
\end{align*}
Combining the preceding bounds with $\delta\lesssim\eta$, we conclude that
\[
    \max_{i\in[R]} \left| \hat\lambda_i^{(t+1)}-|\lambda_i| \right| \lesssim \sigma\sqrt{p_{\max}}.
\]

\subsubsection{Supplement Corollary \ref{corollary_signal_tensor}}

Let
\[
    \eta
    :=
    \sigma\sqrt{p_{\max}},
\]
and work on the high-probability event used in the proof of
Theorem~\ref{thm_local}. Since the corollary assumes $t\geq T+1$,
both $t-1$ and $t$ are at least $T$. Therefore,
\[
    \varepsilon_{t-1}
    \lor
    \varepsilon_t
    \lesssim
    \frac{\eta}{\lambda_{\min}}.
\]
Moreover, the noise terms in \eqref{eq_l1l2norm_bound} corresponding
to the updates from iteration $t-1$ to iteration $t$ and from
iteration $t$ to iteration $t+1$ are both of order at most $\eta$.

Applying \eqref{eq_l1l2norm_bound} to each mode for the update from
iteration $t-1$ to iteration $t$, and using the same two-regime
analysis as in the proof of Supplement Corollary~\ref{corollary_lambda}, we obtain
the following stronger, weighted loading-vector bound.

For every $k\in\{2,\ldots,d\}$ and $r\in[R]$, there exists $s_{k,r}\in\{-1,1\}$ such that, with
\[
    \widetilde a_{k,r}^{(t)} := s_{k,r}\hat a_{k,r}^{(t)},
\]
we have
\begin{equation}\label{eq_weighted_loading_error_signal}
    |\lambda_r| \left\| \widetilde a_{k,r}^{(t)}-a_{k,r} \right\| \lesssim \eta.
\end{equation}
Indeed, if $\hat b_{k,r}^{(t)}$ denotes the unnormalized update that produces $\hat a_{k,r}^{(t)}$, then the argument leading to \eqref{eq_l1l2norm_bound} gives
\[
    \left\| s_{k,r}\hat b_{k,r}^{(t)} - |\lambda_r|a_{k,r} \right\| \lesssim \eta.
\]
Since $s_{k,r}\hat b_{k,r}^{(t)} = \|\hat b_{k,r}^{(t)}\| \widetilde a_{k,r}^{(t)}$, the reverse triangle inequality yields
\begin{align*}
    |\lambda_r| \left\| \widetilde a_{k,r}^{(t)}-a_{k,r} \right\| & \leq \left| |\lambda_r|-\|\hat b_{k,r}^{(t)}\| \right| +
    \left\| s_{k,r}\hat b_{k,r}^{(t)} - |\lambda_r|a_{k,r} \right\|                                                                                       \\
                                                                  & \leq 2 \left\| s_{k,r}\hat b_{k,r}^{(t)} - |\lambda_r|a_{k,r} \right\| \lesssim \eta,
\end{align*}
which proves \eqref{eq_weighted_loading_error_signal}.

For each $k\in\{2,\ldots,d\}$, define
\[
    \mathcal S_{t,k}
    :=
    \diag(s_{k,1},\ldots,s_{k,R}),
    \qquad
    \widetilde A_k^{(t)}
    :=
    \hat A_k^{(t)}\mathcal S_{t,k}.
\]
For each $r\in[R]$, set
\[
    s_r
    :=
    \prod_{k=2}^d s_{k,r},
\]
and define
\[
    \mathcal S_t
    :=
    \prod_{k=2}^d\mathcal S_{t,k}
    =
    \diag(s_1,\ldots,s_R).
\]

Let
\[
    \hat K_t
    :=
    \hat A_d^{(t)}
    \odot\cdots\odot
    \hat A_2^{(t)},
    \qquad
    \widetilde K_t
    :=
    \widetilde A_d^{(t)}
    \odot\cdots\odot
    \widetilde A_2^{(t)}.
\]
Since multiplying the $r$th column of each
$\hat A_k^{(t)}$, $k=2,\ldots,d$, by $s_{k,r}$
multiplies the $r$th Khatri--Rao column by $s_r$, we have
\[
    \widetilde K_t
    =
    \hat K_t\mathcal S_t.
\]
Moreover, since $\mathcal S_t^\top=\mathcal S_t$ and
$\mathcal S_t^2=I_R$,
\[
    \left[
        \widetilde K_t^\top
    \right]^\dagger
    =
    \left[
        \hat K_t^\top
    \right]^\dagger
    \mathcal S_t.
\]
Consequently, the mode-$1$ least-squares update satisfies
\[
\begin{split}
    \hat B_1^{(t+1)}\mathcal S_t
    &=
    \MM_1(\Y)
    \left[
        \hat K_t^\top
    \right]^\dagger
    \mathcal S_t                                                   \\
    &=
    \MM_1(\Y)
    \left[
        \widetilde K_t^\top
    \right]^\dagger.
\end{split}
\]

Let
\[
    K
    :=
    A_d\odot\cdots\odot A_2.
\]
On the same local event used in the proof of
Theorem~\ref{thm_local}, $\widetilde K_t$ has full column rank.
Therefore,
\[
    \widetilde K_t^\top
    \left[
        \widetilde K_t^\top
    \right]^\dagger
    =
    I_R.
\]
Using
\[
    \MM_1(\Y)
    =
    A_1\diag(\lambda)K^\top
    +
    \MM_1(\Z),
\]
we obtain the exact identity
\begin{align*}
    &
    \hat B_1^{(t+1)}\mathcal S_t
    -
    A_1\diag(\lambda)                                           \\
    &\quad=
    A_1\diag(\lambda)
    \left(
        K-\widetilde K_t
    \right)^\top
    \left[
        \widetilde K_t^\top
    \right]^\dagger
    +
    \MM_1(\Z)
    \left[
        \widetilde K_t^\top
    \right]^\dagger.
\end{align*}

Set
\[
    \bar\varepsilon_t
    :=
    \max_{k\in\{2,\ldots,d\}}
    \left\|
        \widetilde A_k^{(t)}-A_k
    \right\|_{1\rightarrow2}.
\]
By \eqref{eq_weighted_loading_error_signal},
\[
    \bar\varepsilon_t
    \lesssim
    \frac{\eta}{\lambda_{\min}}.
\]
Repeating the deterministic and noise estimates used in the
derivation of \eqref{eq_l1l2norm_bound}, now with the fixed compatible
sign matrices $\mathcal S_{t,k}$, gives
\[
    \left\|
        \hat B_1^{(t+1)}\mathcal S_t
        -
        A_1\diag(\lambda)
    \right\|_{1\rightarrow2}
    \lesssim
    \begin{cases}
        \lambda_{\max}(d+R)\bar\varepsilon_t^2+\eta,
        &
        \bar\varepsilon_t>\tau_{\mathrm{ALS}},
        \\[2mm]
        \lambda_{\max}Rd\xi^{d-2}\bar\varepsilon_t+\eta,
        &
        \bar\varepsilon_t\leq\tau_{\mathrm{ALS}}.
    \end{cases}
\]
The same two-regime argument as in the proof of
Supplement Corollary~\ref{corollary_lambda}, together with
$\bar\varepsilon_t\lesssim\eta/\lambda_{\min}$, yields
\[
    \left\|
        \hat B_1^{(t+1)}\mathcal S_t
        -
        A_1\diag(\lambda)
    \right\|_{1\rightarrow2}
    \lesssim
    \eta.
\]

Finally, define
\[
    \widetilde b_{1,r}^{(t+1)}
    :=
    s_r\hat b_{1,r}^{(t+1)},
    \qquad
    r\in[R].
\]
The preceding matrix bound implies
\begin{equation}\label{eq_weighted_B_error_signal}
    \max_{r\in[R]}
    \left\|
        \widetilde b_{1,r}^{(t+1)}
        -
        \lambda_r a_{1,r}
    \right\|
    \lesssim
    \eta.
\end{equation}
Moreover,
\[
    \widetilde b_{1,r}^{(t+1)} \circ \widetilde a_{2,r}^{(t)} \circ\cdots\circ \widetilde a_{d,r}^{(t)} = \hat b_{1,r}^{(t+1)} \circ
    \hat a_{2,r}^{(t)} \circ\cdots\circ \hat a_{d,r}^{(t)},
\]
because $s_r\prod_{k=2}^d s_{k,r}=s_r^2=1$.

We next record a consequence of the coherence assumption.
Set
\[
    \rho_t := \xi+2\varepsilon_t+\varepsilon_t^2.
\]
For every $k\in\{2,\ldots,d\}$ and $r\neq s$,
\[
    \left| \left\langle \widetilde a_{k,r}^{(t)}, \widetilde a_{k,s}^{(t)} \right\rangle \right| \leq \rho_t.
\]
The same bound holds, with $\xi\leq\rho_t$, for the true loading vectors.
Since $\rho_t\leq \xi+3\varepsilon_t$ whenever $\varepsilon_t\leq1$, the denominator bound \eqref{eq_denominator_bound_1} implies
\begin{equation}\label{eq_coherence_sum_signal}
    1+(R-1)\rho_t^{d-1} \lesssim 1.
\end{equation}

We use the following elementary calculation.
Suppose that $\max_r\|e_r\|\lesssim\eta$ and, for each $\ell=1,\ldots,d-1$, the vectors $u_{\ell,r}$ have unit norm and satisfy
\[
    \max_{r\neq s} |\langle u_{\ell,r},u_{\ell,s}\rangle| \leq \rho_t.
\]
Then
\begin{align}
    \left\| \sum_{r=1}^R e_r\circ u_{1,r}\circ\cdots\circ u_{d-1,r} \right\|_F^2 & \leq \sum_{r=1}^R\|e_r\|^2 + \rho_t^{d-1}
    \sum_{r\neq s}\|e_r\|\|e_s\| \notag                                                                                                                    \\
                                                                                 & \lesssim R\eta^2 \left\{ 1+(R-1)\rho_t^{d-1} \right\} \lesssim R\eta^2,
    \label{eq_rank_sum_signal}
\end{align}
where the last inequality follows from \eqref{eq_coherence_sum_signal}.

Finally, use the telescoping decomposition
\begin{align*}
    \hat\X-\X ={} & \sum_{r=1}^R \left( \widetilde b_{1,r}^{(t+1)} - \lambda_r a_{1,r} \right) \circ \widetilde a_{2,r}^{(t)}
    \circ\cdots\circ \widetilde a_{d,r}^{(t)}                                                                                              \\
                  & + \sum_{k=2}^d \sum_{r=1}^R \lambda_r a_{1,r} \circ\cdots\circ a_{k-1,r} \circ \left( \widetilde a_{k,r}^{(t)}-a_{k,r}
    \right) \circ \widetilde a_{k+1,r}^{(t)} \circ\cdots\circ \widetilde a_{d,r}^{(t)}.
\end{align*}
For the first line, the error vectors are bounded by \eqref{eq_weighted_B_error_signal}.
For the term corresponding to each $k\geq2$, the error vectors
\[
    \lambda_r \left( \widetilde a_{k,r}^{(t)}-a_{k,r} \right)
\]
are bounded by \eqref{eq_weighted_loading_error_signal}.
All remaining factor vectors satisfy the coherence bound used in \eqref{eq_rank_sum_signal}.
Applying \eqref{eq_rank_sum_signal} to each of the $d$ groups and using the triangle inequality gives
\[
    \|\hat\X-\X\|_F \lesssim d\eta\sqrt R = d\sigma\sqrt{R p_{\max}}.
\]
This proves Supplement Corollary~\ref{corollary_signal_tensor}.

\subsection{Corollary \ref{corollary_global}}

Fix $\delta\in(0,1/2)$ and let
\[
    \delta_0 = \frac{\delta}{2d}.
\]
Apply Theorem~\ref{thm_tasd} with confidence level $\delta_0$ to each mode.
A union bound shows that, with probability at least $1-\delta/2$, all TASD loading estimates satisfy
\begin{align*}
    \varepsilon_0 \lesssim{} & \sigma \frac{ \sqrt{p_{\max}\lor R^{d-1}} }{ \lambda_{\operatorname{Tucker}} }                                 \\
                             & + \sigma\kappa_T \frac{\lambda_{\max}}{\lambda_{\min}} \frac{ C_{\delta_0}^d d^{3d-5} R^{\frac{5d}{2}-4} (\log
        R)^{\frac{d-2}{2}} (R\lor\log d)^{d-2} \sqrt{p_{\max}\lor R^{d-1}} }{ \lambda_{\min} }.
\end{align*}

The local convergence analysis in Theorem~\ref{thm_local} requires
\[
    \varepsilon_0
    \frac{\lambda_{\max}}{\lambda_{\min}}
    (d+R)
    \lesssim
    1,
\]
together with the coherence condition
\eqref{eq_assumption_2_general_R} and the signal-strength condition
\eqref{eq_assumption_3_general_R}.
The second condition of Corollary~\ref{corollary_global} ensures that
\[
    \sigma
    \frac{
        \sqrt{p_{\max}\lor R^{d-1}}
    }{
        \lambda_{\operatorname{Tucker}}
    }
    \frac{\lambda_{\max}}{\lambda_{\min}}
    (d+R)
    \lesssim
    1.
\]
The third condition ensures that
\begin{align*}
    &\sigma\kappa_T
    \left(
        \frac{\lambda_{\max}}{\lambda_{\min}}
    \right)^2
    \frac{
        C_{\delta_0}^d
        d^{3d-5}
        (d+R)
        R^{\frac{5d}{2}-4}
        (\log R)^{\frac{d-2}{2}}
        (R\lor\log d)^{d-2}
        \sqrt{p_{\max}\lor R^{d-1}}
    }{
        \lambda_{\min}
    }
    \lesssim
    1.
\end{align*}
Consequently, on the preceding TASD event,
condition~\eqref{eq_assumption_1_general_R} of
Theorem~\ref{thm_local} holds.

Moreover, since $d\geq3$, $R\geq2$, $\kappa_T\geq1$,
$C_{\delta_0}\geq1$, and
$\lambda_{\max}/\lambda_{\min}\geq1$, the third condition of
Corollary~\ref{corollary_global}, after increasing its implicit
absolute constant if necessary, also implies
\[
    \lambda_{\min}
    \geq
    C\sigma
    p_{\max}^{1/2}
    d^2
    \log d
\]
and
\[
    \lambda_{\min}
    \geq
    C
    \frac{\lambda_{\max}}{\lambda_{\min}}
    \sigma
    \sqrt{
        dp_{\max}\log d
    }
    (d+R).
\]
Thus, condition~\eqref{eq_assumption_3_general_R} holds.
The fourth condition of Corollary~\ref{corollary_global} implies
condition~\eqref{eq_assumption_2_general_R}.
Therefore, all assumptions needed for the first conclusion of
Theorem~\ref{thm_local} are satisfied.
For the second conclusion, the additional condition
\[
    \frac{\lambda_{\max}}{\lambda_{\min}}
    Rd\xi^{d-2}
    \leq
    c
\]
is imposed explicitly in the second item of
Corollary~\ref{corollary_global}.

The local ALS event in Theorem~\ref{thm_local} has failure probability
at most
\[
    CdR\exp(-cp_{\max}).
\]
Since every loading matrix $A_k\in\RR^{p_k\times R}$ has full column
rank in the setting of Theorem~\ref{thm_tasd}, we have
\[
    R\leq p_{\min}\leq p_{\max}.
\]
Consequently,
\[
    CdR\exp(-cp_{\max})
    \leq
    Cdp_{\max}\exp(-cp_{\max})
    \leq
    C'd\exp(-c'p_{\max})
\]
for some absolute constants $C',c'>0$. Since
$p_{\max}\geq p_{\min}$ and
\[
    p_{\min}
    \geq
    C\log\left(
        \frac{2d^2}{\delta}
    \right),
\]
the absolute constant $C$ can be chosen sufficiently large so that
\[
    CdR\exp(-cp_{\max})
    \leq
    \frac{\delta}{2}.
\]
Intersecting the TASD event with the local ALS event and applying a
union bound therefore gives total probability at least $1-\delta$.
The two iteration bounds and the final statistical error then follow
directly from Theorem~\ref{thm_local}.

\section{Lemmas}

\begin{Lemma}\label{lemma_vector_norm_quadratic_ine}
    For vectors $a,b$ with $\|a\| = \|b\|$, we have
    \[
        |a\T (b - a)| = \frac{1}{2}\|b-a\|^2
    \]
\end{Lemma}
\begin{proof}
    Note that
    \[
        0 = \|b\|^2 - \|a\|^2 = (b-a)\T(b+a) = 2a\T(b-a) + \|b-a\|^2,
    \]
    which implies
    \[
        |a\T (b - a)| = \frac{1}{2}\|b-a\|^2.
    \]
\end{proof}

\begin{Lemma}\label{lemma_kronecker_bound}
    For vectors with norm 1, we have
    \[
        \ln(\hat a_d^{(t)}\otimes \cdots \otimes \hat a_2^{(t)}) - (a_d\otimes \cdots \otimes a_2)\rn^2 = 2 \(1-\prod_{j = 2}^d \( 1
        - \frac{\|a_j - \hat a_j^{(t)}\|^2}{2} \)\).
    \]
\end{Lemma}

\begin{proof}
    \begin{align}
          & \ln(\hat a_d^{(t)}\otimes \cdots \otimes \hat a_2^{(t)}) - (a_d\otimes \cdots \otimes a_2)\rn^2 \notag           \\
        = & 2 - 2\left\langle
                        a_d\otimes\cdots\otimes a_2,
                        \hat a_d^{(t)}\otimes\cdots\otimes\hat a_2^{(t)}
                    \right\rangle \notag     \\
        = & 2 - 2\sum_{i_2, \ldots, i_{d}} (a_2)_{i_2}(\hat a_2^{(t)})_{i_2} \cdots (a_d)_{i_d}(\hat a_d^{(t)})_{i_d} \notag \\
        = & 2 - 2\prod_{j = 2}^d\(\sum_{i_j = 1}^{p_j} (a_j)_{i_j}(\hat a_j^{(t)})_{i_j}\) \notag                           \\
        = & 2 \(1-\prod_{j = 2}^d \left\langle a_j,\hat a_j^{(t)}\right\rangle\) \notag                                                 \\
        = & 2 \(1-\prod_{j = 2}^d \( 1 - \frac{\|a_j - \hat a_j^{(t)}\|^2}{2} \)\) \notag
    \end{align}
\end{proof}

\begin{Lemma}[Norm concentration for independent sub-Gaussian coordinates]
\label{lemma_subgaussian_vector_norm}
Let
\[
    V=(V_1,\ldots,V_p)^\top\in\RR^p
\]
have independent coordinates satisfying
\[
    \EE\exp(tV_i)
    \leq
    \exp\left(
        \frac{\sigma^2t^2}{2}
    \right),
    \qquad
    t\in\RR,\quad i\in[p].
\]
Then, for every $\rho\geq0$,
\[
    \PP\left(
        \|V\|
        >
        \sigma(\sqrt{p}+\rho)
    \right)
    \leq
    2\exp\left(
        -\frac{\rho^2}{2}
    \right).
\]
\end{Lemma}

\begin{proof}
Let $g\sim\mathcal{N}(0,1)$ be independent of $V$. For every
$0<\theta<(2\sigma^2)^{-1}$, the identity
\[
    \exp(\theta x^2)
    =
    \mathbb{E}_g
    \exp\left(
        \sqrt{2\theta}\,gx
    \right)
\]
and the sub-Gaussian moment-generating-function assumption imply
\[
\begin{split}
    \EE\exp(\theta V_i^2)
    &=
    \mathbb{E}_g
    \EE\exp\left(
        \sqrt{2\theta}\,gV_i
    \right)                                                     \\
    &\leq
    \mathbb{E}_g
    \exp\left(
        \sigma^2\theta g^2
    \right)                                                     \\
    &=
    (1-2\sigma^2\theta)^{-1/2}.
\end{split}
\]
Consequently, by independence,
\[
    \EE\exp\left(
        \theta\|V\|^2
    \right)
    \leq
    (1-2\sigma^2\theta)^{-p/2}.
\]
The standard Chernoff calculation for the chi-square
moment-generating function therefore gives, for every $x\geq0$,
\[
    \PP\left(
        \|V\|^2
        >
        \sigma^2
        \left(
            p+2\sqrt{px}+2x
        \right)
    \right)
    \leq
    \exp(-x).
\]
Since
\[
    p+2\sqrt{px}+2x
    \leq
    \left(
        \sqrt{p}+\sqrt{2x}
    \right)^2,
\]
taking $x=\rho^2/2$ yields
\[
    \PP\left(
        \|V\|
        >
        \sigma(\sqrt{p}+\rho)
    \right)
    \leq
    \exp\left(
        -\frac{\rho^2}{2}
    \right),
\]
which proves the stated result.
\end{proof}

\begin{Lemma}\label{lemma_e_net}
    There exists $\varepsilon-$net $\mathcal{N}_{k}$ for $a_k \in \RR^{p_k}, \|a_k\| = 1$ with cardinality $(1+2/\varepsilon)^{p_k}$
\end{Lemma}
\begin{proof}
    See \cite{zhang2018tensor}.
\end{proof}

\begin{Lemma}\label{lemma_norm_transfer_inequality}
    \[
        \|AB\|_{\alpha \rightarrow \beta} \leq \|A\|_{\gamma \rightarrow \beta} \|B\|_{\alpha \rightarrow \gamma}
    \]
\end{Lemma}
\begin{proof}
    By definition: $\|ABx\|_{\beta} \leq \|A\|_{\gamma \rightarrow \beta}\|Bx\|_{\gamma} \leq \|A\|_{\gamma \rightarrow \beta} \|B\|_{\alpha \rightarrow \gamma} \|x\|_\alpha$.
\end{proof}

\begin{Lemma}\label{lemma_generalized_inverse}
    \[
        A^\dagger = A\T (AA\T)^\dagger.
    \]
\end{Lemma}
\begin{proof}
    Let the SVD of $A$ be $A = U S V\T$ where $S$ is a $r\times r$ full-rank diagonal matrix.
    Then we have $A^\dagger = VS\i U\T = V S U\T (U S^2 U\T)^\dagger = A\T (A A\T)^\dagger.$
\end{proof}

\begin{Lemma}\label{lemma_maxZ_bound}
    For an i.i.d.\ sub-Gaussian tensor $\Z$ with variance proxy $1$, we have
    \begin{align*}
         & \PP\Bigg( \max_{\substack{ V_k\in\RR^{p_k\times r_k}                                                     \\
        \|V_k\|\leq1,\;k=2,\ldots,d }} \left\| \MM_1\left( \Z\times_{k=2}^dV_k^\top \right) \right\|                \\
         & \hspace{3cm}\lesssim \sqrt{p_1} + \sqrt{\prod_{k=2}^d r_k} + \sqrt{ \log(d) \sum_{k=2}^d p_kr_k } \Bigg) \\
         & \qquad\geq 1- C\exp\left( -c\log(d) \sum_{k=2}^d p_kr_k \right).
    \end{align*}
\end{Lemma}
\begin{proof}
    Set
    \[
        p_{-1}
        :=
        \prod_{k=2}^d p_k,
        \qquad
        r_{-1}
        :=
        \prod_{k=2}^d r_k,
        \qquad
        G
        :=
        \MM_1(\Z)
        \in
        \RR^{p_1\times p_{-1}}.
    \]
    For matrices
    $V_k\in\RR^{p_k\times r_k}$, $k=2,\ldots,d$, define
    \[
        K(V_2,\ldots,V_d)
        :=
        V_d\otimes\cdots\otimes V_2
        \in
        \RR^{p_{-1}\times r_{-1}}.
    \]
    Under the unfolding convention used in this paper,
    \[
        \MM_1\left(
            \Z\times_{k=2}^d V_k^\top
        \right)
        =
        G K(V_2,\ldots,V_d).
    \]
    Moreover, if $\|V_k\|\leq1$ for every $k=2,\ldots,d$, then
    \[
        \left\|
            K(V_2,\ldots,V_d)
        \right\|
        =
        \prod_{k=2}^d\|V_k\|
        \leq
        1.
    \]

    We first establish a bound for a fixed deterministic matrix
    $K\in\RR^{p_{-1}\times r_{-1}}$ satisfying $\|K\|\leq1$.
    Let $\mathcal{N}_x$ and $\mathcal{N}_y$ be $1/4$-nets of
    $\mathbb{S}^{p_1-1}$ and $\mathbb{S}^{r_{-1}-1}$, respectively,
    satisfying
    \[
        |\mathcal{N}_x|
        \leq
        9^{p_1},
        \qquad
        |\mathcal{N}_y|
        \leq
        9^{r_{-1}}.
    \]
    For fixed $x\in\mathbb{S}^{p_1-1}$ and
    $y\in\mathbb{S}^{r_{-1}-1}$, independence of the entries of $G$
    and the sub-Gaussian moment-generating-function bound imply that,
    for every $s\in\RR$,
    \begin{align*}
        \EE\exp\left(
            s x^\top GKy
        \right)
        &=
        \prod_{i=1}^{p_1}
        \prod_{j=1}^{p_{-1}}
        \EE\exp\left(
            s x_i(Ky)_j G_{ij}
        \right)                                                      \\
        &\leq
        \exp\left[
            \frac{s^2}{2}
            \sum_{i=1}^{p_1}
            \sum_{j=1}^{p_{-1}}
            x_i^2(Ky)_j^2
        \right]                                                      \\
        &=
        \exp\left(
            \frac{s^2}{2}
            \|x\|^2\|Ky\|^2
        \right)                                                      \\
        &\leq
        \exp\left(
            \frac{s^2}{2}
        \right).
    \end{align*}
    Consequently,
    \[
        \PP\left(
            |x^\top GKy|>u
        \right)
        \leq
        2\exp\left(
            -\frac{u^2}{2}
        \right).
    \]
    The standard two-sided net approximation for the spectral norm
    gives
    \[
        \|GK\|
        \leq
        2
        \max_{\substack{
            x\in\mathcal{N}_x\\
            y\in\mathcal{N}_y
        }}
        |x^\top GKy|.
    \]
    Therefore, a union bound over
    $\mathcal{N}_x\times\mathcal{N}_y$ shows that, for every $t\geq0$,
    \begin{equation}\label{eq_fixed_projected_noise_bound}
        \PP\left(
            \|GK\|
            >
            C\left(
                \sqrt{p_1}
                +
                \sqrt{r_{-1}}
                +
                t
            \right)
        \right)
        \leq
        2\exp(-ct^2),
    \end{equation}
    where $C,c>0$ are absolute constants.

    We now make the bound uniform over $V_2,\ldots,V_d$. For
    $k=2,\ldots,d$, let
    \[
        \mathcal{B}_k
        :=
        \left\{
            V\in\RR^{p_k\times r_k}
            :
            \|V\|\leq1
        \right\}.
    \]
    By a standard volumetric argument in the
    $p_kr_k$-dimensional normed space
    $(\RR^{p_k\times r_k},\|\cdot\|)$, there exists an
    $\varepsilon$-net
    $\mathcal{N}_k\subseteq\mathcal{B}_k$ such that
    \[
        |\mathcal{N}_k|
        \leq
        \left(
            1+\frac{2}{\varepsilon}
        \right)^{p_kr_k}.
    \]
    Define
    \[
        M
        :=
        \sup_{\substack{
            V_k\in\mathcal{B}_k\\
            k=2,\ldots,d
        }}
        \left\|
            G
            \left(
                V_d\otimes\cdots\otimes V_2
            \right)
        \right\|,
    \]
    and
    \[
        M_{\mathcal{N}}
        :=
        \max_{\substack{
            \widetilde V_k\in\mathcal{N}_k\\
            k=2,\ldots,d
        }}
        \left\|
            G
            \left(
                \widetilde V_d
                \otimes\cdots\otimes
                \widetilde V_2
            \right)
        \right\|.
    \]

    Fix arbitrary $V_k\in\mathcal{B}_k$ and choose
    $\widetilde V_k\in\mathcal{N}_k$ satisfying
    \[
        \|V_k-\widetilde V_k\|
        \leq
        \varepsilon,
        \qquad
        k=2,\ldots,d.
    \]
    The telescoping identity gives
    \begin{align*}
        &
        V_d\otimes\cdots\otimes V_2
        -
        \widetilde V_d\otimes\cdots\otimes\widetilde V_2             \\
        &\quad=
        \sum_{h=2}^d
        V_d\otimes\cdots\otimes V_{h+1}
        \otimes
        \left(
            V_h-\widetilde V_h
        \right)
        \otimes
        \widetilde V_{h-1}\otimes\cdots\otimes\widetilde V_2,
    \end{align*}
    where empty products are omitted. For every summand, if
    $V_h\neq\widetilde V_h$, then
    \[
        \frac{
            V_h-\widetilde V_h
        }{
            \|V_h-\widetilde V_h\|
        }
    \]
    has spectral norm one, while all the remaining factors have
    spectral norm at most one. Hence, by the definition of $M$,
    \begin{align*}
        &
        \left\|
            G
            \left[
                V_d\otimes\cdots\otimes V_{h+1}
                \otimes
                \left(
                    V_h-\widetilde V_h
                \right)
                \otimes
                \widetilde V_{h-1}
                \otimes\cdots\otimes
                \widetilde V_2
            \right]
        \right\|                                                     \\
        &\qquad\leq
        \|V_h-\widetilde V_h\|M
        \leq
        \varepsilon M.
    \end{align*}
    The same bound is immediate when
    $V_h=\widetilde V_h$. Therefore,
    \[
        M
        \leq
        M_{\mathcal{N}}
        +
        (d-1)\varepsilon M,
    \]
    and thus
    \begin{equation}\label{eq_projected_noise_net_extension}
        M
        \leq
        \frac{
            M_{\mathcal{N}}
        }{
            1-(d-1)\varepsilon
        }
    \end{equation}
    whenever $(d-1)\varepsilon<1$.

    Set
    \[
        S
        :=
        \sum_{k=2}^d p_kr_k.
    \]
    Applying \eqref{eq_fixed_projected_noise_bound} to every tuple
    $(\widetilde V_2,\ldots,\widetilde V_d)$ and taking a union bound
    yield
    \begin{align*}
        &
        \PP\left(
            M_{\mathcal{N}}
            >
            C\left(
                \sqrt{p_1}
                +
                \sqrt{r_{-1}}
                +
                t
            \right)
        \right)                                                      \\
        &\qquad\leq
        2
        \exp\left[
            S
            \log\left(
                1+\frac{2}{\varepsilon}
            \right)
            -
            ct^2
        \right].
    \end{align*}
    Take
    \[
        \varepsilon
        =
        \frac{1}{2(d-1)}
    \]
    and
    \[
        t
        =
        C_0
        \sqrt{
            S\log(4d)
        },
    \]
    where $C_0>0$ is a sufficiently large absolute constant. Then
    \[
        1-(d-1)\varepsilon
        =
        \frac12
    \]
    and
    \[
        \log\left(
            1+\frac{2}{\varepsilon}
        \right)
        =
        \log\left(
            1+4(d-1)
        \right)
        \lesssim
        \log d.
    \]
    Combining this with
    \eqref{eq_projected_noise_net_extension} gives
    \begin{align*}
        &
        \PP\Bigg(
            M
            \lesssim
            \sqrt{p_1}
            +
            \sqrt{
                \prod_{k=2}^d r_k
            }
            +
            \sqrt{
                \log(d)
                \sum_{k=2}^d p_kr_k
            }
        \Bigg)                                                       \\
        &\qquad\geq
        1-
        C\exp\left(
            -c\log(d)
            \sum_{k=2}^d p_kr_k
        \right).
    \end{align*}
    This proves the result.
\end{proof}

\stopcontents[supplement]

\end{document}